\documentclass{llncs}
\usepackage{jb}
\usepackage{epsf}
\usepackage{graphicx}

\newcommand{\bi}{\begin{itemize}}
\newcommand{\ei}{\end{itemize}}
\newcommand{\be}{\begin{enumerate}}
\newcommand{\ee}{\end{enumerate}}
\newcommand{\bd}{\begin{description}}
\newcommand{\ed}{\end{description}}
\newcommand{\tab}{\hspace*{1cm}}

\newcommand{\mca}[2]{\multicolumn{#1}{@{}l@{}}{#2}}
\renewcommand{\.}[1]{\!#1\!}

\newcommand{\tpl}[1]{\langle #1 \rangle}

\newcommand{\ra}{\rightarrow}
\newcommand{\la}{\leftarrow}
\newcommand{\lra}{\leftrightarrow}

\newcommand{\Ra}{\Rightarrow}
\newcommand{\La}{\Leftarrow}
\newcommand{\Lra}{\Leftrightarrow}

\newcommand{\Lraa}{\Longleftrightarrow}

\newcommand{\scs}{\scriptstyle}
\newcommand{\Da}{\Downarrow}
\newcommand{\Ua}{\Uparrow}
\newcommand{\Ma}{\raisebox{-0.10cm}{$\scs \Ua$}}

\renewcommand{\leq}[0]{\leqslant}
\renewcommand{\geq}[0]{\geqslant}

\newcommand{\abbr}[1]{{\rm (#1)}}

\usepackage{amssymb}

\newcommand{\sysyfos}{{\sc Sysyfos}}
\newcommand{\prolog}{{\sc Prolog}}
\newcommand{\korso}{{\sc Korso}}
\newcommand{\lustre}{{\sc Lustre}}
\newcommand{\ttl}{{\sc Ttl}}
\newcommand{\bmft}{{\sc Bmft}}
\newcommand{\fzi}{{\sc Fzi}}
\newcommand{\web}{{\sc Web}}

\newcommand{\refC}{C}

\renewcommand{\arraystretch}{1.1}

\newcommand{\3}{\ss}

\begin{document}

\titlepage

\title{Formale Entwicklung einer Steuerung f\"ur eine Fertigungszelle
	mit SYSYFOS
}

\author{Jochen Burghardt}

\institute{GMD Berlin
\\
jochen@first.gmd.de
\\
http://www.first.gmd.de/persons/Burghardt.Jochen.html
}

\maketitle

\vfill

{\sf

\begin{tabular}[t]{@{}ll@{\hspace*{0.5cm}}l@{}}
Technical Report	\\
Arbeitspapiere der GMD 996	\\
June 1996	\\
ISSN 0723--0508 \\
\end{tabular}
\hfill
\begin{tabular}[t]{@{}ll@{\hspace*{0.5cm}}l@{}}
\mca{2}{GMD -- Forschungszentrum}	\\
\mca{2}{Informationstechnik GmbH}	\\
\mca{2}{D--53754 Sankt Augustin}	\\
Tel. & *49--2241--14--0 \\
Fax & *49--2241--14--2618	\\
Telex & 889469 gmd d	\\
\mca{2}{http://www.gmd.de}	\\
\end{tabular}

}

\newpage

\setcounter{page}{3}

\begin{abstract}
Using the synthesis approach of Manna and Waldinger, a formally
specified and verified control circuitery for a production cell
was developped. 
Building an appropriate formal language level, we could
achieve a requirements specification to the informal
description. 
We demonstrated that the paradigm of deductive synthesis can be
applied to the development of complete verified systems,
including hardware and mechanics. 
We defined two domain--specific logical operators that schematise
frequent patterns in specification and proof and hence allow a
more concise and expressive presentation. 

In \cite{Burghardt.1995a}, an english short version of this
paper, without appendices, can be found. 
\end{abstract}

\vspace*{2cm}

%\GERMAN

\begin{abstract}
Mit Hilfe der deduktiven Programmsynthese nach Manna und
Wal\-din\-ger wird eine formal spezifizierte und verifizierte
Steuerung f\"ur eine Fertigungszelle entwickelt. 
Durch die Erstellung einer geeigneten formalen Sprachebene unter
starker Ausnutzung impliziter Spezifikationstechnik wird
erreicht, da\3 die formale Anforderungsspezifikation einer
satzweisen ``\"Ubersetzung'' der informellen Beschreibung
entspricht. 
Es wird demonstriert, wie das Paradigma der deduktiven Synthese
zur Entwicklung ganzer verifizierter Systeme,
einschlie\3lich Hardware und Mechanik, angewendet werden kann. 
Es werden zwei anwendungsspezifische logische Operatoren
definiert, die eine Schematisierung der in Spezifikation und
Beweis h\"aufig vorkommenden Aussagenmuster darstellen und mit
deren Hilfe sich beide k\"urzer und klarer darstellen lassen. 

In \cite{Burghardt.1995a} findet sich eine englische Kurzfassung dieses
Papiers (ohne Anh\"ange).
\end{abstract}

%\newpage
%
%\tableofcontents
%
%
%
%
%
%
%\clearpage
%
%\listoffigures

\newpage

$\;$

\newpage

\section{Einleitung}
\label{Einleitung}

Die Fallstudie ``Fertigungszelle'' kommt
aus dem Bereich Regeln und Steuern. Aufgabe ist die
Entwicklung verifizierter Steuerungs--Software f\"ur ein
Modell einer Anlage, wie sie bei einer
metallverarbeitenden Firma in Karlsruhe steht.
Sie bearbeitet Metallrohlinge, die auf einem
Zu\-f\"uhr\-f\"or\-der\-band an eine Presse
gelangen. Ein Roboter nimmt die
Metallrohlinge vom Zu\-f\"uhr\-f\"or\-der\-band und legt sie in
die Presse. Der Roboterarm verl\"a\3t die Presse, die Presse
verarbeitet die Metallrohlinge und \"offnet sich wieder. Der
Roboter nimmt das verarbeitete Metallpl\"attchen aus der Presse
und legt es auf ein Ablagef\"orderband.

Ziel der Bearbeitung der Fallstudie war vor allem zu
untersuchen, wie weit der begriffliche Abstand zwischen
vorgegebener informeller Anforderungsbeschreibung und formaler
Spezifikation verringert werden kann. Durch die Erstellung einer
geeigneten formalen Sprachebene unter starker Ausnutzung
impliziter Spezifikationstechnik und unter Einbeziehung auch
mechanischer Aspekte konnte erreicht werden, da\3 die
Spezifikation lokal validierbar ist, d.h.\ durch eine
satzweise ``\"Ubersetzung'' der Anforderungsbeschreibung in
formale Notation entstanden ist. Insbesondere ist das
Spezifikationsziel die formale Entsprechung der Anforderung:
``Wenn ein unbearbeitetes Werkst\"uck auf dem
Zuf\"uhrf\"orderband liegt, erscheint es sp\"ater bearbeitet auf
dem Ablagef\"orderband''. Mit dieser Herangehensweise kann das
Problem der Vertrauensw\"urdigkeit einer Spezifikation
weitgehend entsch\"arft werden.

Die formale Spezifikation zeichnet sich durch folgende Merkmale
aus:
\bi
\item Spezifikation in Pr\"adikatenlogik 1.\ Stufe,
\item modular gegliedert,
\item Zeit als expliziter Parameter,
\item Einbeziehung auch mechanischer und geometrischer Aspekte,
\item Beschreibung der einzelnen Maschinen und ihrer Anordnung,
	\\
	nicht: Programmierung der Steuerung in einer
	Spezifikationssprache.
\ei
Daraus wurden mit Hilfe des Unterst\"utzungswerkzeugs
``{\sysyfos}'' notwendige Zeitbedingungen f\"ur die Steuerung
der Fertigungszelle hergeleitet und schlie\3lich eine
{\ttl}--artige digitale Hardware--Steuerschaltung konstruiert, die
die Zeitbedingungen erf\"ullt. 

Durch die Einbeziehung auch mechanischer und geometrischer
Aspekte in die Modellierung konnte der Ansatz der deduktiven
Synthese von der reinen Steuerungsentwicklung zu einem
methodischen Rahmen f\"ur die integrierte Bearbeitung aller beim
Entwurf der Fertigungszelle anfallenden ingenieurtechnischen
Aspekte ausgeweitet werden. Zum Beispiel wurden auch notwendige
Bedingungen an die Aufstellung der Maschinen (Abst\"ande,
Winkel, usw.) aus der Spezifikation hergeleitet.

Aufgrund der w\"ahrend der Entwicklung
gemachten Erfahrungen wurden im Nachhinein zwei
logische Operatoren definiert, die eine Schematisierung der in
Spezifikation und Beweis h\"aufig vorkommenden Aussagenmuster
darstellen und mit deren Hilfe sich beide
erheblich k\"urzer und klarer darstellen lassen. Beide
Operatoren sind monoton bzw.\ anti--monoton in jeweils beiden
Pr\"adikatargumenten und konnten daher problemlos als
benutzerdefinierte Junktoren in die verwendete
polarit\"atsbasierte Nicht--Klausel--Resolution eingebaut werden.
Zusammen mit einer Hintergrundtheorie \"uber
ihre wichtigsten Eigenschaften bilden sie eine Grundlage f\"ur
die Bearbeitung zustandsorientierter Aspekte auf einem
anwendungsnahen sprachlichen Niveau.

\vspace{0.5cm}

Nach einer informellen Aufgabenbeschreibung der Fertigungszelle in
Abschnitt~\ref{Die Fallstudie ``Fertigungszelle''}
und einem kurzen \"Uberblick \"uber die Methode der deduktiven
Programmsynthese und das Unterst\"utzungswerkzeug {\sysyfos}
in Abschnitt~\ref{Deduktive Programmsynthese}
werden das Vorgehen und die Erfahrungen beim Entwurf der
Steuerungsschaltung in Abschnitt~\ref{Entwurf} diskutiert.
In Abschnitt~\ref{Entwicklung der Robotersteuerung}
wird das Vorgehen anhand der Entwicklung der Robotersteuerung
konkretisiert, ein typischer Teil des maschinengef\"uhrten formalen
Beweises ist in Anhang~\refC\ wiedergegeben.
Abschnitt~\ref{Verwendung hoherer logischer Operatoren} geht auf die 
Verwendung h\"oherer logischer Operatoren ein.
In Abschnitt~\ref{Bewertung}
erfolgt eine Bewertung der Fallstudie nach verschiedenen
Gesichtspunkten.

\begin{figure}
%\begin{center}
%\epsfbox{ap996_Bild1.epsf}
\includegraphics[scale=0.7]{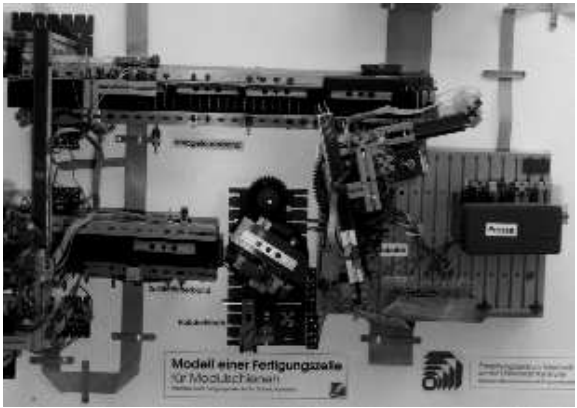}
%\end{center}
\caption{{\fzi}--Spielzeugmodell der Fertigungszelle}
\label{FZI--Spielzeugmodell der Fertigungszelle}
\end{figure}

\section{Die Fallstudie ``Fertigungszelle''}
\label{Die Fallstudie ``Fertigungszelle''}

Die Fallstudie ``Fertigungszelle''
	\footnote{Dieser Abschnitt
        entspricht der deutschen Fassung des zweiten Kapitels
        von \cite{Lewerentz.Lindner.1995} und wurde mir
        freundlicherweise von Thomas Lindner zur Verf\"ugung
        gestellt.}
ist eine Fallstudie
aus dem Bereich Regeln und Steuern. Aufgabe ist die
Entwicklung von verifizierter Steuerungs--Software f\"ur ein
Modell einer Anlage, wie sie bei einer
metallverarbeitenden Firma in Karlsruhe steht.
Am Forschungszetrum Informatik ({\fzi}) in Karlsruhe wurde ein
Spielzeugmodell der Fertigungszelle mit Fischer-Technik
realisiert, das \"uber eine RS232--Schnittstelle angesteuert
werden kann (Abb.~\ref{FZI--Spielzeugmodell der
Fertigungszelle}). 

Die Fertigungszelle (vgl.\ Abb.~\ref{Schematische Darstellung
der Fertigungszelle}) bearbeitet Metallrohlinge, die auf einem
Zu\-f\"uhr\-f\"or\-der\-band \abbr{zfb} an eine Presse
\abbr{prs} gelangen. Ein Roboter \abbr{rob} nimmt die
Metallrohlinge vom Zu\-f\"uhr\-f\"or\-der\-band und legt sie in
die Presse. Der Roboterarm verl\"a\3t die Presse, die Presse
verarbeitet die Metallrohlinge und \"offnet sich wieder. Der
Roboter nimmt das verarbeitete Metallpl\"attchen aus der Presse
und legt es auf ein Ablagef\"orderband \abbr{afb}. 

Dieser grundlegende Ablauf wird durch weitere Mechanismen kompliziert:
\bi
\item Um eine bessere Auslastung der Presse zu erzielen,
	wurde der Roboter mit zwei Armen ausgestattet. So kann
	der zweite Arm w\"ahrend des Pressens bereits einen neuen
	Rohling aufnehmen.
	Beide Arme stehen im unver\"anderlichen Winkel von 90$^\circ$
	zueinander und k\"onnen gemeinsam gedreht werden.
	Jeder der Arme kann horizontal ein- und ausgefahren werden.
	Die horizontale Beweglichkeit ist wegen der
	unterschiedlichen
	Abst\"ande zum Drehzentrum des Roboters beim Be- und Entladen
	n\"otig.
\item Die beiden Roboterarme befinden sich nicht auf der
	gleichen H\"ohe. Au\3erdem sind sie nicht vertikal
	beweglich. Deshalb wurde im Anschlu\3 an das
	Zuf\"uhrf\"orderband ein Hubdrehtisch \abbr{hub} eingef\"ugt.
	Er hat
	die Aufgabe, die Metallpl\"attchen auf die H\"ohe des ersten
	Roboterarmes anzuheben und um
	etwa 45$^\circ$ zu drehen, so da\3 sie im richtigen Winkel von
	ihm aufgenommen und in die Presse gelegt werden
	k\"onnen.
	Der Greifer des Roboterarms ist selbst nicht drehbar.
\item Ebenfalls wegen des unterschiedlichen Niveaus der
	Roboterarme hat die Presse nicht nur zwei, sondern drei
	Zust\"ande: ge\"offnet zur Entladung durch den unteren Arm (2),
	ge\"offnet zur Beladung durch den oberen Arm (1), geschlossen
	(pressend).
\item Um das Modell bei Demonstrationen einsetzen zu
	k\"onnen, soll der Fertigungsvorgang
	be\-die\-nungs\-un\-ab\-h\"an\-gig
	ablaufen k\"onnen. Aus diesem Grund werden die
	``verarbeiteten'' Metallplatten (von der Modell--Presse
	unver\"andert gelassen) vom Ablagef\"orderband durch ein
	Handhabungsger\"at \abbr{han} wieder zum Zuf\"uhrf\"orderband
	gebracht
	und der Gesamtvorgang dadurch zyklisch gemacht.
	Das Handhabungsger\"at besitzt einen als Elektromagnet
	realisierten Greifer, der horizontal und vertikal
	beweglich ist. Die horizontale Beweglichkeit dient zur
	Bew\"altigung der Strecken zwischen den beiden
	F\"orderb\"andern, die vertikale Beweglichkeit ist
	notwendig, da die B\"ander unterschiedlich hoch sind.
\ei

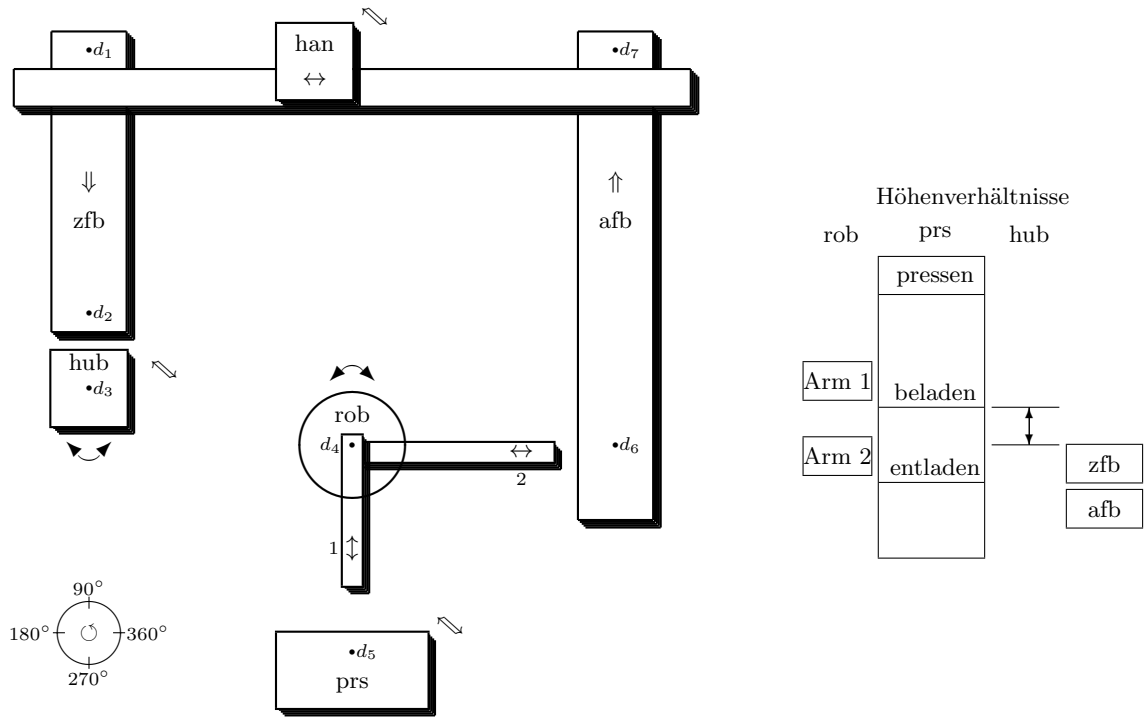
\begin{figure}
\begin{center}
\begin{picture}(15,9.4)
	%\put(0,0){\makebox(0,0){+}}
%
\put(12.750,6.700){\makebox(0.000,0.000)[b]{H\"ohenverh\"altnisse}}
\put(12.250,6.200){\makebox(0.000,0.000)[b]{\small prs}}
	\put(11.500,2.000){\framebox(1.400,4.000){}}
	\put(11.500,5.500){\line(1,0){1.400}}
	\put(12.250,5.600){\makebox(0.000,0.000)[b]{\small pressen}}
	\put(11.500,4.000){\line(1,0){1.400}}
	\put(12.250,4.100){\makebox(0.000,0.000)[b]{\small beladen}}
	\put(11.500,3.000){\line(1,0){1.400}}
	\put(12.250,3.100){\makebox(0.000,0.000)[b]{\small entladen}}
\put(11.000,6.200){\makebox(0.000,0.000)[b]{\small rob}}
	\put(10.500,4.100){\framebox(0.900,0.500){\small Arm 1}}
	\put(10.500,3.100){\framebox(0.900,0.500){\small Arm 2}}
\put(13.500,6.200){\makebox(0.000,0.000)[b]{\small hub}}
	\put(13.000,4.000){\line(1,0){0.900}}
	\put(13.000,3.500){\line(1,0){0.900}}
	\put(13.500,3.500){\vector(0,1){0.500}}
	\put(13.500,4.000){\vector(0,-1){0.500}}
\put(14.000,3.000){\framebox(1.000,0.500){\small zfb}}
\put(14.000,2.400){\framebox(1.000,0.500){\small afb}}
\put(1.000,1.000){\circle{0.800}}
\put(1.000,1.000){\makebox(0,0){$\circlearrowleft$}}
\put(1.320,1.000){\line(1,0){0.160}}
\put(0.520,1.000){\line(1,0){0.160}}
\put(1.000,1.320){\line(0,1){0.160}}
\put(1.000,0.520){\line(0,1){0.160}}
\put(1.5,1){\makebox(0,0)[l]{\scriptsize $360^\circ$}}
\put(0.5,1){\makebox(0,0)[r]{\scriptsize $180^\circ$}}
\put(1,1.5){\makebox(0,0)[b]{\scriptsize $90^\circ$}}
\put(1,0.5){\makebox(0,0)[t]{\scriptsize $270^\circ$}}
\thicklines
\put(0.5,5){\line(1,0){1}}				% zfb
	\put(0.5,9){\line(1,0){1}}
	\put(0.5,5){\line(0,1){3}}
	\put(1.5,5){\line(0,1){3}}
	\put(0.5,8.5){\line(0,1){0.5}}
	\put(1.5,8.5){\line(0,1){0.5}}
	\put(1,6.5){\makebox(0,0){zfb}}
	\put(1,7){\makebox(0,0){$\Downarrow$}}
	\put(1,8.75){\circle*{0.07}}
	\put(1.050,8.750){\makebox(0.000,0.000)[l]{\scriptsize $d_1$}}
	\put(1,5.25){\circle*{0.07}}
	\put(1.050,5.250){\makebox(0.000,0.000)[l]{\scriptsize $d_2$}}
\put(0.500,3.750){\framebox(1.000,1.000){}}		% hub
	\put(1.000,4.50){\makebox(0.000,0.000)[b]{hub}}
	\put(1.0,4.25){\circle*{0.07}}
	\put(1.050,4.250){\makebox(0.000,0.000)[l]{\scriptsize $d_3$}}
	\put(1.000,3.300){\makebox(0.000,0.000){$\smile$}}
	\put(1.100,3.350){\vector(1,1){0.200}}
	\put(0.900,3.350){\vector(-1,1){0.200}}
	\put(2.000,4.500){\makebox(0,0)
			{{\scriptsize $\nwarrow\!\!\!\!\searrow$}}}
\put(4.5,3.5){\circle{2}}				% rob
	\put(4.5,3.5){\circle*{0.07}}
	\put(4.350,3.500){\makebox(0.000,0.000)[r]{\scriptsize $d_4$}}
	\put(4.5,3.8){\makebox(0,0)[b]{rob}}
	\put(4.5,4.5){\makebox(0,0){$\frown$}}
	\put(4.6,4.55){\vector(1,-1){0.2}}
	\put(4.4,4.55){\vector(-1,-1){0.2}}
	\put(4.375,1.625){\framebox(0.250,2.000){}}	% rob arm1
	\put(4.325,2.125){\makebox(0.000,0.000)[r]{{\scriptsize 1}}}
	\put(4.500,2.125){\makebox(0.000,0.000){$\updownarrow$}}
	\put(4.675,3.275){\framebox(2.500,0.250){}}	% rob arm2
	\put(6.750,3.125){\makebox(0.000,0.000)[t]{{\scriptsize 2}}}
	\put(6.750,3.400){\makebox(0.000,0.000){$\leftrightarrow$}}
\put(3.5,0){\framebox(2.0,1.0){}}			% prs
	\put(4.5,0.2){\makebox(0,0)[b]{prs}}
	\put(4.5,0.75){\circle*{0.07}}
	\put(4.550,0.750){\makebox(0.000,0.000)[l]{\scriptsize $d_5$}}
	\put(5.800,1.100){\makebox(0,0)
			{{\scriptsize $\nwarrow\!\!\!\!\searrow$}}}
\put(7.5,2.5){\line(1,0){1}}				% afb
	\put(7.5,9.0){\line(1,0){1}}
	\put(8.5,2.5){\line(0,1){5.5}}
	\put(8.5,8.5){\line(0,1){0.5}}
	\put(7.5,2.5){\line(0,1){5.4}}
	\put(7.500,8.500){\line(0,1){0.500}}
	\put(8,6.5){\makebox(0,0){afb}}
	\put(8,7){\makebox(0,0){$\Uparrow$}}
	\put(8.0,3.5){\circle*{0.07}}
	\put(8.050,3.500){\makebox(0.000,0.000)[l]{\scriptsize $d_6$}}
	\put(8.0,8.75){\circle*{0.07}}
	\put(8.050,8.750){\makebox(0.000,0.000)[l]{\scriptsize $d_7$}}
\put(0.0,8.0){\line(1,0){9}}				% han
	\put(0.0,8.5){\line(1,0){3.5}}
	\put(4.5,8.5){\line(1,0){4.5}}
	\put(0.0,8.0){\line(0,1){0.5}}
	\put(9.0,8.0){\line(0,1){0.5}}
	\put(4,8.85){\makebox(0,0){han}}
	\put(3.500,8.100){\framebox(1.000,1.000){}}
	\put(4,8.35){\makebox(0,0){$\leftrightarrow$}}
	\put(4.800,9.200){\makebox(0,0)
			{{\scriptsize $\nwarrow\!\!\!\!\searrow$}}}
\put(0.525,4.975){\line(1,0){1.000}}			% zfb
	\put(1.525,4.975){\line(0,1){3.000}}
	\put(1.525,8.500){\line(0,1){0.475}}
\put(0.525,3.725){\line(1,0){1.000}}			% hub
	\put(1.525,3.725){\line(0,1){1.000}}
\put(4.400,1.600){\line(1,0){0.250}}			% rob arm1
	\put(4.650,1.600){\line(0,1){2.000}}
\put(4.650,3.250){\line(1,0){2.550}}			% rob arm2
	\put(7.200,3.250){\line(0,1){0.250}}
\put(3.525,-0.025){\line(1,0){2.000}}			% prs
	\put(5.525,-0.025){\line(0,1){1.000}}
\put(7.525,2.475){\line(1,0){1.000}}			% afb
	\put(8.525,2.475){\line(0,1){5.500}}
	\put(8.525,8.500){\line(0,1){0.475}}
\put(0.025,7.975){\line(1,0){9.000}}			% han
	\put(9.025,7.975){\line(0,1){0.500}}
	\put(3.525,8.075){\line(1,0){1.000}}
	\put(4.525,8.075){\line(0,1){1.000}}
\put(0.550,4.950){\line(1,0){1.000}}			% zfb
	\put(1.550,4.950){\line(0,1){3.000}}
	\put(1.550,8.500){\line(0,1){0.450}}
\put(0.550,3.700){\line(1,0){1.000}}			% hub
	\put(1.550,3.700){\line(0,1){1.000}}
\put(4.425,1.575){\line(1,0){0.250}}			% rob arm1
	\put(4.675,1.575){\line(0,1){2.000}}
\put(4.675,3.225){\line(1,0){2.550}}			% rob arm2
	\put(7.225,3.225){\line(0,1){0.250}}
\put(3.550,-0.050){\line(1,0){2.000}}			% prs
	\put(5.550,-0.050){\line(0,1){1.000}}
\put(7.550,2.450){\line(1,0){1.000}}			% afb
	\put(8.550,2.450){\line(0,1){5.500}}
	\put(8.550,8.500){\line(0,1){0.450}}
\put(0.050,7.950){\line(1,0){9.000}}			% han
	\put(9.050,7.950){\line(0,1){0.500}}
	\put(3.550,8.050){\line(1,0){1.000}}
	\put(4.550,8.050){\line(0,1){1.000}}
\put(0.575,4.925){\line(1,0){1.000}}			% zfb
	\put(1.575,4.925){\line(0,1){3.000}}
	\put(1.575,8.500){\line(0,1){0.425}}
\put(0.575,3.675){\line(1,0){1.000}}			% hub
	\put(1.575,3.675){\line(0,1){1.000}}
\put(4.450,1.550){\line(1,0){0.250}}			% rob arm1
	\put(4.700,1.550){\line(0,1){2.000}}
\put(4.650,3.200){\line(1,0){2.600}}			% rob arm2
	\put(7.250,3.200){\line(0,1){0.250}}
\put(3.575,-0.075){\line(1,0){2.000}}			% prs
	\put(5.575,-0.075){\line(0,1){1.000}}
\put(7.575,2.425){\line(1,0){1.000}}			% afb
	\put(8.575,2.425){\line(0,1){5.500}}
	\put(8.575,8.500){\line(0,1){0.425}}
\put(0.075,7.925){\line(1,0){9.000}}			% han
	\put(9.075,7.925){\line(0,1){0.500}}
	\put(3.575,8.025){\line(1,0){1.000}}
	\put(4.575,8.025){\line(0,1){1.000}}
\put(0.600,4.900){\line(1,0){1.000}}			% zfb
	\put(1.600,4.900){\line(0,1){3.000}}
	\put(1.600,8.500){\line(0,1){0.400}}
\put(0.600,3.650){\line(1,0){1.000}}			% hub
	\put(1.600,3.650){\line(0,1){1.000}}
\put(4.475,1.525){\line(1,0){0.250}}			% rob arm1
	\put(4.725,1.525){\line(0,1){2.000}}
\put(4.675,3.175){\line(1,0){2.600}}			% rob arm2
	\put(7.275,3.175){\line(0,1){0.250}}
\put(3.600,-0.100){\line(1,0){2.000}}			% prs
	\put(5.600,-0.100){\line(0,1){1.000}}
\put(7.600,2.400){\line(1,0){1.000}}			% afb
	\put(8.600,2.400){\line(0,1){5.500}}
	\put(8.600,8.500){\line(0,1){0.400}}
\put(0.100,7.900){\line(1,0){9.000}}			% han
	\put(9.100,7.900){\line(0,1){0.500}}
	\put(3.600,8.000){\line(1,0){1.000}}
	\put(4.600,8.000){\line(0,1){1.000}}
%
% \put(0.5,5){\line(1,0){1}}				% zfb
% 	\put(1.5,5){\line(0,1){3}}
% 	\put(1.5,8.5){\line(0,1){0.5}}
% \put(0.500,3.750){\line(1,0){1.000}}			% hub
% 	\put(1.500,3.750){\line(0,1){1.000}}
% \put(4.375,1.625){\line(1,0){0.250}}			% rob arm1
% 	\put(4.625,1.625){\line(0,1){2.000}}
% \put(4.575,3.275){\line(1,0){2.600}}			% rob arm2
% 	\put(7.175,3.275){\line(0,1){0.250}}
% \put(3.500,0.000){\line(1,0){2.000}}			% prs
% 	\put(5.500,0.000){\line(0,1){1.000}}
% \put(7.5,2.5){\line(1,0){1}}				% afb
% 	\put(8.5,2.5){\line(0,1){5.5}}
% 	\put(8.5,8.5){\line(0,1){0.5}}
% \put(7.500,2.500){\line(1,0){1.000}}			% afb
% 	\put(8.500,2.500){\line(0,1){6.500}}
% \put(0.000,8.000){\line(1,0){9.000}}			% han
% 	\put(9.000,8.000){\line(0,1){0.500}}
% 	\put(3.500,8.100){\line(1,0){1.000}}
% 	\put(4.500,8.100){\line(0,1){1.000}}
\end{picture}
\caption{Schematische Darstellung der Fertigungszelle}
\label{Schematische Darstellung der Fertigungszelle}
\end{center}
\end{figure}

Der generelle Ablauf ist (sequentialisiert aus der Sicht
eines Metallpl\"attchens):
\bi
\item \"Uber das Zuf\"uhrf\"orderband gelangt das
	Metallpl\"attchen auf den 
	Hubdrehtisch.
\item Der Hubdrehtisch wird in die dem aufnehmenden ersten Roboterarm
	angemessene Position gebracht.
\item Der erste Roboterarm nimmt das Metallpl\"attchen auf.
\item Der Roboter dreht sich, so da\3 Arm 1 in die ge\"offnete Presse
        zeigt, legt das Metallpl\"attchen dort ab und verl\"a\3t die
        Presse.
\item Die Presse verarbeitet das Metallpl\"attchen und \"offnet sich
	wieder.
\item Der Roboter nimmt mit seinem zweiten Arm das
	Metallpl\"attchen auf, 
        dreht sich weiter und legt das Metallpl\"attchen auf das
        Ablagef\"orderband.
\item \"Uber das Ablagef\"orderband gelangt das Metallpl\"attchen zum
	Handhabungsger\"at.
\item Das Handhabungsger\"at nimmt das Metallpl\"attchen auf, f\"ahrt
        zum Zuf\"uhrf\"orderband und legt das Metallpl\"attchen dort
        wieder ab.
\ei

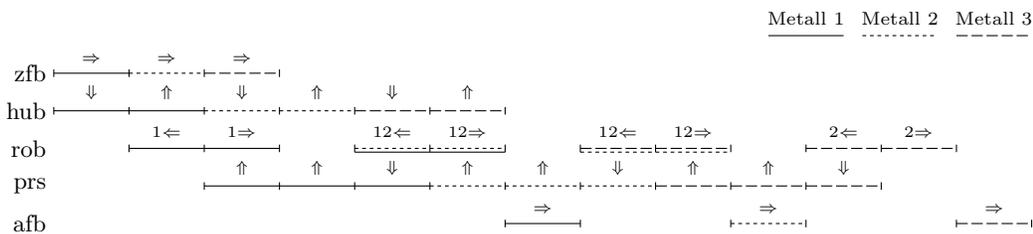
\begin{figure}
\begin{center}
\begin{picture}(13.7,3.1)
	%\put(0,0){\makebox(0,0){+}}
%
\put(0.500,2.200){\makebox(0.000,0.000)[r]{zfb}}
\put(0.500,1.700){\makebox(0.000,0.000)[r]{hub}}
\put(0.500,1.200){\makebox(0.000,0.000)[r]{rob}}
\put(0.500,0.700){\makebox(0.000,0.000)[r]{prs}}
\put(0.500,0.200){\makebox(0.000,0.000)[r]{afb}}
\put(10.600,2.850){\makebox(0.000,0.000)[b]{{\scriptsize Metall 1}}}
\put(11.850,2.850){\makebox(0.000,0.000)[b]{{\scriptsize Metall 2}}}
\put(13.100,2.850){\makebox(0.000,0.000)[b]{{\scriptsize Metall 3}}}
\put(10.100,2.700){\line(1,0){1.000}}
\multiput(11.350,2.700)(0.100,0.000){10}{\line(1,0){0.050}}
\multiput(12.600,2.700)(0.200,0.000){5}{\line(1,0){0.150}}
	\put(0.600,2.150){\line(0,1){0.100}}	% efb
\put(0.600,2.200){\line(1,0){1.000}}
	\put(1.100,2.350){\makebox(0.000,0.000)[b]{$\scs \Ra$}}
	\put(1.600,2.150){\line(0,1){0.100}}
\multiput(1.600,2.200)(0.100,0.000){10}{\line(1,0){0.050}}
	\put(2.100,2.350){\makebox(0.000,0.000)[b]{$\scs \Ra$}}
	\put(2.600,2.150){\line(0,1){0.100}}
\multiput(2.600,2.200)(0.200,0.000){5}{\line(1,0){0.150}}
	\put(3.100,2.350){\makebox(0.000,0.000)[b]{$\scs \Ra$}}
	\put(3.600,2.150){\line(0,1){0.100}}
	\put(0.600,1.650){\line(0,1){0.100}}	% hdt
\put(0.600,1.700){\line(1,0){1.000}}
	\put(1.100,1.850){\makebox(0.000,0.000)[b]{$\scs \Da$}}
	\put(1.600,1.650){\line(0,1){0.100}}
\put(1.600,1.700){\line(1,0){1.000}}
	\put(2.100,1.850){\makebox(0.000,0.000)[b]{$\scs \Ua$}}
	\put(2.600,1.650){\line(0,1){0.100}}
\multiput(2.600,1.700)(0.100,0.000){10}{\line(1,0){0.050}}
	\put(3.100,1.850){\makebox(0.000,0.000)[b]{$\scs \Da$}}
	\put(3.600,1.650){\line(0,1){0.100}}
\multiput(3.600,1.700)(0.100,0.000){10}{\line(1,0){0.050}}
	\put(4.100,1.850){\makebox(0.000,0.000)[b]{$\scs \Ua$}}
	\put(4.600,1.650){\line(0,1){0.100}}
\multiput(4.600,1.700)(0.200,0.000){5}{\line(1,0){0.150}}
	\put(5.100,1.850){\makebox(0.000,0.000)[b]{$\scs \Da$}}
	\put(5.600,1.650){\line(0,1){0.100}}
\multiput(5.600,1.700)(0.200,0.000){5}{\line(1,0){0.150}}
	\put(6.100,1.850){\makebox(0.000,0.000)[b]{$\scs \Ua$}}
	\put(6.600,1.650){\line(0,1){0.100}}
	\put(1.600,1.150){\line(0,1){0.100}}	% rob
\put(1.600,1.200){\line(1,0){1.000}}
	\put(2.100,1.350){\makebox(0.000,0.000)[b]{$\scs 1 \La$}}
	\put(2.600,1.150){\line(0,1){0.100}}
\put(2.600,1.200){\line(1,0){1.000}}
	\put(3.100,1.350){\makebox(0.000,0.000)[b]{$\scs 1 \Ra$}}
	\put(3.600,1.150){\line(0,1){0.100}}
	\put(4.600,1.150){\line(0,1){0.100}}
\put(4.600,1.150){\line(1,0){1.000}}
\multiput(4.600,1.200)(0.100,0.000){10}{\line(1,0){0.050}}
	\put(5.100,1.350){\makebox(0.000,0.000)[b]{$\scs 12 \La$}}
	\put(5.600,1.150){\line(0,1){0.100}}
\put(5.600,1.150){\line(1,0){1.000}}
\multiput(5.600,1.200)(0.100,0.000){10}{\line(1,0){0.050}}
	\put(6.100,1.350){\makebox(0.000,0.000)[b]{$\scs 12 \Ra$}}
	\put(6.600,1.150){\line(0,1){0.100}}
	\put(7.600,1.150){\line(0,1){0.100}}
\multiput(7.600,1.150)(0.100,0.000){10}{\line(1,0){0.050}}
\multiput(7.600,1.200)(0.200,0.000){5}{\line(1,0){0.150}}
	\put(8.100,1.350){\makebox(0.000,0.000)[b]{$\scs 12 \La$}}
	\put(8.600,1.150){\line(0,1){0.100}}
\multiput(8.600,1.150)(0.100,0.000){10}{\line(1,0){0.050}}
\multiput(8.600,1.200)(0.200,0.000){5}{\line(1,0){0.150}}
	\put(9.100,1.350){\makebox(0.000,0.000)[b]{$\scs 12 \Ra$}}
	\put(9.600,1.150){\line(0,1){0.100}}
	\put(10.600,1.150){\line(0,1){0.100}}
\multiput(10.600,1.200)(0.200,0.000){5}{\line(1,0){0.150}}
	\put(11.100,1.350){\makebox(0.000,0.000)[b]{$\scs 2 \La$}}
	\put(11.600,1.150){\line(0,1){0.100}}
\multiput(11.600,1.200)(0.200,0.000){5}{\line(1,0){0.150}}
	\put(12.100,1.350){\makebox(0.000,0.000)[b]{$\scs 2 \Ra$}}
	\put(12.600,1.150){\line(0,1){0.100}}
	\put(2.600,0.650){\line(0,1){0.100}}	% prs
\put(2.600,0.700){\line(1,0){1.000}}
	\put(3.100,0.850){\makebox(0.000,0.000)[b]{$\scs \Ma$}}
	\put(3.600,0.650){\line(0,1){0.100}}
\put(3.600,0.700){\line(1,0){1.000}}
	\put(4.100,0.850){\makebox(0.000,0.000)[b]{$\scs \Ua$}}
	\put(4.600,0.650){\line(0,1){0.100}}
\put(4.600,0.700){\line(1,0){1.000}}
	\put(5.100,0.850){\makebox(0.000,0.000)[b]{$\scs \Da$}}
	\put(5.600,0.650){\line(0,1){0.100}}
\multiput(5.600,0.700)(0.100,0.000){10}{\line(1,0){0.050}}
	\put(6.100,0.850){\makebox(0.000,0.000)[b]{$\scs \Ma$}}
	\put(6.600,0.650){\line(0,1){0.100}}
\multiput(6.600,0.700)(0.100,0.000){10}{\line(1,0){0.050}}
	\put(7.100,0.850){\makebox(0.000,0.000)[b]{$\scs \Ua$}}
	\put(7.600,0.650){\line(0,1){0.100}}
\multiput(7.600,0.700)(0.100,0.000){10}{\line(1,0){0.050}}
	\put(8.100,0.850){\makebox(0.000,0.000)[b]{$\scs \Da$}}
	\put(8.600,0.650){\line(0,1){0.100}}
\multiput(8.600,0.700)(0.200,0.000){5}{\line(1,0){0.150}}
	\put(9.100,0.850){\makebox(0.000,0.000)[b]{$\scs \Ma$}}
	\put(9.600,0.650){\line(0,1){0.100}}
\multiput(9.600,0.700)(0.200,0.000){5}{\line(1,0){0.150}}
	\put(10.100,0.850){\makebox(0.000,0.000)[b]{$\scs \Ua$}}
	\put(10.600,0.650){\line(0,1){0.100}}
\multiput(10.600,0.700)(0.200,0.000){5}{\line(1,0){0.150}}
	\put(11.100,0.850){\makebox(0.000,0.000)[b]{$\scs \Da$}}
	\put(11.600,0.650){\line(0,1){0.100}}
	\put(6.600,0.150){\line(0,1){0.100}}	% afb
\put(6.600,0.200){\line(1,0){1.000}}
	\put(7.100,0.350){\makebox(0.000,0.000)[b]{$\scs \Ra$}}
	\put(7.600,0.150){\line(0,1){0.100}}
	\put(9.600,0.150){\line(0,1){0.100}}
\multiput(9.600,0.200)(0.100,0.000){10}{\line(1,0){0.050}}
	\put(10.100,0.350){\makebox(0.000,0.000)[b]{$\scs \Ra$}}
	\put(10.600,0.150){\line(0,1){0.100}}
	\put(12.600,0.150){\line(0,1){0.100}}
\multiput(12.600,0.200)(0.200,0.000){5}{\line(1,0){0.150}}
	\put(13.100,0.350){\makebox(0.000,0.000)[b]{$\scs \Ra$}}
	\put(13.600,0.150){\line(0,1){0.100}}
\end{picture}
\end{center}
\caption{Zeitdiagramm f\"ur den Durchlauf dreier Metallpl\"attchen} 
\label{Zeitdiagramm fur den Durchlauf dreier Metallplattchen} 
\end{figure}

Dies ist eine vereinfachte Beschreibung. Dabei besteht
die Vereinfachung zum einen in der groben Beschreibung
der einzelnen Einheiten. Zum anderen ist die Anlage so
ausgerichtet, da\3 mehrere Metallplatten gleichzeitig
verarbeitet und transportiert werden; das soll gerade so
geschehen, da\3 die Anlage optimal ausgenutzt ist.
Abbildung~\ref{Zeitdiagramm fur den Durchlauf dreier Metallplattchen} 
zeigt ein Zeitdiagramm f\"ur
einen Durchlauf dreier Metallpl\"attchen durch
die Fertigungszelle, das Handhabungsger\"at ist dabei weggelassen.

\subsection{Aktoren}
\label{Aktoren}

Die Steuerung hat folgende Aktionsm\"oglichkeiten:
\bi
\item	Bewegung des unteren Teils der Presse (Elektromotor)
\item	Ein- und Ausfahren des Roboterarmes 1 (Elektromotor)
\item	Ein- und Ausfahren des Roboterarmes 2 (Elektromotor)
\item	Aufnehmen eines Metallteils durch Arm 1 (Elektromagnet)
\item	Aufnehmen eines Metallteils durch Arm 2 (Elektromagnet)
\item	Drehung des Roboters (Elektromotor)
\item	Drehung des Hubdrehtisches (Elektromotor)
\item	H\"ohenverstellung des Hubdrehtisches (Elektromotor)
\item	Horizontale Bewegung des Greifers des Handhabungsger\"ates
	(Elektromotor)
\item	Vertikale Bewegung des Greifers des Handhabungsger\"ates
	(Elektromotor)
\item	Aufnehmen eines Metallpl\"attchens durch den Greifer des
	Handhabungsger\"ates (Elektromagnet)
\item	Ein- und Ausschalten des Zuf\"uhrf\"orderbandes (Elektromotor)
\item	Ein- und Ausschalten des Ablagef\"orderbandes (Elektromotor)
\ei

\subsection{Sensoren}
\label{Sensoren}

Die Steuerung erh\"alt Informationen \"uber folgende Sensoren:
\bi
\item	Ist die Presse in unterer Position? (Schalter)
\item	Ist die Presse in mittlerer Position? (Schalter)
\item	Ist die Presse in oberer Position? (Schalter)
\item	Wie weit ist Arm 1 ausgefahren? (Potentiometer)
\item	Wie weit ist Arm 2 aufgefahren? (Potentiometer)
\item	Wie weit ist der Roboter gedreht? (Potentiometer)
\item	Ist der Hubdrehtisch in unterer Position? (Schalter)
\item	Ist der Hubdrehtisch in oberer Position? (Schalter)
\item	Wie weit ist der Hubdrehtisch gedreht? (Potentiometer)
\item	Befindet sich das Handhabungsger\"at \"uber dem
	Zuf\"uhrf\"orderband? (Lichtschranke)
\item	Befindet sich das Handhabungsger\"at \"uber dem
	Ablagef\"orderband? (Lichtschranke)
\item	In welcher vertikalen Position befindet sich der Greifer?
	(Potentiometer)
\item	Befindet sich ein Metallteil am \"au\3ersten Ende des
	Ablagef\"orderbandes? (Lichtschranke)
\ei

W\"ahrend Lichtschranken und Schalter
ja/nein--Informationen liefern, besteht die Information
eines Potentiometers aus einer Spannung, die im Beispiel
der Drehung proportional zum Winkel ist.

\subsection{Sicherheitsanforderungen}
\label{Sicherheitsanforderungen}

Die Steuerung soll verschiedenen
Sicherheitsanforderungen gen\"ugen. Diese
Sicherheitsanforderungen dienen unterschiedlichen
Zwecken: zum einen gew\"ahrleisten sie, da\3 die Anlage sich
nicht selbst zerst\"ort (z.B.\ die Presse einen in ihr
befindlichen Roboterarm), zum anderen sch\"utzen sie im
Bereich der Anlage t\"atige Arbeiter.

\be
\item Die Presse wird nur geschlossen, wenn sich keiner
	der Roboterarme in ihr befindet.
\item Ein Roboterarm wird nur vor die Presse gedreht,
	wenn der Arm eingefahren ist oder die Presse in oberer
	oder unterer Stellung ist.
\item Der Roboter wird nicht weiter als n\"otig nach
	links und rechts gedreht, da er sonst m\"oglicherweise
	andere Teile der Anlage (z.B.\ den Hubdrehtisch)
	besch\"adigen k\"onnte.
\item S\"amtliche Elektromotoren werden sofort gestoppt,
	wenn ein dadurch bewegtes Ger\"at an den Rand seiner
	Beweglichkeit ger\"at. Zum Beispiel wird das
	Handhabungsger\"at in seiner horizontalen Bewegung
	gestoppt, sobald einer der beiden Schalter signalisiert,
	da\3 es \"uber einem Band steht.
\item Der Hubdrehtisch wird nicht unter das Niveau des
	F\"orderbandes bewegt. Er wird in unteren Stellungen
	nicht gedreht.
\item Das Handhabungsger\"at st\"o\3t nicht seitlich
	gegen ein F\"orderband. Dies geschieht zum Beispiel,
	wenn das Handhabungsger\"at auf dem Niveau, auf dem es
	vom Ablagef\"orderband aufnimmt, ohne vertikale Bewegung
	hin zum Zuf\"uhrf\"orderband f\"ahrt. Au\3erdem st\"o\3t
	das Handhabungsger\"at nicht von oben gegen Zuf\"uhr-
	oder Ablagef\"orderband.
\item Ein Elektromagnet wird nur
	deaktiviert, wenn der zugeh\"orige Roboterarm bzw.\ Greifer in
	einer Position ist, in der gefahrlos abgeladen werden
	kann.
	Gefahrlos abgeladen werden kann \"uber
	F\"orderb\"andern, in der Presse und \"uber dem
	Hubdrehtisch, vorausgesetzt der Abstand zwischen Magnet
	und Ablagefl\"ache ist gen\"ugend klein.
\item Das Ablagef\"orderband transportiert h\"ochstens
	dann, wenn sich kein Fertigteil im Aufnahmebereich des
	Handhabungsger\"ats befindet.
\item Damit ein Metallteil vom Zuf\"uhrf\"orderband auf
	den Hubdrehtisch \"ubergeht, mu\3 der Tisch in
	entsprechender Position sein.
\item (Das Zuf\"uhrf\"orderband darf nur dann Metallpl\"attchen
	transportieren, wenn der Hubdrehtisch leer ist.)
\ee

%\clearpage
\section{Deduktive Programmsynthese}
\label{Deduktive Programmsynthese}

In diesem Abschnitt wird die zur Entwicklung der Fertigungszelle
verwendete formale Methode der deduktiven Programmsynthese nach Manna
und Waldinger \cite{Manna.Waldinger.1980} kurz vorgestellt.
Ihre grundlegende Vorgehensweise besteht darin, die Erf\"ullbarkeit
einer gegebenen pr\"adikatenlogischen Spezifikation konstruktiv
zu beweisen.
Dabei entsteht aus den auftretenden
Antwort--Substitutionen gleichzeitig das zu
synthetisierende funktionale Programm, das die
Spezifikation erf\"ullt.

Die Beweisregeln umfassen Resolution auch
f\"ur Formeln, die nicht in Klausel--Normalform vorliegen, sowie die in
\cite{Manna.Waldinger.1986}
beschriebenen Verallgemeinerungen von E--Resolution und
Paramodulation. Es gibt eine $\vee$--Split--Regel f\"ur Goals und eine
$\wedge$--Split--Regel f\"ur Assertions, aber nicht umgekehrt. Rekursive
Programme lassen sich mit Hilfe struktureller Induktion erzeugen, wobei
die Terminierungsordnung vom Benutzer angegeben werden mu\3. Es kann
gezeigt werden, da\3 dieses Ableitungssystem f\"ur Logik erster Stufe
vollst\"andig ist.

\begin{figure}
\begin{center}
\em
\begin{picture}(14,9)
	%\put(0,0){\makebox(0,0){+}}
\put(  0.000,  2.500){\dashbox{  0.100}( 12.000,  5.000){}}
\put(  1.500,  1.250){\line(0,1){  5.750}}
\put(  1.500,  3.500){\vector(1,0){  2.750}}
\put(  1.500,  4.000){\vector(1,0){  3.000}}
\put(  1.500,  4.500){\vector(1,0){  3.000}}
\put(  1.500,  5.000){\vector(1,0){  3.000}}
\put(  1.500,  5.500){\vector(1,0){  3.000}}
\put(  1.500,  6.000){\vector(1,0){  3.000}}
\put(  1.500,  6.500){\vector(1,0){  3.000}}
\put(  1.500,  7.000){\vector(1,0){  3.000}}
\put(  5.500,  3.000){\vector(1,0){  2.500}}
\put(  5.750,  3.500){\vector(1,0){  2.250}}
\put(  6.000,  4.000){\vector(1,0){  2.000}}
\put(  6.000,  4.500){\vector(1,0){  2.000}}
\put(  6.000,  5.000){\vector(1,0){  2.000}}
\put(  6.000,  5.500){\vector(1,0){  2.000}}
\put(  6.000,  6.000){\vector(1,0){  2.000}}
\put(  6.000,  6.500){\vector(1,0){  2.000}}
\put(  6.000,  7.000){\vector(1,0){  2.000}}
\put(  9.500,  5.000){\line(1,0){  1.000}}
\put( 10.500,  5.000){\vector(0,-1){  3.750}}
\put(  3.750,  2.000){\line(0,1){  0.375}}
\put(  3.750,  2.375){\line(1,0){  0.500}}
\put(  4.250,  2.375){\vector(0,1){  0.375}}
\put(  5.625,  2.000){\vector(0,1){  1.250}}
\put(  6.750,  2.000){\line(0,1){  0.375}}
\put(  6.750,  2.375){\line(-1,0){  0.875}}
\put(  5.875,  2.375){\vector(0,1){  1.375}}
\put(  0.000,  0.000){\dashbox{  0.100}( 12.000,  2.250){}}
\put( 12.000,  1.125){\bf \Large $\Lraa$}
\put( 12.900,  0.750){\line(0,1){  0.750}}
\put( 13.900,  0.750){\line(0,1){  0.750}}
\put( 13.400,  1.500){\oval(  1.000,  0.500)}
\put( 13.400,  0.750){\oval(  1.000,  0.500)[b]}
\put( 13.400,  1.000){\makebox(0,0){Datei}}
\thicklines
\put(  0.500,  8.100){\framebox( 11.000,  1.000){\shortstack{
			Benutzerschnittstelle\\Pretty Printing}}}
\put(  5.900,  7.700){\bf \Large $\Updownarrow$}
\put(  4.000,  2.750){\framebox(  1.500,  0.500){hy}}
\put(  4.250,  3.250){\framebox(  1.500,  0.500){co}}
\put(  4.500,  3.750){\framebox(  1.500,  0.500){tf}}
\put(4.500,4.250){\framebox(1.500,0.500){sp}}
\put(4.500,4.750){\framebox(1.500,0.500){un}}
\put(4.500,5.250){\framebox(1.500,0.500){rm/1}}
\put(4.500,5.750){\framebox(1.500,0.500){rm/2}}
\put(4.500,6.250){\framebox(1.500,0.500){rp}}
\put(4.500,6.750){\framebox(1.500,0.500){rs}}
\put(  8.000,  2.750){\framebox(  1.500,  4.500){simp}}
\put(  0.500,  0.250){\framebox( 11.000,  1.000)
				{Nr $\cdot$ Assertions $\cdot$ Goals
				$\cdot$ Output $\cdot$ Orig}}
\put(  2.500,  1.500){\framebox(  1.500,  0.500){Specs}}
\put(  4.500,  1.500){\framebox(  1.500,  0.500){Types}}
\put(  6.500,  1.500){\framebox(  1.500,  0.500){Tfms}}
\put( 11.900,  0.050){\makebox(0,0)[br]{\scriptsize \em Datenbasis}}
\end{picture}
\end{center}
\caption{Struktur des {\sysyfos}--Systems}
\label{Struktur des sysyfos--Systems}
\end{figure}
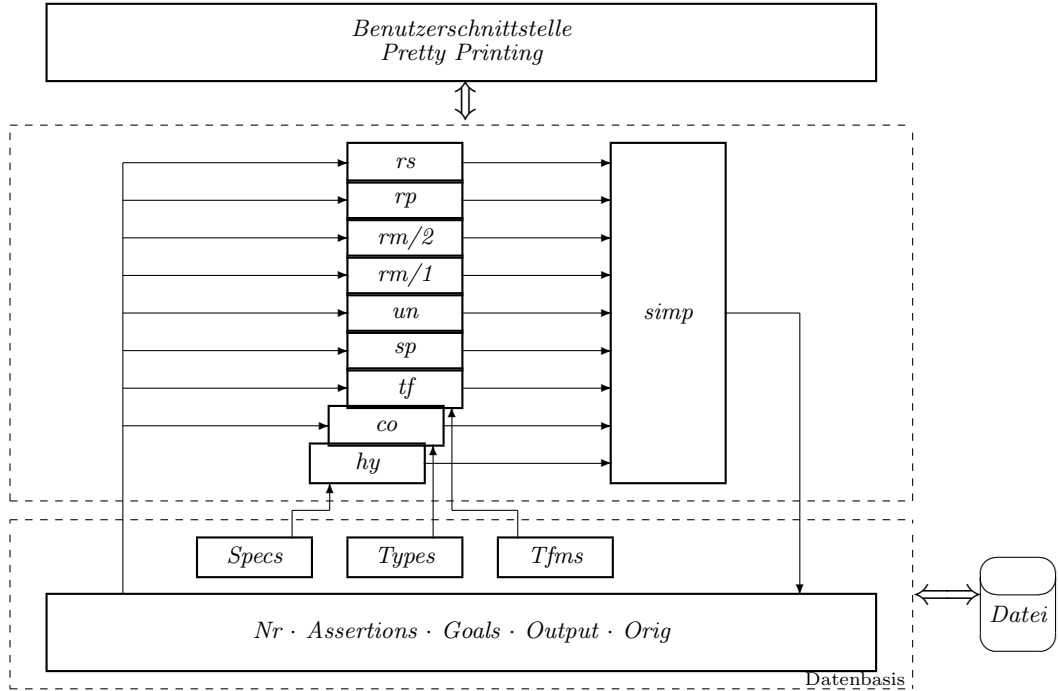

Das nach diesem Ansatz prototypisch in {\prolog} implementierte
{\sysyfos}--System (siehe Abb.~\ref{Struktur des sysyfos--Systems})
arbeitet auf einer Datenbasis
aus Formeln ({\em Assertions\/},
{\em Goals\/}), denen teilweise ein Term als Antwortsubstitution ({\em
Output\/}) zugeordnet ist. Jede Formel ist eindeutig numeriert ({\em
Nr\/}).
Unter {\em Orig\/} wird die Beweisoperation abgespeichert, aus der die
Formel entstanden ist.
Folgende Operationen sind auf dieser Datenbasis definiert:
\bi
\item Nicht--Klausel--Resolution ({\em rs\/}),
\item Paramodulation bzw.\ allgemeiner: Relation Replacement
	({\em rp\/}), 
\item E--Resolution bzw.\ allgemeiner: Relation Matching ({\em rm/2\/}),
\item Monotonie--Regel ({\em rm/1\/}),
\item Unifikation zweier Teilformeln bzw.\ explizite Faktorisierung
	({\em un\/}),
\item Aufspalten in Teilformeln ({\em sp\/}),
\item Logische Transformationen ({\em tf\/}),
\item Konkretisierung von Skolem--Konstanten entsprechend ihres Typs
       ({\em co\/}),
\item Generierung einer Induktionshypothese aus einer Spezifikation
	({\em hy\/}).
\ei
Das Ergebnis einer Operation wird von einem Formelvereinfacher
({\em simp\/})
auf eine m\"oglichst einfache Form
gebracht, bevor es in die Datenbasis abgespeichert wird.
Im Folgenden werden die f\"ur die Fallstudie ben\"otigten
Operationen und Komponenten kurz erl\"autert.

\begin{figure}
\begin{center}
\begin{tabular}{@{}|r|ll|l|@{}}
\hline
& Assertions & Goals & \\
\hline \hline
& \multicolumn{2}{@{}|l|}{\bf Resolution:} &	\\
1 && $F[p^m]$ &	\\
2 & $G[p^{-m}]$	&&		\\
3 && $F[\neg m\.\cdot G[(-m) \.\cdot true]]$ & rs(1,2) \\
% \cline{2-3}
& \multicolumn{2}{@{}|l|}{\bf Beispiel:} &	\\
4 && $rob(R) \wedge T_0 \.\leq t \.\leq T_1 \ra drv(R,t)$ &	\\
5 & \multicolumn{2}{l|}{$val(C_7,T) = 1 \ra drv(r(C_7),T)$} &	\\
6 && $rob(r(C_7)) \wedge T_0 \.\leq t \.\leq T_1
	\ra \neg (val(C_7,t) = 1 \ra \neg true)$ & rs(4,5) \\
7 && $rob(r(C_7)) \wedge T_0 \.\leq t \.\leq T_1 \ra val(C_7,t) = 1$
	& simp(6) \\
\hline \hline
& \multicolumn{2}{@{}|l|}{\bf Paramodulation:} &	\\
8 & $F[s \preceq t]$ &&	\\
9 && $G \langle t^+,s^- \rangle$ &	\\
10 && $\neg F[false] \wedge G \langle s^+,t^- \rangle$ & rp(8,9) \\
% \cline{2-3}
& \multicolumn{2}{@{}|l|}{\bf Beispiel:} &	\\
11 & \multicolumn{2}{l|}{$ha_1(R,S,T) \ra
	win(S,T) = wxy(x,ps_1(R,T))-90$} &	\\
12 && $win(s_0,t) = 180$ &	\\
13 && $\neg (ha_1(R,s_0,t) \ra false)
	\wedge wxy(x,ps_1(R,t))-90 = 180$ & rp(11,12)	\\
14 && $ha_1(R,s_0,t) \wedge wxy(x,ps_1(R,t))-90 = 180$ & simp(13)\\
\hline \hline
& \multicolumn{2}{@{}|l|}{\bf E--Resolution:} &	\\
15 && $F[P(s^+)^m]$ &	\\
16 & $G[P(t)^{-m}]$  &&	\\
17 && $F[\neg m\.\cdot G[(-m) \.\cdot (t \preceq s)]]$ & rm(15,16) \\
% \cline{2-3}
& \multicolumn{2}{@{}|l|}{\bf Beispiel:} &	\\
18 && $rob \wedge win(s_0,T) = w_0$ &	\\
19 & \multicolumn{2}{l|}{$(T_0 \.\leq t \.\leq t_1 \ra drv(t))
	\ra wxy(t_1) = A$} &	\\
20 && $rob \wedge \neg ((T_0 \.\leq t \.\leq t_1
	\ra drv(t)) \ra \neg win(s_0,T) = wxy(t_1)$ & rm(18,19) \\
21 && $rob \wedge (T_0 \.\leq t \.\leq t_1
	\ra drv(t)) \wedge win(s_0,T) = wxy(t_1)$ & simp(20)	\\
\hline \hline
& \multicolumn{2}{@{}|l|}{\bf Monotonie--Regel:} &	\\
22 && $f(s^+,t^-) \preceq f(u^+,v^-)$ &	\\
23 && $s \preceq u \wedge v \preceq t$ & rm(22)	\\
% \cline{2-3}
& \multicolumn{2}{@{}|l|}{\bf Beispiel:} &	\\
24 && $wxy(x_1,x_2) = wxy(x_3,x_4)$ &	\\
25 && $x_1 = x_3 \wedge x_2 = x_4$ & rm(24)	\\
\hline
\end{tabular}
\vspace*{0.3cm}
\par
Dabei sei $m$ bzw. $-m$ die Polarit\"at von $p$, angedeutet durch
$p^m$ bzw. $p^{-m}$.
F\"ur Terme $s$ und $t$ bedeute $G\langle s^+,t^- \rangle$,
da\3 $G$ bzgl.\ $\preceq$ im ersten Argument
monoton wachsend und im zweiten monoton fallend ist.
In den Beispielen ist die Relation ``$\preceq$'' die Gleichheit.
Gro\3geschriebene Namen bezeichenen Variablen, kleingeschriebene
Konstanten.
Die rechte Seite gibt an, wie die aktuelle Formel entstanden ist.
\caption{Verwendete Beweisregeln}
\label{Verwendete Beweisregeln}
\end{center}
\end{figure}

\subsection{Nicht--Klausel--Resolution}
\label{Nicht--Klausel--Resolution}

Die seinerzeit von Robinson eingef\"uhrte Resolutionsmethode
\cite{Robinson.1965} setzt voraus, da\3 die zu resolvierenden
Formeln in konjunktive Normalform und weiter in
Klauselnormalform \"uberf\"uhrt worden sind. Demgegen\"uber
geben Manna und Waldinger in \cite{Manna.Waldinger.1980} eine
Form der Resolution an, die auf beliebigen Formeln arbeitet
(Nicht--Klausel--Resolution). Dadurch wird die
Lesbarkeit von Zwischenergebnissen im Beweis stark erh\"oht, was
f\"ur ein interaktives System eine unbedingte Notwendigkeit ist.
Die Lesbarkeit wird ebenfalls durch die Schreibweise in zwei
Spalten (Assertions, Goals) gesteigert, wobei
logisch gesehen die Formeln in der Goal--Spalte implizit
negiert sind. Das {\sysyfos}--System arbeitet nicht mit der in
\cite{Manna.Waldinger.1980} verwendeten Resolutionsregel,
sondern mit der etwas leistungsf\"ahigeren
Regel von Schmerl \cite{Schmerl.1988}, die in \cite{Haase.1992}
entsprechend angepa\3t wurde. 

Abbildung~\ref{Verwendete Beweisregeln} zeigt in den Schritten 1
bis 3 und in den Schritten 4 bis 7 ein konkretes Beispiel der
Nicht--Klausel--Resolution. Der dabei wie auch im folgenden
verwendete Begriff der Polarit\"at einer Teilformel $p$
innerhalb einer Formel $F[p]$ gibt an, ob $p$ unter einer
geraden (Polarit\"at ``$+$'') oder ungeraden (``$-$'') Anzahl
von Negationen vorkommt. Die Multiplikation zweier Polarit\"aten
sowie einer Polarit\"at mit einer Teilformel ist entsprechend
definiert, vgl.\ \cite{Schutte.1960}.

\subsection{Paramodulation}
\label{Paramodulation}

In \cite{Manna.Waldinger.1986}
definieren Manna und Waldinger unter dem Namen ``Relation
Replacement'' ({\em rp\/} in Abb.~\ref{Struktur des sysyfos--Systems})
eine Verallgemeinerung der Paramodulation,
%\cite{Duffy.1991}
die erstens
keine Klauselform voraussetzt und zweitens nicht nur bzgl.\ einer
\"Aquivalenzrelation ``$=$'', sondern bzgl.\ beliebiger Relationen
``$\preceq$'' arbeitet, sofern die entsprechenden Funktionen und
Pr\"adikate monoton sind (Schritte 8 bis 10 in
Abb.~\ref{Verwendete Beweisregeln}).
Im {\sysyfos}--System wird zus\"atzlich gefordert, da\3
$\preceq$ transitiv ist, dann d\"urfen in $G$ auch
mehrere Ersetzungen von $s$ und $t$ vorgenommen werden.

\subsection{E--Resolution}
\label{E--Resolution}

Ebenfalls in \cite{Manna.Waldinger.1986}
wird eine entsprechende Verallgemeinerung der
E--Resolution 
%\cite{Duffy.1991}
auf Nicht--Klausel--Form und beliebige Relationen definiert
und ``Relation Replacement'' genannt ({\em rm/2\/} in
Abb.~\ref{Struktur des sysyfos--Systems}).
Das {\sysyfos}--System arbeitet hier mit einer f\"ur E--Resolution
modifizierten Version der Schmerl--Regel (Schritte 15 bis 17 in
Abb.~\ref{Verwendete Beweisregeln}, vgl.\ \cite{Burghardt.1989a}).
Daneben ist noch eine Operation zur Anwendung der
Monotonie--Regel definiert, ebenfalls f\"ur beliebige
transitive Relationen (Schritte 22 bis 23 in Abb.~\ref{Verwendete
Beweisregeln}).

\subsection{Vereinfacher}
\label{Vereinfacher}

Nach jedem Beweisschritt im System wird das Ergebnis
durch einen Vereinfacher geschickt, wobei nicht nur
die Idempotenz-, sondern auch die Absorptionsgesetze
angewendet werden, um mehrfach vorkommende Teilformeln zu
vermeiden (Faktorisierung, vgl.\ \cite{Schmerl.1988}).
Zus\"atzlich kann der Benutzer 
durch eine entsprechende Option die \"Uberf\"uhrung
in eine linksassoziative Negationsnormalform veranlassen, was 
sich w\"ahrend der Fallstudie als
positiv f\"ur die Lesbarkeit herausgestellt hat.

In \cite{Mohaupt.1991} wurde als Anwendungsfallstudie mit Hilfe des
{\sysyfos}--Systems ein
Algorithmus zum Finden der gemeinsamen Teilterme eines Terms
synthetisiert,
der k\"unftig als verifizierte Komponente innerhalb des
Formelvereinfachers (Faktorisierung) des Systems selbst
eingesetzt werden soll.

\subsection{Benutzeroberfl\"ache}
\label{Benutzeroberflache}

W\"ahrend der Bearbeitung der Fallstudie Fertigungszelle wurde
parallel das {\sysyfos}--System
weiterentwickelt und u.a.\ um eine halbgraphische
Benutzeroberfl\"ache zur Darstellung von Formel- und
Beweisb\"aumen und Auswahl von Teilb\"aumen f\"ur
Beweisoperationen erg\"anzt, die weitgehend unabh\"angig vom
unterliegenden Fenstersystem und vom eigentlichen Beweissystem
verwendbar ist. Sie erlaubt neben einer flexiblen
Layout--Festlegung f\"ur Formelb\"aume die wahlweise Aus- und
Einblendung der Argumente von Skolemfunktionen, die Navigation
innerhalb eines Formelbaums und die gezielte Suche in einer
Menge von Formeln (z.B. Spezifikationsaxiomen). 

\begin{figure}
%\epsfbox{ap996_screen.epsf}
\includegraphics[scale=0.7]{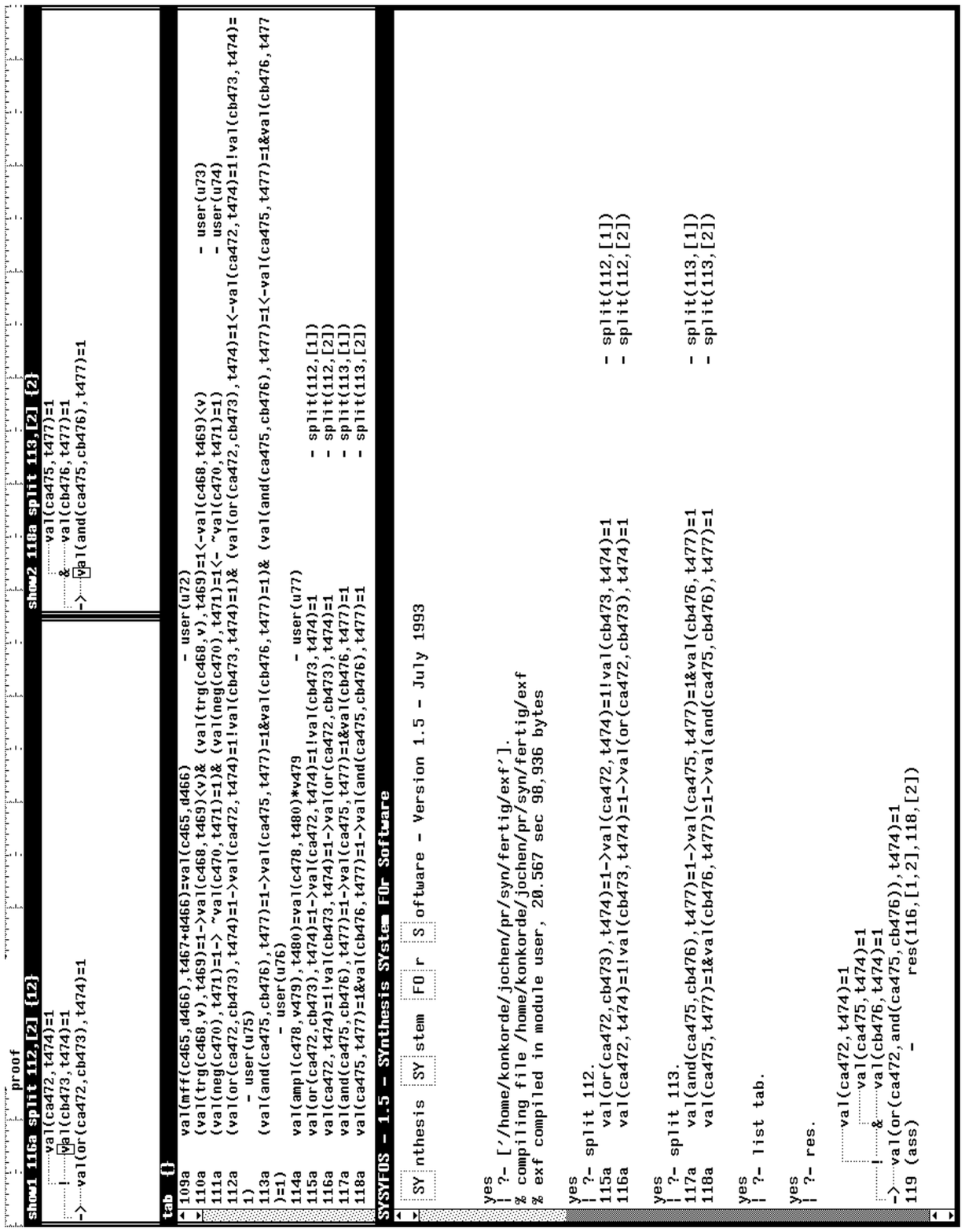}
\caption{Die {\sysyfos}--Oberfl\"ache}
\label{Die sysyfos--Oberflache}
\end{figure}

Abbildung~\ref{Die sysyfos--Oberflache} zeigt als Beispiel die
{\sysyfos}--Benutzeroberfl\"ache w\"ahrend des ersten
Resolutionsschritts aus Anhang~\refC. In den Fenstern ``show1''
und ``show2'' werden die aktuellen Elternformeln in
halbgraphischer Notation angezeigt, das Fenster ``tab'' zeigt
eine Tabelle aller bisher abgeleiteten Formeln. Im Hauptfenster
(``SYSYFOS'') wurde der Resolutionsbefehl eingegeben, nachdem in
den Fenstern ``show1'' und ``show2'' die zu resolvierenden
Teilformeln durch entsprechende Navigationskommandos
ausgew\"ahlt wurden. Die Angaben in der Titelzeile des Fensters
``show1'' besagen, da\3 die angezeigte Formel die Nummer 116
hat, eine Assertion ist, aus der Opration $split \;\; 112,[2]$
entstanden ist, und da\3 in ihr die Teilformel $1.2$
ausgew\"ahlt wurde (siehe auch die Cursor--Position im Fenster).
Entsprechendes gilt f\"ur das Fenter ``show2''. F\"ur die
Formelsyntax siehe Abb.~\ref{PROLOG--Notation fur Relationen und
Junktoren}. Die vorkommenden Variablennamen weichen von
Anhang~\refC\ ab, da letztere aus Darstellungsgr\"unden
systematisch umnumeriert wurden.

\subsection{Beweiswiederholung}
\label{Beweiswiederholung}

Das {\sysyfos}--System verf\"ugt \"uber einen Mechanismus zum
Nachspielen bereits gefundener Beweisteile, der auch noch nach
leichten \"Anderungen der Spezifkation verwendet werden kann.
Beweisteile k\"onnen als Terme abgespeichert werden, wobei jeder
Beweisoperation aus Abb.~\ref{Struktur des sysyfos--Systems} ein
Funktionssymbol entspricht und jedem Axiom sein eindeutiger
benutzerdefinierter Name als Konstante.
Abbildung~\ref{Termdarstellung des Beweises aus Abb.
Synthesebeweis der Schaltung aus Abb. Einfache
Steuerungsschaltung} zeigt die Termdarstellung des Beweises aus
Abb.~\ref{Synthesebeweis der Schaltung aus Abb. Einfache
Steuerungsschaltung}. Der Name eines Axioms ist von seiner
Formelnummer zu unterscheiden: w\"ahrend sich letztere zwischen
verschiedenen Sitzungen am {\sysyfos}-System \"andern kann
(z.B.\ durch Umordnung der Axiome), bleibt ersterer stets
gleich, was f\"ur das Beweis--Nachspielen unumg\"anglich ist.
Weiterhin wurden neue, abgeleitete Beweisoperationen definiert,
die weniger empfindlich gegen \"Anderungen der Ausgangsformeln
sind. So wurde z.B.\ die in \cite{Manna.Waldinger.1980}
vorgesehene explizite Instanziierung einer Formel ersetzt durch
die Unifikation zweier Teilformeln, da letztere unempfindlich
gegen gebundene Umbenennung ist, wie sie h\"aufig durch die
automatisch erzeugten Variablennamen vorkommt.

\clearpage

\section{Entwurf}
\label{Entwurf}

\subsection{Anforderungsspezifikation}
\label{Anforderungsspezifikation}

Es ist bekannt, da\3 der \"Ubergang von einer informellen
Anforderungsbeschreibung zu einer formalen Spezifikation der
bzgl.\ der Korrektheit problematischste Schritt innerhalb des
formalen Vorgehensmodells ist, da die formale Spezifikation
naturgem\"a\3 nicht gegen die informelle Beschreibung
verifiziert werden kann. In der vorliegenden Fallstudie haben
wir versucht, einen Ansatz zu verfolgen, der dieses Problem so
weit als m\"oglich entsch\"arft. Wir haben eine formale
Sprachebene geschaffen, in der die informelle Beschreibung aus
Abschnitt~\ref{Die Fallstudie
``Fertigungszelle''} quasi 1:1 ausgedr\"uckt und dadurch
leicht validiert werden kann. Die so erhaltene Spezifikation ist
eine {\em Anforderungs--}Spezifikation, keine
Entwurfsspezifikation. Durch ihren hohen Grad an Implizitheit
gestattet sie weder ein ``rapid prototyping'' noch die
unmittelbare schrittweise Verfeinerung in ausf\"uhrbaren Code.

Zuerst wurde eine geeignete Terminologie, bestehend aus
Pr\"adikaten- und Funktionssymbolen und ihren informellen
Bedeutungen, festgelegt (siehe Anhang~\ref{Informelle
Beschreibung der verwendeten Bezeichner}). Die Zeit wurde mit
Hilfe expliziter Parameter modelliert, um innerhalb der
Pr\"adikatenlogik erster Ordnung zu bleiben und um explizite
Aussagen \"uber Zeitpunkte machen zu k\"onnen. Der Raum wurde
durch dreidimensionale kartesische Koordinatenvektoren
modelliert, wobei Transformationen von und nach Polarkoordinaten
soweit als notwendig axiomatisiert wurden. Das gew\"unschte
``Programm'' soll aus einer asynchronen Hardware--Schaltung
bestehen. Sie wird aus {\ttl}--artigen Komponenten aufgebaut, die
formal durch zeitabh\"angige Funktionen dargestellt sind, wobei
die Schaltzeiten ignoriert werden. Explizite R\"uckkoppelungen
in der Schaltung sind nicht erlaubt, da sie auf der formalen
Seite zu unendlichen Termen f\"uhren w\"urden, die vom
Beweiswerkzeug nicht unterst\"utzt werden. Stattdessen wurden
die notwendigen R\"uckkoppelungen in Bauelementen wie Flip--Flops
eingekapselt.

Auf dieser Grundlage konnte in
einem n\"achsten Schritt eine Sammlung offensichtlicher
Fakten \"uber das Verhalten der einzelnen Maschinen formalisiert
werden, siehe
Anhang~\ref{Formale Spezifikation der Fertigungszelle}.
Diese formale Spezifikation besteht aus vier Teilen:
\bi
\item der von jedem einzelnen Maschinentyp verlangten
	Verhaltensbeschreibung,
\item der Verhaltensbeschreibung der einzelnen Hardware--Bauelemente,
\item den ben\"otigten Hintergrundfakten aus Geometrie, Arithmetik und
	Physik und
\item der eigentlichen Spezifikation der Aufgabe der Fertigungszelle.
\ei

Die Spezifikation ist {\em lokal verstehbar\/} in dem Sinne, da\3
es zur Validation eines Axioms gen\"ugt, nur dieses eine Axiom
mit Hilfe der Terminologiedefinitionen gegen seine informelle
Beschreibung zu \"uberpr\"ufen.

Die Spezifikation wurde in der naheliegenden Weise
modularisiert. F\"ur jeden Maschinentyp existiert ein Modul, das
das von ihm verlangte Verhalten formal beschreibt, daneben gibt
es drei weitere Spezifikationsmodule, in denen das Verhalten der
Hardware--Bauelemente, der Gesamtaufbau der Fertigungszelle sowie
die ben\"otigten Hintergrundfakten aus Mathematik und Physik
enthalten sind. Man beachte, da\3 keines dieser Module zu einem
Teil der Implementierung korrespondiert in dem Sinne, da\3
letzterer durch eine Folge von Entwicklungsschritten aus
ersterem gewonnen werden k\"onnte. Die Spezifikationsmodule
beschreiben verschiedene Aspekte der modellierten Realit\"at,
nicht der Implementierung.

Diese Herangehensweise machte auch formal die Sinnlosigkeit
einer ``Fertigungs''--Zelle deutlich, deren einziger Zweck darin
besteht, Metallpl\"attchen in einem Kreis zu bewegen, da es
nicht m\"oglich war, eine Spezifikationsformel daf\"ur
anzugeben, die nicht auch von einer leeren Zelle ohne alle
Maschinen erf\"ullt worden w\"are. Daher versahen wir das
Handhabungsger\"at mit der zus\"atzlichen F\"ahigkeit,
Metallpl\"attchen zu konsumieren, d.h.\ sie in den
unbearbeiteten Zustand zur\"uck zu \"uberf\"uhren, und stellten
zwei separate Spezifikationsformeln auf: eine f\"ur den
Konsumenten, das Handhabungsger\"at, und eine f\"ur den
Produzenten, den Rest der Fertigungszelle. Letztere besteht in
einer formalen \"Ubersetzung der Anforderung: ``Wenn ein
unbearbeitetes Metallpl\"attchen auf dem Zuf\"uhrf\"orderband
liegt, erscheit es irgendwann sp\"ater in bearbeitetem Zustand
auf dem Ablagef\"orderband.''

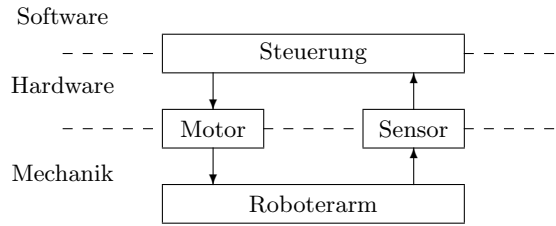
\begin{figure}
\begin{center}
\begin{picture}(7.5,3.0)
	%\put(0,0){\makebox(0,0){+}}
\put(2.133,0.000){\framebox(4.000,0.500){{Roboterarm}}}
\put(2.133,1.000){\framebox(1.333,0.500){{Motor}}}
\put(4.800,1.000){\framebox(1.333,0.500){{Sensor}}}
\put(2.133,2.000){\framebox(4.000,0.500){{Steuerung}}}
\put(2.800,2.000){\vector(0,-1){0.500}}
\put(2.800,1.000){\vector(0,-1){0.500}}
\put(5.467,0.500){\vector(0,1){0.500}}
\put(5.467,1.500){\vector(0,1){0.500}}
\put(0.800,2.749){\makebox(0.000,0.000){{Software}}}
\put(0.800,1.833){\makebox(0.000,0.000){{Hardware}}}
\put(0.800,0.667){\makebox(0.000,0.000){{Mechanik}}}
\multiput(0.800,1.250)(0.267,0.000){5}{\line(1,0){0.133}}
\multiput(3.467,1.250)(0.267,0.000){5}{\line(1,0){0.133}}
\multiput(6.133,1.250)(0.267,0.000){5}{\line(1,0){0.133}}
\multiput(0.800,2.250)(0.267,0.000){5}{\line(1,0){0.133}}
\multiput(6.133,2.250)(0.267,0.000){5}{\line(1,0){0.133}}
\end{picture}
\end{center}
\caption{Steuerschleife mit mechanischer R\"uckkoppelung}
\label{Steuerschleife mit mechanischer Ruckkoppelung}
\end{figure}

\subsection{Einbeziehung der Mechanik}
\label{Einbeziehung der Mechanik}

Die Verwendung von Pr\"adikatenlogik mit expliziter Zeit als
Spezifikationssprache erlaubt es, die vorgegebenen
technisch--physikalischen Anforderungen mit in die Spezifikation
einzubeziehen und dadurch auch Systeme mit
R\"uckkoppelungsschleifen zu behandeln, die teilweise au\3erhalb
des Bereichs der Hardware bzw.\ Software liegen. Zum Beispiel
f\"ahrt die Robotersteuerung in einigen Situationen einen der
Arme soweit aus, bis er eine bestimmte L\"ange erreicht, siehe
Abb.~\ref{Steuerschleife mit mechanischer Ruckkoppelung}. Es
ist unm\"oglich zu beweisen, da\3 der Roboterarm tats\"achlich
die gew\"unschte L\"ange erreichen und dann anhalten wird, ohne
die mechanischen Eigenschaften des Arms miteinzubeziehen. Das
Gleiche gilt f\"ur die gesamte Fertigungszelle: um das oberste
Spezifikationsziel zu beweisen, da\3 sie bearbeitete
Me\-tall\-pl\"att\-chen aus unbearbeiteten produziert, ist die formale
Behandlung ihres mechanichen Verhaltens im Beweis notwendig; es
gen\"ugt nicht, den Beweis auf die reinen Software- bzw.\
Hardware--Aspekte zu beschr\"anken.

Dar\"uber hinaus ist es m\"oglich, notwendige Bedingungen formal
herzuleiten, die die Konfiguration au\3erhalb des
Hardware/Software--Bereichs betreffen. So wurde z.B.\ gezeigt,
da\3 der vom Anfangspunkt des Ablagebands, dem Drehzentrum des
Roboters und der Presse gebildete Winkel notwendig 90$^\circ$
betragen mu\3, damit die bearbeiteten Metallpl\"attchen in der
richtigen Ausrichtung auf dem Band abgelegt werden. Der
deduktive Ansatz konnte somit zu einem methodischen Rahmen f\"ur
die Entwicklung der gesamten Fertigungszelle, einschlie\3lich
mechanischer Aspekte, erweitert werden. Es ist denkbar, da\3 ein
Ingenieur in Zukunft eine formale Anforderungsbeschreibung der
gew\"unschten Fertigungszelle vom Auftraggeber erh\"alt und
eine formale Verhaltensbeschreibung der Maschinen von deren
Hersteller. Daraus k\"onnte er dann einen verifizierten
Gesamtentwurf der Zelle erstellen, einschlie\3lich ihrer
Steuerungs--Software und ihres mechanischen Aufbaus. Die
deduktive Synthese dient dabei als der Rahmen, innerhalb dessen
klassischer mechanischer Entwurf und Software--Entwurf integriert
wird.

\subsection{Zeitmodellierung}
\label{Zeitmodellierung}

Schlie\3lich soll eine eher \"uberraschende Erfahrung mit der
Zeitmodellierung erw\"ahnt werden, die zeigt, wieviel Sorgfalt
die Formalisierung des Hintergrundwissens f\"ur die
Anforderungsspezifikation erfordert. Wir beziehen uns wieder auf
die Steuerschleife aus Abb.~\ref{Steuerschleife mit
mechanischer Ruckkoppelung}. An einer bestimmten Stelle im
Beweis ist es notwendig zu zeigen, da\3 es einen Zeitpunkt $t_2$
gibt, an dem der Roboterarm die gew\"unschte L\"ange erreicht,
sofern seine L\"ange zu einem gegebenen Zeitpunkt $t_1$ kleiner
war. Aus der formalen Verhaltensbeschreibung des Roboters wissen
wir, da\3 der Arm irgendwann jede vorgegebene L\"ange (innerhalb
seiner Grenzen) erreichen wird, wenn der Ausfahrmotor nur lange
genug l\"auft.

F\"ur den Korrektheitsbeweis der obigen Steuerschleife
ben\"otigen wir jedoch die Existenz eine {\em fr\"uhesten\/}
Zeitpunkts, an dem die gew\"unschte L\"ange erreicht wird, um
den Ausfahrmotor genau zu diesem Zeitpunkt anzuhalten. Es
gen\"ugt daher nicht, die Zeit durch rationale Zahlen zu
modellieren, da diese nicht abgeschlossen gegen Infima sind.
Wenn z.B.\ die gew\"unschte L\"ange zuf\"allig so gew\"ahlt ist,
da\3 sie erreicht wird, wenn $(t_2-t_1)^2 = 2$ ist, dann ist sie
zu jedem Zeitpunkt $t_2 > t_1+\sqrt{2}$ erreicht und
\"uberschritten, aber es gibt kein minimales (rationales) $t_2$.
Dieses Problem wurde umgangen, indem in den
Spezifikationsaxiomen zus\"atzlich die Existenz fr\"uhester
Zeitpunkte gefordert wurde, vgl.\ z.B.\ Axiom $u16$ in
Anhang~\ref{Formale Spezifikation der Fertigungszelle}.

\begin{figure}
\begin{center}
\begin{tabular}[b]{@{}l@{\hspace*{0.5cm}}}
L\"ange des Roboterarms 1:	\\
(Presse in oberer Position)	\\
\\
\\
\\
\\
\end{tabular}
\begin{picture}(6.4,3.5)
	%\put(0,0){\makebox(0,0){+}}
	\put(1.565,2.078){\line(2,1){0.360}}
	\put(1.925,2.258){\line(4,1){0.750}}
	\put(2.675,2.445){\line(6,1){0.750}}
\put(3.425,2.569){\line(6,1){0.700}}
\put(4.175,2.680){\circle{0.105}}
\put(4.175,3.280){\circle*{0.105}}
\put(4.225,3.270){\line(1,0){0.600}}
\multiput(1.025,2.925)(0.1,0){42}{\line(1,0){0.050}}
%\put(4.175,1.675){\line(0,1){0.150}}
\put(1.575,0.550){\line(1,0){2.555}}
\put(4.220,0.160){\line(1,0){0.600}}
\put(4.175,0.565){\circle{0.105}}
\put(4.175,0.175){\circle*{0.105}}
\put(1.425,0.362){\makebox(0.000,0.000)[r]{\em f\"ahrt\_aus}}
\put(5.350,0.560){\makebox(0.000,0.000)[l]{$true$}}
\put(5.350,0.165){\makebox(0.000,0.000)[l]{$false$}}
\thicklines
\put(0.800,1.825){\vector(1,0){4.400}}
\put(1.175,1.450){\vector(0,1){2.000}}
\put(5.350,1.825){\makebox(0.000,0.000)[l]{$t$}}
%\put(1.175,4.650){\makebox(0.000,0.000)[b]{$l$}}
\put(1.025,3.000){\makebox(0.000,0.000)[r]{$d_s$}}
\end{picture}
\end{center}
\caption{Verletzung einer Sicherheitsanforderung durch unstetige
	Bewegung}
\label{Verletzung einer Sicherheitsanforderung durch unstetige
	Bewegung}
\end{figure}
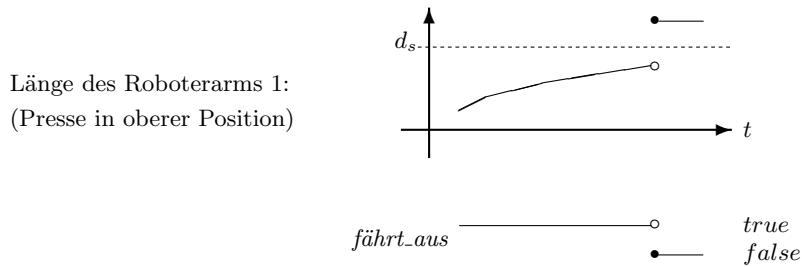

Eine der Hauptschwierigkeiten beim Finden des Beweises bestand
darin, die notwendigen Voraussetzungen \"uber die Stetigkeit
bestimmter Funktionen in einfachen R\"uckkoppelungsschleifen
explizit zu machen. In ersten Versionen der Spezifikation waren
sie vergessen worden, was erst aufgrund der Analyse
fehlgeschlagener Beweisversuche bemerkt wurde. Eine der
Sicherheitsanforderungen besagt z.B., da\3 der erste Roboterarm
nur dann in den Bereich der Presse gelangen darf, wenn diese in
ihrer mittleren Position steht. Angenommen, die Steuerschaltung
stoppt den Roboterarm, bevor er sich auf einen Mindestabstand
$d_s$ der Presse n\"ahert, sofern diese nicht in ihrer mittleren
Position steht, und stellt auch sicher, da\3 die Presse in der
mittleren Position verharrt, solange der Roboterarm innerhalb
des Abstands $d_s$ bleibt. Der Beweis daf\"ur, da\3 eine solche
Schaltung die obige Sicherheitsanforderung erf\"ullt, ben\"otigt
unerwarteterweise den Zwischenwertsatz aus der Analysis.
Abbildung~\ref{Verletzung einer Sicherheitsanforderung durch
unstetige Bewegung} zeigt ein Gegenbeispiel, falls die Bewegung
des Roboterarms nicht stetig ist, die Presse stehe dabei in
ihrer oberen Position. Daher mu\3te f\"ur jede Funktion, deren
Stetigkeit verlangt wird, eine entsprechende Instanz des
Zwischenwertsatzes zur Spezifikation hinzugef\"ugt werden.

\begin{figure}
\begin{center}
\newcommand{\eqd}[1]{\mbox{\bf #1.}}
$\begin{array}{@{}l@{\hspace*{0.5cm}}cl@{}}
\eqd{u20} &&
 \forall [ r,x,t,d]\; ( \\& &
   roboter(r,x)
\\&  \wedge  & dist\_xy(x,pos_{1}(r,t)) \.\leq  d  \.\leq maxlg_{1}
\\&  \ra	 & (	\forall [ t_{1}]\; (t \.\leq  t_{1} <tra_{1}(r,d,t) \ra	 faehrt\_aus_{1}(r,t_{1}))
    \ra	 dist\_xy(x,pos_{1}(r,tra_{1}(r,d,t)))=d )
\\&  \wedge  &  \forall [ t_{3}]\; (t \.\leq  t_{3} <tra_{1}(r,d,t) \ra	dist\_xy(x,pos_{1}(r,t_{3}))<d) )
\\
[0.1cm]
\eqd{u21} &&
 \forall [ r,x,t,t_{2}]\; ( \\& &
   roboter(r,x)
\\&  \wedge  & t \.\leq t_{2}
\\&  \wedge  & dist\_xy(x,pos_{1}(r,t))<dist\_xy(x,pos_{1}(r,t_{2}))
\\&  \ra	 &  \exists [ t_{1}]\; (t< t_{1} <t_{2} \wedge	faehrt\_aus_{1}(r,t_{1})) )
\\
[0.1cm]
\eqd{u30} &&
 \forall [ c_{1},c_{2},c_{3},c_{4},c_{5},c_{6},c_{7},c_{8},s_{1},s_{2},s_{3},x,t]\; ( \\& &
   roboter(r(c_{1},c_{2},c_{3},c_{4},c_{5},c_{6},c_{7},c_{8},s_{1},s_{2},s_{3}),x)
\\&  \ra	 & (faehrt\_aus_{1}(r(c_{1},c_{2},c_{3},c_{4},c_{5},c_{6},c_{7},c_{8},s_{1},s_{2},s_{3}),t)  \lra val(c_{1},t)=1)
\\&  \wedge  & (faehrt\_ein_{1}(r(c_{1},c_{2},c_{3},c_{4},c_{5},c_{6},c_{7},c_{8},s_{1},s_{2},s_{3}),t)	\lra val(c_{2},t)=1)
\\&  \wedge  & (faehrt\_aus_{2}(r(c_{1},c_{2},c_{3},c_{4},c_{5},c_{6},c_{7},c_{8},s_{1},s_{2},s_{3}),t)	\lra val(c_{3},t)=1)
\\&  \wedge  & (faehrt\_ein_{2}(r(c_{1},c_{2},c_{3},c_{4},c_{5},c_{6},c_{7},c_{8},s_{1},s_{2},s_{3}),t)	\lra val(c_{4},t)=1)
\\&  \wedge  & (greift_{1}(r(c_{1},c_{2},c_{3},c_{4},c_{5},c_{6},c_{7},c_{8},s_{1},s_{2},s_{3}),t)	\lra val(c_{5},t)=1)
\\&  \wedge  & (greift_{2}(r(c_{1},c_{2},c_{3},c_{4},c_{5},c_{6},c_{7},c_{8},s_{1},s_{2},s_{3}),t)	\lra val(c_{6},t)=1)
\\&  \wedge  & (dreht\_vor(r(c_{1},c_{2},c_{3},c_{4},c_{5},c_{6},c_{7},c_{8},s_{1},s_{2},s_{3}),t)     \lra val(c_{7},t)=1)
\\&  \wedge  & (dreht\_zurueck(r(c_{1},c_{2},c_{3},c_{4},c_{5},c_{6},c_{7},c_{8},s_{1},s_{2},s_{3}),t)\lra val(c_{8},t)=1)
\\&  \wedge  & (dist\_xy(x,pos_{1}(r(c_{1},c_{2},c_{3},c_{4},c_{5},c_{6},c_{7},c_{8},s_{1},s_{2},s_{3}),t))  =val(s_{1},t))
\\&  \wedge  & (dist\_xy(x,pos_{2}(r(c_{1},c_{2},c_{3},c_{4},c_{5},c_{6},c_{7},c_{8},s_{1},s_{2},s_{3}),t))  =val(s_{2},t))
\\&  \wedge  & (winkel\_xy(x,pos_{1}(r(c_{1},c_{2},c_{3},c_{4},c_{5},c_{6},c_{7},c_{8},s_{1},s_{2},s_{3}),t))=val(s_{3},t)))
\\
[0.1cm]
\eqd{u47b} &&
dist\_xy(d_4,d_3) \leq maxlg_1
\\
[0.1cm]
\eqd{u73} &&
 \forall [ c,v,t]\; ( \\& &
    val(trigger(c,v),t)=1
\\&  \lra  & val(c,t)<v )
\\
[0.1cm]
\eqd{r11} &&
roboter(r(c_1,c_2,c_3,c_4,c_5,c_6,c_7,c_8,s_1,s_2,s_3),d_4)	\\
\end{array}$
\end{center}
\caption{Spezifikation der einfachen Steuerungsschaltung}
\label{Spezifikation der einfachen Steuerungsschaltung}
\end{figure}

\begin{figure}
\begin{center}
\renewcommand{\labelitemi}{}
\newcommand{\eqr}[1]{{\bf #1}}
\newcommand{\eqd}[1]{{\bf #1}}
%\eqs{50}

Finde eine Steuerungsschaltung, die den ersten Roboterarm auf eine
vorgegebene L\"ange $d_{34}$ ausf\"ahrt.

\begin{tabular}{@{}l@{\hspace*{0.5cm}}l@{}}
Formal:
    & $\exists r_0: \; \forall t_0: \; \exists t\;\;$
    \begin{tabular}[t]{@{}l@{$\;$}l@{}}
    & $dist\_xy(d_4,pos1(r_0,t_0)) \.\leq d_{34}$	\\
    $\ra$ & $dist\_xy(d_4,pos1(r_0,t)) \.= d_{34}$	\\
    \end{tabular}
\\
& mit $d_{34}=dist\_xy(d_4,d_3)$
\\[0.2cm]
\mca{2}{Beweis (Skolem--Funktionen markiert mit ``$^{\$}$''):}	
\\[0.2cm]
\eqr{Vor}:
&
$dist\_xy(d_4,pos1(r_0,t_0^{\$})) \.\leq d_{34}$	
\\[0.2cm]
\eqr{Beh}:
&
$dist\_xy(d_4,pos1(r_0,t)) \.= d_{34}$	
\\[0.2cm]
\mca{2}{\eqd{51} = \eqr{u20} rs \eqr{Vor} , \eqr{r11} , \eqr{u47b}:}
\\
&
\begin{tabular}[t]{@{}r@{$\;$}l@{}l@{$\;$}l@{}}
& $($ && $(t_0^{\$} \.\leq t_1 \.< t_2^{\$}
	\;\ra\; faehrt\_aus1(r_0,t_1))$ \\
	&& $\ra$ & $dist\_xy(d_4,pos1(r_0,t_2^{\$})) \.= d_{34})$ \\
$\wedge$ &&& $t_0^{\$} \.\leq t_3 \.< t_2^{\$}
	\;\ra\; dist\_xy(d_4,pos1(r_0,t_3)) \.< d_{34}$	\\[0.2cm]
\end{tabular}
\\[0.2cm]
\eqd{52} = sp \eqr{51}:
&
\begin{tabular}[t]{@{}r@{$\;$}l@{}}
& $(t_0^{\$} \.\leq t_1 \.< t_2^{\$} \;\ra\; faehrt\_aus1(r_0,t_1))$ \\
$\ra$ & $dist\_xy(d_4,pos1(r_0,t_2^{\$})) \.= d_{34}$	\\[0.2cm]
\end{tabular}
\\[0.2cm]
\eqd{53} = sp \eqr{51}:
&
$t_0^{\$} \.\leq t_3 \.< t_2^{\$}
	\;\ra\; dist\_xy(d_4,pos1(r_0,t_3)) \.< d_{34}$
\\[0.2cm]
\eqd{54} = \eqr{52} rs \eqr{u30}:
&
\begin{tabular}[t]{@{}r@{$\;$}l@{}}
& $(t_0^{\$} \.\leq t_1 \.< t_2^{\$} \;\ra\; val(c_1,t_1) \.= 1)$ \\
$\ra$ & $dist\_xy(d_4,pos1(r_0,t_2^{\$})) \.= d_{34}$	\\[0.2cm]
\end{tabular}
\\[0.2cm]
\eqd{55} = \eqr{54} rs \eqr{u73}:
&
\begin{tabular}[t]{@{}r@{$\;$}l@{}}
& $(t_0^{\$} \.\leq t_1 \.< t_2^{\$} \;\ra\; val(c,t_1) \.< d_{34})$ \\
$\ra$ & $dist\_xy(d_4,pos1(r_0,t_2^{\$})) \.= d_{34}$	\\
\end{tabular}
\\
& mit
$r_0=r(trigger(c,d_{34}),c_2,c_3,\ldots,c_8,s_1,s_2,s_3)$
\\[0.2cm]
\eqd{56} = \eqr{55} rp \eqr{u30}:
&
\begin{tabular}[t]{@{}r@{$\;\;$}l@{}}
& $(t_0^{\$} \.\leq t_1 \.< t_2^{\$}
	\;\ra\; dist\_xy(d_4,pos1(r_0,t_1)) \.< d_{34})$	\\
$\ra$ & $dist\_xy(d_4,pos1(r_0,t_2^{\$})) \.= d_{34}$ \\
\end{tabular}
\\
& mit
$r_0=r(trigger(s_1,d_{34}),c_2,c_3,\ldots,c_8,s_1,s_2,s_3)$
\\[0.2cm]
\eqd{57} = \eqr{56} rs \eqr{53}:
&
$dist\_xy(d_4,pos1(r_0,t_2^{\$}))=d_{34}$
\\
& mit
$r_0=r(trigger(s_1,d_{34}),c_2,c_3,\ldots,c_8,s_1,s_2,s_3)$
\\
\end{tabular}
\end{center}
\caption{Synthesebeweis der Schaltung aus Abb.\
	\protect\ref{Einfache Steuerungsschaltung}}
\label{Synthesebeweis der Schaltung aus Abb.
	Einfache Steuerungsschaltung}
\end{figure}

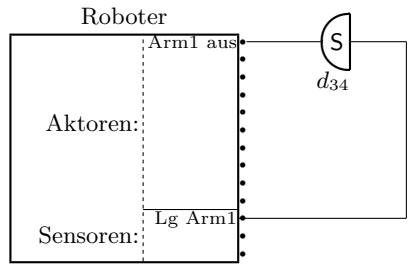
\begin{figure}
\begin{center}
\begin{picture}(5.25,3.5)
	%\put(0,0){\makebox(0,0){+}}
\put(1.500,3.150){\makebox(0.000,0.000)[b]{Roboter}}
\put(1.700,0.344){\makebox(0.000,0.000)[r]{{\footnotesize Sensoren:}}}
\put(1.700,1.844){\makebox(0.000,0.000)[r]{{\footnotesize Aktoren:}}}
\put(3.000,2.920){\makebox(0.000,0.000)[r]{{\scriptsize Arm1 aus}}}
\put(3.000,0.550){\makebox(0.000,0.000)[r]{{\scriptsize Lg Arm1}}}
\multiput(1.75,0)(0,.1){30}{\line(0,1){.05}}
\put(3.000,0.687){\line(-1,0){1.250}}
	\put(3.000,0.570){\line(1,0){2.250}}
	\put(5.250,0.570){\line(0,1){2.350}}
	\put(5.250,2.920){\line(-1,0){0.750}}
	\put(4.125,2.920){\line(-1,0){1.000}}
\put(4.325,2.920){\makebox(0.000,0.000){{\small\sf S}}}
\put(4.275,2.400){\makebox(0.000,0.000){$d_{34}$}}
\multiput(3.075,0.100)(0.000,0.235){13}{\circle*{0.075}}
\thicklines
\put(0.000,0.000){\framebox(3.000,3.000){}}
\put(4.5,2.92){\oval(.75,.75)[l]}
\put(4.500,2.545){\line(0,1){0.750}}
%\put(4.500,2.545){\line(-2,1){0.750}}
%\put(4.500,3.295){\line(-2,-1){0.750}}
\end{picture}
\end{center}
\caption{Einfache Steuerungsschaltung}
\label{Einfache Steuerungsschaltung}
\end{figure}

\subsection{Verifikation}
\label{Verifikation}

Die Steuerschaltung wurde nicht wirklich synthetisiert in dem
Sinne, da\3 aus einem aktuellen Teilbeweisziel viele
Informationen \"uber die zu synthetisierende Schaltung gewonnen
worden w\"aren. Stattdessen wurde eher eine vorher unabh\"angig
vom Beweis gefundene Schaltung verifiziert. Dar\"uber hinaus ist
die Wiederverwendung fr\"uherer Beweisteile erheblich einfacher,
wenn die Beweise vorw\"arts (bottom--up) durchgef\"uhrt werden,
w\"ahrend eine echte Synthese R\"uckw\"artsbeweise (top--down)
verlangt. Aus diesem Grund wurden gro\3e Teile des Beweises
r\"uckw\"arts gef\"uhrt, vgl.\ Abb.~\ref{Synthesebeweis der
Schaltung aus Abb. Einfache Steuerungsschaltung} und
Anhang~\refC. 

Es wurden zwei verschiedene Ans\"atze untersucht, eine
Steuerungsschaltung f\"ur die Fertigungszelle zu synthetisieren.
Der erste Ansatz verwendete ausschlie\3lich Pr\"adikatenlogik
erster Ordnung. Er ging aus von der oben beschriebenen
Spezifikation und bewies ihre Erf\"ullbarkeit. Da Spezifikation
und Verifikation der vollen Fertigungszellensteuerung bereits
sehr kompliziert und un\"ubersichtlich sind, soll hier als
Beispiel f\"ur einen Synthesebeweis die stark vereinfachte
fiktive Aufgabe aus Abb.~\ref{Steuerschleife mit mechanischer
Ruckkoppelung} dienen, einen Roboterarm auf eine bestimmte
L\"ange auszufahren. Abbildung~\ref{Spezifikation der einfachen
Steuerungsschaltung} fa\3t die daf\"ur ben\"otigten
Spezifikationsaxiome aus Anhang~\ref{Formale Spezifikation der
Fertigungszelle} zusammen, Abb.~\ref{Synthesebeweis der
Schaltung aus Abb. Einfache Steuerungsschaltung} zeigt den
Synthesebeweis, Abb.~\ref{Einfache Steuerungsschaltung}
zeigt die synthetisierte Schaltung selbst. 

In Abschnitt~\ref{Entwicklung der Robotersteuerung} wird die
formale Entwicklung der Robotersteuerung gezeigt. In
Abschnitt~\ref{Verwendung hoherer logischer Operatoren} wird der
zweite Ansatz vorgestellt, der aufgrund der gemachten
Erfahrungen mit der Modellierung des Anwendungsgebiets neue
logische Operatoren definierte und dadurch Spezifikation und
Beweis k\"urzer und \"ubersichtlicher zu gestalten erlaubte.

% \begin{figure}
% \begin{center}
% \footnotesize
% \begin{verbatim}
% /* 57 = */      rs(
% /* 56 = */          rp(
% /* 55 = */              rs(
% /* 54 = */                  rs(
% /* 52 = */                      sp(
%                                     lab(
%                                         l1,
% /* 51 = */                              rs(
%                                             rs(
%                                                 rs(
%                                                     user(u20),
%                                                     user(r11)
%                                                 ),
%                                                 user(Vor),
%                                             ),
%                                             user(u47b)
%                                         )
%                                     )
%                                 ),
%                                 user(u30)
%                             ),
%                             user(u73)
%                         ),
%                         user(u30)
%                     ),
% /* 53 = */          sp(
%                         ref(l1)
%                     )
%                 )
% \end{verbatim}
% \end{center}
% \caption{Termdarstellung des Beweises aus Abb.\
%         \protect\ref{Synthesebeweis der Schaltung aus Abb.
%         Einfache Steuerungsschaltung}}
% \label{Termdarstellung des Beweises aus Abb. Synthesebeweis 
%         der Schaltung aus Abb. Einfache Steuerungsschaltung}
% \end{figure}

\begin{figure}
\begin{center}

$$
\begin{array}{ll}

rs(rp(rs(rs( 
	& sp(lab(l1,rs(rs(rs(user(u20),user(r11)),
	user(Vor)),user(u47b)))),	\\
& user(u30) \;\; ), \;\; user(u73) \;\; ), \;\; user(u30) \;\; ), 
	\;\; sp(ref(l1)) \;\; )	\\

\end{array}
$$

\end{center}
\caption{Termdarstellung des Beweises aus Abb.\
        \protect\ref{Synthesebeweis der Schaltung aus Abb.
        Einfache Steuerungsschaltung}}
\label{Termdarstellung des Beweises aus Abb. Synthesebeweis 
        der Schaltung aus Abb. Einfache Steuerungsschaltung}
\end{figure}

\clearpage
\section{Entwicklung der Robotersteuerung}
\label{Entwicklung der Robotersteuerung}

Im folgenden soll das Vorgehen beim Entwurf
anhand der Steuerung des Roboters erl\"autert werden, die wegen der
Koordinationsprobleme der beiden Arme die schwierigste von allen
vorkommenden Maschinensteuerungen darstellt.
Beim Roboter lassen sich vier teilweise ineinander verzahnte
Bewegungsphasen unterscheiden:
\be
\item Arm 1 nimmt ein Metallpl\"attchen vom Hubdrehtisch auf und
	transportiert es in die Presse.
\item Arm 2 nimmt ein bearbeitetes Metallpl\"attchen von der Presse
	auf und
	transportiert es auf das Ablagef\"orderband.
\item Arm 1 f\"ahrt von der Presse (leer) zur\"uck zum Hubdrehtisch.
\item Arm 2 f\"ahrt vom Ablagef\"orderband
	(leer) zur\"uck zur Presse.
\ee

Aufgrund der Konstruktion des Roboters k\"onnen die Phasen 1.\ und 2.
sowie die Phasen 3.\ und 4.\ immer nur gleichzeitig ablaufen.

\begin{figure}
\begin{center}
\begin{tabular}[t]{@{}|r||c|cc|cc||c|cc|c|@{}}
\hline
Ph & \multicolumn{5}{l||}{Beginn}
			     & \multicolumn{4}{l|}{Ende Rob.bewegung} \\
   & Rob     & \multicolumn{2}{c|}{Hub}
		     & \multicolumn{2}{c||}{Prs}
			     & Rob     & \multicolumn{2}{c|}{Prs}
					       &Afb \\
\hline \hline
1. & 150,240 & o & r &   &   & 270,360 & m & - &    \\
   &         & o & - &   &   &         & m & r &    \\ \hline
2. & 180,270 &   &   & u & b & 270,360 &   &   & -  \\
   &         &   &   & u & - &         &   &   & b  \\ \hline
3. & 270,360 &   & r &   &   & 150,240 &   &   &    \\
   &         &   &   &   &   &         &   &   &    \\ \hline
4. & 270,360 &   &   &   &b,r& 180,270 &   &   &    \\
   &         &   &   &   &   &         &   &   &    \\ \hline
\end{tabular}

\vspace{0.3cm}

\parbox{10cm}{
Die obere Zeile zu jeder Phase enth\"alt ihre Vorbedingungen, die
untere ihre Resultate.
Die Position des Roboters wird durch die Winkel seiner beiden Arme
angegeben (vgl.\ Abb.~\ref{Schematische Darstellung der
Fertigungszelle}).  }

\vspace{0.3cm}

Abk\"urzungen:

\begin{tabular}[t]{@{}ll@{}}
o & obere Position	\\
m & mittlere Position	\\
u & untere Position	\\
b & bearbeitetes Metallpl\"attchen	\\
r & unbearbeitetes Metallpl\"attchen	\\
- & leer	\\
\end{tabular}
\caption{Bewegungsphasen des Roboters (schematisch)}
\label{Bewegungsphasen des Roboters (schematisch)}
\end{center}
\end{figure}

Abbildung~\ref{Bewegungsphasen des Roboters (schematisch)} zeigt
die Bewegungsphasen des Roboters schematisch. Man sieht, da\3
die Voraussetzungen f\"ur die Phasen 1.\ und 2. bzw.\ 3.\ und
4.\ jeweils einzeln oder f\"ur gemeinsam erf\"ullt sein
k\"onnen. Die Steuerung mu\3 sicherstellen, da\3 Vor- und
R\"ucklaufphasen des Roboters nie gleichzeitig ablaufen.
W\"ahrend des Ablaufs von Phase 4.\ (R\"ucklauf von Arm 2 zur
Presse) kann der Fall eintreten, da\3 auf dem Hubdrehtisch ein
neues unbearbeitetes Metallpl\"attchen ankommt. In diesem Fall
soll Phase 3.\ zugeschaltet werden, so da\3 der Roboter
seinen Arm 1 zun\"achst bis zum Hubdrehtisch zur\"uckf\"ahrt,
dort das Metallpl\"attchen aufnimmt (Phase 1.) und unterwegs
Phase 2.\ zuschaltet, um die Presse durch Arm 2 zu leeren.

Der Grobentwurf der Steuerung teilt diese in Module f\"ur die
einzelnen Bewegungsphasen auf.
Sie lassen sich, wie in Abb.~\ref{Bewegungsphasen des Roboters}
gezeigt, informell beschreiben, die Ein- und
Ausgabekan\"ale sind in Abb.~\ref{Steuerungsmodule mit Ein- und
Ausgabekanalen} gezeigt.

\begin{figure}
\begin{center}
\be
\item	\begin{tabular}[t]{@{}ll@{}}
	Aufgabe: & transportiere ein Metallpl\"attchen vom Hubdrehtisch
		in die Presse	\\
	Voraussetzungen: & Roboterarm 1 in Position \"uber dem
		Hubdrehtisch (150,240)	\\
		& Hubdrehtisch in oberer Position im richtigen Winkel \\
		& unbearbeitetes Metallpl\"attchen liegt auf
			dem Hubdrehtisch	\\
	\end{tabular}
\item	\begin{tabular}[t]{@{}ll@{}}
	Aufgabe: & transportiere ein Metallpl\"attchen von der Presse
		auf das Ablagef\"orderband	\\
	Voraussetzungen: & Roboterarm 2 in Position in der Presse
		(180,270)	\\
		& Presse in unterer Position	\\
		& bearbeitetes Metallpl\"attchen liegt in der Presse \\
	\end{tabular}
\item	\begin{tabular}[t]{@{}ll@{}}
	Aufgabe: & fahre Arm 1 zur\"uck zum Hubdrehtisch	\\
	Voraussetzungen: & Roboterarm 1 in Position in der Presse
		(270,360)	\\
		& \tab oder Phase 4 l\"auft bereits	\\
		& unbearbeitetes Metallpl\"attchen liegt auf
			dem Hubdrehtisch	\\
	\end{tabular}
\item	\begin{tabular}[t]{@{}ll@{}}
	Aufgabe: & fahre Arm 2 zur\"uck zur Presse	\\
	Voraussetzungen: & Roboterarm 2 \"uber dem
		Ablagef\"orderband (270,360)	\\
		& Metallpl\"attchen (bearbeitet oder unbearbeitet)
			in der Presse \\
	\end{tabular}
\ee
\end{center}
\caption{Bewegungsphasen des Roboters}
\label{Bewegungsphasen des Roboters}
\end{figure}

\begin{figure}
\begin{center}
\begin{tabular}[t]{@{}|l||l|l|l|l|@{}}
\hline
Modul/Phase		& 1. & 2. & 3. & 4. \\
\hline \hline
Roboter Arm 1 L\"ange   & *  &    & *  &    \\
Roboter Arm 2 L\"ange   &    & *  &    & *  \\
Roboter Winkel          & *  & *  & *  & *  \\
\hline
Presse unten            &    &    & *  & *  \\
Presse mitte            & *  &    & *  & *  \\
Ablagef\"orderband frei &    & *  &    &    \\
\hline \hline
Roboter Arm 1 ausfahren &    &    & *  &    \\
Roboter Arm 1 einfahren & *  &    &    &    \\
Roboter Arm 1 greifen   & *  &    &    &    \\
Roboter Arm 2 ausfahren &    & *  &    &    \\
Roboter Arm 2 einfahren &    &    &    & *  \\
Roboter Arm 2 greifen   &    & *  &    &    \\
Roboter vor drehen      & +  & +  &    &    \\
Roboter zur\"uck drehen &    &    & +  & +  \\
\hline
\end{tabular}
\vspace{0.2cm}

\parbox{10cm}{
(Die mit ``+'' gekennzeichneten Ausgaben werden von mehreren Modulen
beeinflu\3t.)
}

\caption{Steuerungsmodule mit Ein- und Ausgabekan\"alen}
\label{Steuerungsmodule mit Ein- und Ausgabekanalen}
\end{center}
\end{figure}

Die Steuerung wird als {\ttl}--\"ahnliche Digitalschaltung konzipiert.
Es mu\3 sichergestellt werden, da\3 Ablauf der Phase 1.\ durch einen
eventuellen gleichzeitigen Ablauf der Phase 2.\ nicht gest\"ort wird
und umgekehrt;
analog f\"ur die Phase 3.\ und 4.

Dazu wird die Spezifikation z.B.\ des Moduls f\"ur Phase 1.\ so
ausgelegt, da\3 keine Voraussetzungen gemacht werden, die durch die
gleichzeitige Aktivierung von Phase 2.\ ung\"ultig werden.
Es wird nur vorausgesetzt, da\3 in Phase 1.\
nicht gleichzeitig die Steuerung f\"ur
das Zur\"uckdrehen des Roboters oder f\"ur das Ausfahren seines ersten
Armes aktiviert wird, was beides in Phase 2.\ nicht erfolgt.
Dadurch k\"onnen
die von mehreren Modulen gemeinsam beeinflu\3ten Ausgaben
jeweils durch ein Oder--Gatter zusammengefa\3t werden.
Abbildung~\ref{Formale Spezifikation des Moduls fur Phase 1}
zeigt die formale Spezifikation des Moduls f\"ur Phase 1.
Die verwendeten Bezeichner sind in Anhang~\ref{Informelle Beschreibung
der verwendeten Bezeichner} informell erl\"autert und in
Anhang~\ref{Formale Spezifikation der Fertigungszelle} formal
definiert.

\begin{figure}
\begin{center}
\begin{tabular}[t]{@{}lll@{}}
\mca{2}{$\forall r,s_0,t_0,ci \; \exists t,co \;\; ($}	\\
& $up(ci,t_0)$	\\
$\wedge$ & $ort(s_0,t_0)=d_3$	\\
$\wedge$ & $pos_1(r,t_0)=d_3$	\\
$\wedge$ & \mca{2}{$winkel(s_0,t_0)=winkel\_xy(d_4,d_3)-90$}	\\
$\wedge$ & \mca{2}{$dist\_xy(d_4,d_5) 
	\leq dist\_xy(d_4,pos_1(r,t_0))$}	\\
$\wedge$ & $minlg_1 \leq dist\_xy(d_4,d_5)$	\\
$\wedge$ & $winkel\_xy(d_4,d_5) \leq 270$	\\
$\wedge$ & \mca{2}{$winkel\_xy(d_4,pos_1(r,t_0))
	\leq winkel\_xy(d_4,d_5)$}	\\
$\wedge$ & $\forall t_1,r \;\;
	(t_0 \leq t_1 \leq trv_1(r,270,t_0) \ra$
	& $\neg val(cfa_1,t_1)=1 \wedge \neg val(cfe_1,t_1)=1 \wedge$ \\
	&& $\neg val(cdrv,t_1)=1 \wedge \neg val(cgr_1,t_1)=1)$ \\
$\wedge$ & \mca{2}{$roboter(r(c_1,c_2,c_3,c_4,c_5,c_6,c_7,c_8,
	s_1,s_2,s_3),d_4)$}	\\
$\ra$ & $proj\_xy(pos_1(r,t))=proj\_xy(d_5)$	\\
$\wedge$ & $ort(s_0,t)=pos_1(r,t)$	\\
$\wedge$ & $winkel(s_0,t)=180$	\\
$\wedge$ & $up(co,t)$	\\
\end{tabular}
\end{center}
\caption{Formale Spezifikation des Moduls f\"ur Phase 1}
\label{Formale Spezifikation des Moduls fur Phase 1}
\end{figure}

Die Steuerungsschaltung enth\"alt Schleifen (R\"uckkoppelungen),
kann also nicht als geschlossener endlicher Term dargestellt
werden. Die Programmsynthese beruht jedoch wesentlich auf
Unifikation, also dem L\"osen von Gleichungen in der freien
Algebra der (endlichen) Terme. Daher ist es nicht m\"oglich, die
Schaltung durch Synthese zu konstruieren, sondern sie mu\3 im
Vorhinein ---~durch ein Gleichungssystem, das einen unendlichen
Term als L\"osung hat~--- angegeben und dann verifiziert werden.

Abbildung~\ref{Steuerungsschaltung des Roboters} zeigt die
Steuerungsschaltung des Roboters unter Verwendung von Modulen f\"ur die
Phasen 1.\ bis 4.
Die in Abb.~\ref{Steuerungsmodule mit Ein- und Ausgabekanalen}
aufgef\"uhrten Ein- und Ausgabekan\"ale, die in direkter Verbindung mit
Sensoren bzw.\ Motorsteuerungen stehen, sind dabei weggelassen.
Es wird au\3erdem
angenommen, da\3 folgende Signale bereits zur Verf\"ugung stehen
(vgl.\ Abb.~\ref{Schematische Darstellung der Fertigungszelle}):
\bi
\item $(150,240)$ liefert den Wert $1$ genau dann, wenn sich Roboterarm
	1 \"uber dem Hubdrehtisch befindet.
\item $(180,270)$ liefert den Wert $1$ genau dann, wenn sich Roboterarm
	2 in der Presse befindet.
\item $(270,360)$ liefert den Wert $1$ genau dann, wenn sich Roboterarm
	1 in der Presse und Arm 2 \"uber dem Ablagef\"orderband
	befindet.
\item $(prs \;\; u,b)$ liefert den Wert $1$ genau dann, wenn sich die
	Presse in unterer Postition befindet und ein bearbeitetes
	Metallpl\"attchen enth\"alt.
\item $(prs \;\; *,rb)$ liefert den Wert $1$ genau dann, wenn die Presse
	ein bearbeitetes  oder unbearbeitetes
	Metallpl\"attchen enth\"alt.
\item $(hub \;\; o,r)$ liefert den Wert $1$ genau dann, wenn sich der
	Hubdrehtisch in oberer Position und im richtigen Winkeln zur
	Entladung durch Roboterarm 1 befindet und ein unbearbeitetes
	Me\-tall\-pl\"att\-chen darauf liegt.
\item $(hub \;\; *,r)$ liefert den Wert $1$ genau dann, wenn ein
	unbearbeitetes Metallpl\"attchen auf dem Hubdrehtisch liegt.
\ei

Diese Signale sind leicht aus den Sensoren und den
Steuerungsmodulen f\"ur die anderen Maschinen zu gewinnen,
die Vorgehensweise dazu wird anhand der Synthese der Schaltung
f\"ur Modul 1.\ deutlich.

\begin{figure}
\begin{center}
\begin{picture}(7,6)
	%\put(0,0){\makebox(0,0){+}}
\put(1.000,0.200){\circle{0.080}}			% 150,240
	\put(1.000,0.200){\line(-1,0){0.625}}
	\put(0.375,0.200){\line(0,1){3.000}}
\put(1.000,0.700){\circle{0.080}}			% hub o,r
	\put(1.000,0.700){\line(-1,0){0.500}}
	\put(0.500,0.700){\line(0,1){2.500}}
\put(2.500,1.700){\circle{0.080}}			% prs u,b
	\put(2.500,1.700){\line(-1,0){0.625}}
	\put(1.875,1.700){\line(0,1){1.500}}
\put(2.500,2.200){\circle{0.080}}			% 180,270
	\put(2.500,2.200){\line(-1,0){0.500}}
	\put(2.000,2.200){\line(0,1){1.000}}
\put(4.500,0.200){\circle{0.080}}			% prs *,rb
	\put(4.500,0.200){\line(1,0){2.125}}
	\put(6.625,0.200){\line(0,1){3.000}}
\put(4.500,0.700){\circle{0.080}}			% 270,360
	\put(4.500,0.700){\line(1,0){0.625}}
	\put(5.125,0.700){\line(0,1){1.700}}
	\put(5.125,1.450){\circle*{0.080}}
	\put(5.125,1.450){\line(1,0){1.375}}
	\put(6.500,1.450){\line(0,1){1.750}}
\put(4.500,1.200){\circle{0.080}}			% hub *,r
	\put(4.500,1.200){\line(1,0){0.375}}
	\put(4.875,1.200){\line(0,1){2.000}}
\put(0.500,5.450){\line(0,1){0.500}}			% output 1
	\put(0.500,5.950){\line(1,0){2.625}}
	\put(3.125,5.950){\line(0,-1){1.200}}
\put(2.000,5.450){\line(0,1){0.250}}			% output 2
	\put(2.000,5.700){\line(1,0){0.875}}
	\put(2.875,5.700){\line(0,-1){0.950}}
\put(5.000,5.450){\line(0,1){0.250}}			% output 3
	\put(5.000,5.700){\line(-1,0){0.875}}
	\put(4.125,5.700){\line(0,-1){0.950}}
\put(6.500,5.450){\line(0,1){0.500}}			% output 4
	\put(6.500,5.950){\line(-1,0){2.625}}
	\put(5.750,5.950){\circle*{0.080}}
	\put(5.750,5.950){\line(0,-1){4.000}}
	\put(5.750,1.950){\line(-1,0){0.375}}
	\put(5.375,1.950){\line(0,1){0.450}}
	\put(3.875,5.950){\line(0,-1){1.200}}
\put(3.000,3.950){\line(0,-1){1.000}}			% nor 1,2
	\put(3.000,2.950){\line(1,0){3.375}}
	\put(5.000,2.950){\circle*{0.080}}
	\put(5.000,2.950){\line(0,1){0.250}}
	\put(6.375,2.950){\line(0,1){0.250}}
\put(4.000,3.950){\line(0,-1){1.250}}			% nor 3,4
	\put(4.000,2.700){\line(-1,0){3.375}}
	\put(2.125,2.700){\circle*{0.080}}
	\put(2.125,2.700){\line(0,1){0.500}}
	\put(0.625,2.700){\line(0,1){0.500}}
\put(5.250,2.450){\line(0,1){0.250}}			% or(inp 3)
	\put(5.250,2.700){\line(-1,0){0.125}}
	\put(5.125,2.700){\line(0,1){0.500}}
\put(0.500,3.450){\line(0,1){0.500}}			% input 1
\put(2.000,3.450){\line(0,1){0.500}}			% input 2
\put(5.000,3.450){\line(0,1){0.500}}			% input 3
\put(6.500,3.450){\line(0,1){0.500}}			% input 4
\put(4.400,0.200){\makebox(0.000,0.000)[r]{prs *,rb}}	% Eing\"ange
\put(4.400,0.700){\makebox(0.000,0.000)[r]{270,360}}
\put(4.400,1.200){\makebox(0.000,0.000)[r]{hub *,r}}
\put(1.100,0.200){\makebox(0.000,0.000)[l]{150,240}}
\put(1.100,0.700){\makebox(0.000,0.000)[l]{hub o,r}}
\put(2.600,1.700){\makebox(0.000,0.000)[l]{prs u,b}}
\put(2.600,2.200){\makebox(0.000,0.000)[l]{180,270}}
\thicklines
\put(0.000,3.950){\framebox(1.000,1.500){\large\bf 1}}	% modul 1
	\put(0.500,3.950){\makebox(0.000,0.000)[b]
		{$\scriptstyle \blacktriangle$}}
	\put(0.500,4.200){\makebox(0.000,0.000)[b]
		{\scriptsize ci}}
	\put(0.500,5.500){\makebox(0.000,0.000)[t]
		{$\scriptstyle \blacktriangle$}}
	\put(0.500,5.250){\makebox(0.000,0.000)[t]
		{\scriptsize co}}
\put(1.500,3.950){\framebox(1.000,1.500){\large\bf 2}}	% modul 2
	\put(2.000,3.950){\makebox(0.000,0.000)[b]
		{$\scriptstyle \blacktriangle$}}
	\put(2.000,5.500){\makebox(0.000,0.000)[t]
		{$\scriptstyle \blacktriangle$}}
\put(4.500,3.950){\framebox(1.000,1.500){\large\bf 3}}	% modul 3
	\put(5.000,3.950){\makebox(0.000,0.000)[b]
		{$\scriptstyle \blacktriangle$}}
	\put(5.000,5.500){\makebox(0.000,0.000)[t]
		{$\scriptstyle \blacktriangle$}}
\put(6.000,3.950){\framebox(1.000,1.500){\large\bf 4}} % modul 4
	\put(6.500,3.950){\makebox(0.000,0.000)[b]
		{$\scriptstyle \blacktriangle$}}
	\put(6.500,5.500){\makebox(0.000,0.000)[t]
		{$\scriptstyle \blacktriangle$}}
\put(0.250,3.200){\line(1,0){0.500}}			% and input 1
	\put(0.500,3.200){\oval(0.500,0.500)[t]}
\put(1.750,3.200){\line(1,0){0.500}}			% and input 2
	\put(2.000,3.200){\oval(0.500,0.500)[t]}
\put(4.750,3.200){\line(1,0){0.500}}			% and input 3
	\put(5.000,3.200){\oval(0.500,0.500)[t]}
\put(5.000,2.200){\line(1,0){0.500}}			% or input 3
	\put(5.250,2.200){\oval(0.500,0.500)[t]}
\put(6.250,3.200){\line(1,0){0.500}}			% and input 4
	\put(6.500,3.200){\oval(0.500,0.500)[t]}
\put(2.750,4.950){\line(1,0){0.500}}			% or 1,2
	\put(3.000,4.950){\oval(0.500,0.500)[b]}
\put(3.000,4.700){\line(0,-1){0.500}}
\put(2.750,4.200){\line(1,0){0.500}}			% neg or(1,2)
	\put(3.000,4.200){\oval(0.500,0.500)[b]}
	\put(3.000,3.950){\circle*{0.100}}
\put(3.750,4.950){\line(1,0){0.500}}			% or 3,4
	\put(4.000,4.950){\oval(0.500,0.500)[b]}
\put(4.000,4.700){\line(0,-1){0.500}}
\put(3.750,4.200){\line(1,0){0.500}}			% neg or(3,4)
	\put(4.000,4.200){\oval(0.500,0.500)[b]}
	\put(4.000,3.950){\circle*{0.100}}
\end{picture}
\vspace{0.2cm}
\par
Termdarstellung der Schaltung:
\par
\begin{tabular}[t]{@{}l@{$\;$}l@{}}
$c1$ & $=modul1(and(rob\_150\_240,hub\_o\_r,neg(or(c3,c4))))$\\
$c2$ & $=modul2(and(rob\_180\_270,prs\_u\_b,neg(or(c3,c4))))$ \\
$c3$ & $=modul3(and(or(rob\_270\_360,c4),hub\_r,neg(or(c1,c2))))$ \\
$c4$ & $=modul4(and(rob\_270\_360,prs\_rb,neg(or(c1,c2))))$ \\
\end{tabular}
\caption{Steuerungsschaltung des Roboters}
\label{Steuerungsschaltung des Roboters}
\end{center}
\end{figure}

Bei der Synthese der Schaltungen f\"ur die Module 1.\ bis 4.\
wurden jeweils zun\"achst die notwendigen Zeitbedingungen f\"ur
die Steuerung hergeleitet, danach wurde eine {\ttl}--artigen
Steuerschaltung konstruiert, die diese Zeitbedingungen
erf\"ullt. 
Anhang~\refC\ zeigt den Beweisbaum f\"ur den Hauptteil der
Verifikation des Moduls 1 (Roboter transportiert ein
Metallpl\"attchen vom Hubdrehtisch in die Presse). Er entspricht
im Wesentlichen der von {\sysyfos} erzeugten Ausgabe, wurde aber
aus Lesbarkeitsgr\"unden manuell nachbearbeitet (Umbruch von
Formeln, Umbenennung von Variablen). 

Jeder Knoten beginnt mit der Formelnummer, unmittelbar gefolgt
von der Operation, aus der die Formel entstanden ist (siehe
Abb.~\ref{Struktur des sysyfos--Systems} f\"ur die verwendeten
Abk\"urzungen). F\"ur Axiome wurde dabei ihr eindeutiger Name
angegeben; die Angabe ``{\tt **}'' steht f\"ur eine bereits
fr\"uher verwendete Formel. Schlie\3lich folgt entweder ``{\tt
$F$ , -}'' f\"ur Assertions oder ``{\tt - , $F$}'' f\"ur Goals,
wobei $F$ die aktuelle Formel in {\prolog}--Notation ist (siehe
Abb.~\ref{PROLOG--Notation fur Relationen und Junktoren}). Ein
``{\tt \$}'' nach einem Namen deutet an, da\3 es sich um eine
Skolemfunktion handelt, deren Argumente nicht angezeigt werden;
ein ``{\tt ()}'' deutet an, da\3 die Argumente manuell entfernt
wurden. Man beachte, da\3 \"Aquivalenzen vom System in
Konjunktionen von Implikationen aufgel\"ost werden, um
eindeutige Polarit\"aten zu erreichen. 

Jede Formel ergibt sich aus der (bzw.\ den beiden) dar\"uber liegenden
um zwei Spalten nach rechts einger\"uckten Ausgangsformel(n), je nach
Stelligkeit der angewendeten Operation.
Weiter entfernte Ausgangsformeln werden durch senkrechte Striche
verbunden.

Der Beweisbaum wird automatisch soweit als m\"oglich linearisiert.
Dazu werden in manchen F\"allen die beiden Ausgangsformeln einer
zweistelligen Operation vertauscht, was durch Gro\3schreibung der
entsprechenden Abk\"urzung angedeutet ist.
Zum Beispiel ist die Formel $158$ in Anhang~\refC\ entstanden
durch die Operation $rp(114,157)$.
Man beachte, da\3 an der Wurzel des Beweisbaums die noch offene
Rest-Beweisverpflichtung steht (Formel $325$).

\begin{figure}
\begin{center}
\begin{tabular}[t]{@{}l@{\hspace*{0.5cm}}l@{\hspace*{0.5cm}}c@{}}
{\tt =} & $=$ & (h\"ochste Bindungsst\"arke)	\\
{\tt =<} & $\leq$	\\
{\tt <} & $<$	\\
{\tt \verb+~+} & $\neg$ & .	\\
{\tt \&} & $\wedge$ & .	\\
{\tt !} & $\vee$ & .	\\
{\tt ->} & $\ra$	\\
{\tt <-} & $\la$ & (niedrigste Bindungsst\"arke)	\\
\end{tabular}
\end{center}
\caption{{\prolog}--Notation f\"ur Relationen und Junktoren}
\label{PROLOG--Notation fur Relationen und Junktoren}
\end{figure}

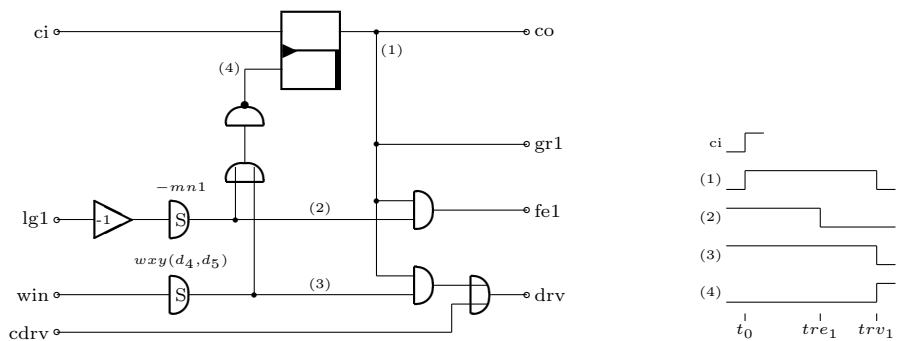
\begin{figure}
\begin{center}
\begin{picture}(12.0,4.7)
	%\put(0,0){\makebox(0,0){+}}
\put(0.800,0.750){\line(1,0){1.500}}			% win
	\put(0.800,0.750){\circle{0.080}}
	\put(0.700,0.750){\makebox(0.000,0.000)[r]{\scriptsize win}}
\put(2.550,0.750){\line(1,0){3.000}}			% trig(win)
	\put(3.425,0.750){\circle*{0.080}}
	\put(3.425,0.750){\line(0,1){1.700}}
	\put(4.300,0.800){\makebox(0.000,0.000)[b]{\tiny (3)}}
\put(0.800,1.750){\line(1,0){0.500}}			% lg1
	\put(0.800,1.750){\circle{0.080}}
	\put(0.700,1.750){\makebox(0.000,0.000)[r]{\scriptsize lg1}}
\put(1.800,1.750){\line(1,0){0.500}}			% ampl(lg1)
\put(2.550,1.750){\line(1,0){3.000}}			% trig(ampl)
	\put(3.175,1.750){\circle*{0.080}}
	\put(3.175,1.750){\line(0,1){0.700}}
	\put(4.300,1.800){\makebox(0.000,0.000)[b]{\tiny (2)}}
\put(3.300,2.500){\line(0,1){0.500}}			% or(trig,trig)
\put(3.300,3.250){\line(0,1){0.500}}			% neg(or)
	\put(3.300,3.750){\line(1,0){0.500}}
	\put(3.250,3.750){\makebox(0.000,0.000)[r]{\tiny (4)}}
\put(0.800,4.250){\line(1,0){3.000}}			% ci
	\put(0.800,4.250){\circle{0.080}}
	\put(0.700,4.250){\makebox(0.000,0.000)[r]{\scriptsize ci}}
\put(4.550,4.250){\line(1,0){2.500}}			% dff
	\put(5.050,4.250){\circle*{0.080}}
	\put(7.050,4.250){\circle{0.080}}
	\put(7.150,4.250){\makebox(0.000,0.000)[l]{\scriptsize co}}
	\put(5.100,4.000){\makebox(0.000,0.000)[l]{\tiny (1)}}
	\put(5.050,4.250){\line(0,-1){3.250}}
	\put(5.050,2.750){\circle*{0.080}}
	\put(5.050,2.000){\circle*{0.080}}
	\put(5.050,2.000){\line(1,0){0.500}}
	\put(5.050,1.000){\line(1,0){0.500}}
\put(5.050,2.750){\line(1,0){2.000}}			% gr1
	\put(7.050,2.750){\circle{0.080}}
	\put(7.150,2.750){\makebox(0.000,0.000)[l]{\scriptsize gr1}}
\put(5.800,1.875){\line(1,0){1.250}}			% fe1
	\put(7.050,1.875){\circle{0.080}}
	\put(7.150,1.875){\makebox(0.000,0.000)[l]{\scriptsize fe1}}
\put(6.550,0.750){\line(1,0){0.500}}			% drv
	\put(7.050,0.750){\circle{0.080}}
	\put(7.150,0.750){\makebox(0.000,0.000)[l]{\scriptsize drv}}
\put(5.800,0.875){\line(1,0){0.725}}			% and(dff,trig)
\put(0.800,0.250){\line(1,0){5.250}}			% drv*
	\put(0.800,0.250){\circle{0.080}}
	\put(0.700,0.250){\makebox(0.000,0.000)[r]{\scriptsize cdrv}}
	\put(6.050,0.250){\line(0,1){0.375}}
	\put(6.050,0.625){\line(1,0){0.475}}
\thicklines
\put(2.300,0.500){\line(0,1){0.500}}			% trig win
	\put(2.300,0.750){\oval(0.500,0.500)[r]}
	\put(2.450,1.750){\makebox(0.000,0.000){\scriptsize S}}
	\put(2.450,1.100){\makebox(0.000,0.000)[b]
			{$\scriptscriptstyle wxy(d_4,d_5)$}}
\put(2.300,1.500){\line(0,1){0.500}}			% trig lg1
	\put(2.300,1.750){\oval(0.500,0.500)[r]}
	\put(2.450,0.750){\makebox(0.000,0.000){\scriptsize S}}
	\put(2.450,2.100){\makebox(0.000,0.000)[b]
			{$\scriptscriptstyle -mn1$}}
\put(5.550,1.625){\line(0,1){0.500}}			% and fe1
	\put(5.550,1.875){\oval(0.500,0.500)[r]}
\put(5.550,0.625){\line(0,1){0.500}}			% and drv
	\put(5.550,0.875){\oval(0.500,0.500)[r]}
\put(6.300,0.500){\line(0,1){0.500}}			% or drv
	\put(6.300,0.750){\oval(0.500,0.500)[r]}
\put(3.050,2.250){\line(1,0){0.500}}			% or trig,trig
	\put(3.300,2.250){\oval(0.500,0.500)[t]}
\put(3.050,3.000){\line(1,0){0.500}}			% neg or
	\put(3.300,3.000){\oval(0.500,0.500)[t]}
	\put(3.300,3.275){\circle*{0.100}}
\put(1.300,1.500){\line(0,1){0.500}}			% ampl lg1
	\put(1.300,1.500){\line(2,1){0.500}}
	\put(1.300,2.000){\line(2,-1){0.500}}
	\put(1.425,1.750){\makebox(0.000,0.000){\tiny -1}}
\put(3.800,3.500){\framebox(0.750,1.000){}}		% dff
	\put(3.800,4.000){\line(1,0){0.750}}
	\put(3.775,4.000){\makebox(0.000,0.000)[l]
			{$\blacktriangleright$}}
	\put(4.500,3.500){\makebox(0.000,0.000)[bl]
			{\rule{0.05cm}{0.5cm}}}
\thinlines
\put(9.700,0.650){\line(1,0){2.000}}		% neg(or(trig,trig))
	\put(11.700,0.650){\line(0,1){0.250}}
	\put(11.700,0.900){\line(1,0){0.250}}
	\put(9.650,0.775){\makebox(0.000,0.000)[r]{\tiny (4)}}
\put(9.700,1.400){\line(1,0){2.000}}		% trig(win)
	\put(11.700,1.400){\line(0,-1){0.250}}
	\put(11.700,1.150){\line(1,0){0.250}}
	\put(9.650,1.275){\makebox(0.000,0.000)[r]{\tiny (3)}}
\put(9.700,1.900){\line(1,0){1.250}}		% trig(ampl(lg1))
	\put(10.950,1.900){\line(0,-1){0.250}}
	\put(10.950,1.650){\line(1,0){1.000}}
	\put(9.650,1.775){\makebox(0.000,0.000)[r]{\tiny (2)}}
\put(9.700,2.150){\line(1,0){0.250}}		% dff
	\put(9.950,2.150){\line(0,1){0.250}}
	\put(9.950,2.400){\line(1,0){1.750}}
	\put(11.700,2.400){\line(0,-1){0.250}}
	\put(11.700,2.150){\line(1,0){0.250}}
	\put(9.650,2.275){\makebox(0.000,0.000)[r]{\tiny (1)}}
\put(9.700,2.650){\line(1,0){0.250}}		% start
	\put(9.950,2.650){\line(0,1){0.250}}
	\put(9.950,2.900){\line(1,0){0.250}}
	\put(9.650,2.775){\makebox(0.000,0.000)[r]{\tiny ci}}
\put(9.950,0.500){\line(0,-1){0.100}}
	\put(10.950,0.500){\line(0,-1){0.100}}
	\put(11.700,0.500){\line(0,-1){0.100}}
	\put(9.950,0.350){\makebox(0.000,0.000)[t]{$\scriptstyle t_0$}}
	\put(10.950,0.350){\makebox(0.000,0.000)[t]
				{$\scriptstyle tre_1$}}
	\put(11.700,0.350){\makebox(0.000,0.000)[t]
				{$\scriptstyle trv_1$}}
\end{picture}
\caption{Synthetisierte Schaltung f\"ur Modul 1}
\label{Synthetisierte Schaltung fur Modul 1}
\end{center}
\end{figure}

Die Argumente der Funktion $r$ wurden aus Lesbarkeitsgr\"unden
fast \"uberall von Hand entfernt. Ausnahmen bilden nur die
Formeln $185$, $203$, $238$, $276$ und $325$, die neben Formel
$163$ die Wurzeln der gro\3en Teilbeweisb\"aume bilden.
Abbildung~\ref{Termdarstellung der Robotersteuerung fur Modul 1}
zeigt die Termdarstellung der Robotersteuerung f\"ur Modul 1,
wie sie aus Formel $325$ in Anhang~\refC\ entnommen werden kann.
Abbildung~\ref{Synthetisierte Schaltung fur Modul 1} zeigt die
entsprechende Schaltung unter Verwendung der \"ublichen Symbole.
Die Oder--Gatter mit dem Eingang $cgr_1$ bzw.\ $cfe_1$ wurden
bereits weggelassen, da sie sich nach der Synthese aller vier
Robotersteuerungs--Module als \"uberfl\"ussig herausstellen.

\begin{figure}
\begin{center}
\footnotesize
\begin{verbatim}
r(
  /* Arm 1 ausfahren: */     cfa1,
  /* Arm 1 einfahren: */     or(cfe1,
                                and(dff(ci,
                                        neg(or(trg(ampl(s22,
                                                        -1
                                                       ),
                                                   dxy(d4,d5)*-1
                                                  ),
                                               trg(s10,
                                                   wxy(d4,d5)
                                        )  )  )   ),
                                    trg(ampl(s22,
                                             -1
                                            ),
                                        dxy(d4,d5)*-1
                               )   )   ),
  /* Arm 2 ausfahren: */     c31,
  /* Arm 2 einfahren: */     c32,
  /* Arm 1 greifen: */       or(cgr1,
                                dff(ci,
                                    neg(or(trg(ampl(s22,
                                                    -1
                                                   ),
                                               dxy(d4,d5)*-1
                                              ),
                                           trg(s10,
                                               wxy(d4,d5)
                               )   )   )  )   ),
  /* Arm 2 greifen: */       c34,
  /* vordrehen: */           or(cdrv,
                                and(dff(ci,
                                        neg(or(trg(ampl(s22,
                                                        -1
                                                       ),
                                                   dxy(d4,d5)*-1
                                                  ),
                                               trg(s10,
                                                   wxy(d4,d5)
                                       )   )  )   ),
                                    trg(s10,
                                        wxy(d4,d5)
                               )   )   ),
  /* zurueckdrehen: */       c36,
  /* Laenge Arm 1: */        s22,
  /* Laenge Arm 2: */        s23,
  /* Winkel Arm 1: */        s10
)
\end{verbatim}
\end{center}
\caption{Termdarstellung der Robotersteuerung f\"ur Modul 1}
\label{Termdarstellung der Robotersteuerung fur Modul 1}
\end{figure}

\clearpage

\section{Verwendung h\"oherer logischer Operatoren}
\label{Verwendung hoherer logischer Operatoren}

Ein zweiter Ansatz verwendete die bisher gemachten Erfahrungen und
identifizierte h\"ohere Sprachkonzepte, die es erlaubten, sowohl
die Spezifikation als auch den Verifikationsbeweis auf eine
h\"o\-he\-re Ausdrucksebene zu heben. Es wurden zwei neue
dreistellige logische Operatoren definiert, indem sie auf eine
eingeschr\"ankte Form der Pr\"adikatenlogik zweiter Ordnung
zur\"uckgef\"uhrt wurden, siehe Abb.~\ref{Anwendungsspezifische
logische Operatoren}. 

\begin{figure}
\begin{center}
\begin{tabular}[t]{@{}l@{$\;$}l@{$\;$}l@{$\;$}l@{}}
$unt(t_0,P,Q)$ & $\Lra$
	& $\forall t_1: \;\; t_1 \.< t_0$
	& $\;\vee\; (\exists t:\;\; t_0 \.\leq t \.\leq t_1 \wedge Q(t))
		\;\vee\; P(t_1)$	\\
$ldt(t_0,P,Q)$ & $\Lra$
	& $\exists t_1: \;\; t_0 \.\leq t_1$
	& $\;\wedge\; (\forall t: \;\;
		(t_0 \.\leq t \.\leq t_1 \ra P(t)) \;\ra\; Q(t_1))$ \\
	&&& $\;\wedge\; (\forall t: \;\;
		t_0 \.\leq t \.\leq t_1 \ra \neg Q(t))$	\\
\end{tabular}
\end{center}
\caption{Anwendungsspezifische logische Operatoren}
\label{Anwendungsspezifische logische Operatoren}
\end{figure}

Die Konzepte wurden von der Sprache {\em Unity\/} von Misra und
Chandy \cite{Chandy.Misra.1988} \"ubernommen. Die Formel
$unt(t_0,P,Q)$ besagt, da\3 ab dem Zeitpunkt $t_0$ das
einstellige Pr\"adikat $Q$ solange wahr ist, bis das einstellige
Pr\"adikat $P$ wahr wird, ggf.\ f\"ur immer (``$P$ until
$Q$''). Die Formel $ldt(t_0,P,Q)$ besagt, da\3, wenn ab dem
Zeitpunkt $t_0$ das Pr\"adikat $P$ nur lange genug gilt, $Q$
irgendwann wahr wird, und da\3 es daf\"ur einen fr\"uhesten Zeitpunkt
$t_1$ gibt (``$P$ leads to $Q$'').

\begin{figure}
\begin{center}
\begin{tabular}[t]{@{}ll@{\hspace*{0.5cm}}l@{}}
$\forall t_0: unt(t_0,P,Q)
	\lra \forall t: t_0 \leq t
	\wedge (\forall t_1:
	t_0 \leq t_1 <t \ra \neg Q(t_1) )
	\ra P(t)$	\\
$\forall t_0: P(t_0) \ra unt(t_0,P,Q)$	\\
$\forall t_0: unt(t_0,P,Q) \ra P(t_0)$	\\
$\forall t_0: P(t_0) \vee Q(t_0) 
	\lra unt(t_0,P,Q) \vee unt(t_0,Q,P)$	\\
$\forall t_0: unt(t_0,P, \neg P)$	\\
$\forall t_0: unt(t_0,P,Q) \vee unt(t_0, \neg Q, \neg P)$	\\
$\forall t_0: unt(t_0,P,Q \vee R)
	\lra unt(t_0,P,Q) \vee unt(t_0,P,R)$	\\
$\forall t_0: unt(t_0,P,false)
	\lra \forall t:(t_0 \leq t \ra P(t))$	\\
$\forall t_0: unt(t_0,P,Q) \wedge unt(t_0,R,S)
	\ra unt(t_0,P \vee R,Q \wedge S)$	\\
$\forall t_0: unt(t_0,P,Q) \wedge unt(t_0,R,S)
	\ra unt(t_0,P \wedge R,Q \vee S)$	\\
$\forall t_0: unt(t_0,P,Q) \wedge unt(t_0,R, \neg P)
	\ra unt(t_0,P \wedge R,Q)$	\\
$\forall t_0: unt(t_0,true,Q)$	\\
$\forall t_0: unt(t_0,false,Q) 
	\lra ( \forall t: t_0 \leq t
	\ra \exists t_1: t_0 \leq t_1<t
	\wedge Q(t_1))$	\\
$\forall t_0: (Q \ra \neg R)
	\ra (ldt(t_0,P \wedge R,Q) \lra ldt(t_0,P,Q))$	\\
$\forall t_0: t_0 \leq t_2 \wedge Q(t_2)
	\ra ldt(t_0,true,Q)$	\\
$\forall t_0: ldt(t_0,P,Q) \wedge ( \forall t:R(t))
	\ra ldt(t_0,P,Q \wedge R)$	\\
$\forall t_0: ldt(t_0,true,P)
	\lra \forall t_0: \exists t_1:
	t_0 \leq t_1 \wedge P(t_1)$ \\
$\forall t_0: ldt(t_0,\neg P,P)
	\ra \exists t: t_0 \leq t \wedge P(t)$	\\
$\forall t_0: ldt(t_0,P,Q) \wedge unt(t_0, \neg R,Q)
	\ra ldt(t_0,P,Q \vee R)$	\\
$\forall t_0: ldt(t_0,P \wedge Q,R) \wedge unt(t_0,P,R) 
	\ra ldt(t_0,Q,R)$	\\
$\forall t_0: unt(t_0,P,Q) \wedge t_0 \leq t_1
	\wedge \neg P(t_1)
	\ra ldt(t_0,P,Q)$	\\
\end{tabular}
\end{center}
\caption{Hintergrundtheorie zu $unt$ und $ldt$}
\label{Hintergrundtheorie zu $unt$ und $ldt$}
\end{figure}

Es wurde eine Hintergrundtheorie mit n\"utzlichen Eigenschaften
von $unt$ und $ldt$ bewiesen, einschlie\3lich der Monotonie von
$unt$ im zweiten und dritten und von $ldt$ im dritten Argument,
sowie der Anti--Monotonie von $ldt$ im zweiten Argument, so da\3
sich beide Operatoren problemlos in die in
Abschnitt~\ref{Nicht--Klausel--Resolution} erl\"auterte
polarit\"atsbasierte Nicht--Klausel--Resolution einbeziehen
lassen. Abbildung~\ref{Hintergrundtheorie zu $unt$ und $ldt$}
zeigt die wichtigsten Axiome der Hintergrundtheorie, dabei sind
$P$, $Q$, $R$ und $S$ Variablen f\"ur einstellige Pr\"adikate. 

\begin{figure}
\begin{center}
\newcommand{\eqr}[1]{{\bf #1}}
\newcommand{\eqd}[1]{{\bf #1}}
%\eqs{10}
\begin{tabular}{@{}l@{\hspace*{0.5cm}}l@{}}
\eqd{u20'}:
&
\begin{tabular}[t]{@{}r@{$\;\;$}l@{$\;$}l@{}}
$\forall r,x,t,d:$
	& \mca{2}{$robot(r,x)$}	\\
	$\wedge$ & \mca{2}{$dist\_xy(x,pos1(r,t)) \leq d
		\leq maxlg_1$} \\
$\ra$ & $ldt(t,$ & $\lambda t_1 \!: faehrt\_aus1(r,t_1),$	\\
	&& $\lambda t_2 \!: dist\_xy(x,pos1(r,t_2)) \.\geq d)$ \\[0.2cm]
\end{tabular}	
\\[0.2cm]
\eqd{61}:
&
$ldt(t_0,\neg P,P) \;\ra\; \exists t \;\; t_0 \.\leq t \wedge P(t)$ 
\\[0.2cm]
\mca{2}{\eqd{71} = \eqr{u20'} rs \eqr{Vor} , \eqr{r11} , \eqr{u47b}:}
\\[0.2cm]
&
\begin{tabular}[t]{@{}l@{$\;\;$}l@{}}
$ldt(t_0,$ & $\lambda t_1 \!: faehrt\_aus1(r,t_1),$	\\
	& $\lambda t_2 \!: dist\_xy(x,pos1(r,t_2)) \.\geq d_{34})$ 
	\\[0.2cm]
\end{tabular}	
\\[0.2cm]
\eqd{72} = \eqr{u30} rs \eqr{u73}:
& $dist\_xy(x,pos1(r_0,t)) \.< d_{34} \;\ra\; faehrt\_aus1(r_0,t_1)$ \\
& mit
$r_0=r(trigger(s_1,d_{34}),c_2,c_3,\ldots,s_3)$
wie in Abb.~\ref{Synthesebeweis der Schaltung aus Abb. Einfache
Steuerungsschaltung}
\\[0.2cm]
\eqd{73} = \eqr{71} rs \eqr{72}:
&
\begin{tabular}[t]{@{}l@{$\;\;$}l@{}}
$ldt(t_0,$ & $\lambda t_1 \!: dist\_xy(x,pos1(r_0,t_1)) \.< d_{34},$ \\
	& $\lambda t_2 \!: dist\_xy(x,pos1(r,t_2)) \.\geq d_{34})$ 
	\\[0.2cm]
\end{tabular}
\\[0.2cm]
\eqd{74} = \eqr{73} rs \eqr{61}:
& $\exists t_2: \;\; dist\_xy(x,pos1(r,t_2)) \.\geq d_{34}$	\\
\end{tabular}
\end{center}
\caption{Synthesebeweis der Schaltung aus Abb.\
	\protect\ref{Einfache Steuerungsschaltung}
	unter Verwendung h\"oherer logischer Operatoren}
\label{Synthesebeweis der Schaltung aus Abb.
	Einfache Steuerungsschaltung
	unter Verwendung hoherer logischer Operatoren}
\end{figure}

Da die Operatoren $unt$ und $ldt$ h\"aufig vorkommenden Mustern
in der Spezifikation und im Beweis entsprechen, konnten beide
durch ihre Verwendung k\"urzer und leichter verst\"andlich
gestaltet werden. Abbildung~\ref{Synthesebeweis der Schaltung
aus Abb. Einfache Steuerungsschaltung unter Verwendung hoherer
logischer Operatoren} zeigt das Analogon zum Beweis in
Abb.~\ref{Synthesebeweis der Schaltung aus Abb. Einfache
Steuerungsschaltung} unter der Verwendung von $ldt$. Ein Axiom
({\bf 61}) aus der Hintergrundtheorie \"uber $unt$ und $ldt$ wurde
verwendet.

\begin{figure}
\begin{center}
\begin{tabular}{@{}l@{$\;$}l@{}}
& $pos(s,t_0) = d_5$	\\
$\ra$ & $ldt(t_0,presse\_hoch,\lambda t_1\!:
	ldt(t_1,presse\_nieder,\lambda t_2\!:
	bearbeitungszustand(s,t_2) = bearbeitet))$	\\
\end{tabular}
\par
\begin{picture}(9.6,1.5)
	%\put(0,0){\makebox(0,0){+}}
\thicklines
\put(0.900,0.600){\circle*{0.100}}
	\put(0.900,0.900){\makebox(0.000,0.000)[b]{$t_0$}}
	\put(0.900,0.400){\makebox(0.000,0.000)[t]{$s$ in Presse}}
\put(1.000,0.600){\vector(1,0){3.800}}
	\put(2.900,0.800){\makebox(0.000,0.000)[b]{$presse\_hoch$}}
\put(4.900,0.600){\circle*{0.100}}
	\put(4.900,0.900){\makebox(0.000,0.000)[b]{$t_1$}}
	\put(4.900,0.400){\makebox(0.000,0.000)[t]{pressen}}
\put(5.000,0.600){\vector(1,0){3.800}}
	\put(6.900,0.800){\makebox(0.000,0.000)[b]{$presse\_nieder$}}
\put(8.900,0.600){\circle*{0.100}}
	\put(8.900,0.900){\makebox(0.000,0.000)[b]{$t_2$}}
	\put(8.900,0.400){\makebox(0.000,0.000)[t]{$s$ bearbeitet}}
\end{picture}
\end{center}
\caption{Modellierung von Zustands\"uberg\"angen durch $ldt$--Ketten}
\label{Modellierung von Zustandsubergangen durch ldt--Ketten}
\end{figure}
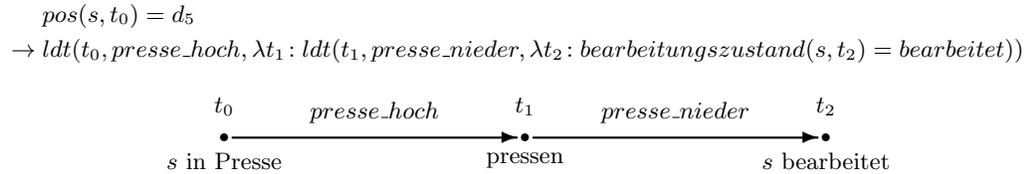

Man beachte, da\3 Ketten von $ldt$s Zustands\"uberg\"ange wie bei
endlichen Automaten modellieren k\"onnen, Abb.~\ref{Modellierung von
Zustandsubergangen durch ldt--Ketten} zeigt ein
Beispiel.
Innerhalb des {\bmft}--Projekts ``Korrekte Software ({\korso})''
wurde die Fertigungszelle u.a.\ mit Hilfe
der Sprache {\lustre} als eine Art endlicher Automat modelliert, so da\3
die wichtigsten Anforderungen vollautomatisch mit Hilfe von Binary
Decision Diagrams verifiziert werden konnten \cite{Holenderski.1995}.
Ein solcher Ansatz kann jedoch nur Aspekte behandeln, die sich als
Automateneigenschaften darstellen lassen. Es w\"are interessant zu
untersuchen, ob sich durch den $unt$/$ldt$--Ansatz eine vertikale
Dekomposition des Modells erreichen l\"a\3t in dem Sinne, da\3 auf der
oberen Ebene nur Automateneigenschaften betrachtet werden m\"ussen,
w\"ahrend alle anderen Eigenschaften auf der unteren Ebene behandelt
werden.

\clearpage
\section{Bewertung}
\label{Bewertung}

\subsection{Modellierbare Eigenschaften}
\label{Modellierbare Eigenschaften}

Der hier vorgestellte Ansatz macht es leicht, alle verlangten
Lebendigkeits- und Sicherheitseigenschaften zu formulieren und
nachzuweisen. Die Lebendigkeitseigenschaft besagt, da\3 jedes
unbearbeitete Metallpl\"attchen, das in die Fertigungszelle
hineingegeben wird, sie irgendwann im bearbeiteten Zustand
wieder verl\"a\3t; siehe die Diskussion am Ende des
Abschnitts~\ref{Anforderungsspezifikation}. Die
Sicherheitseigenschaften k\"onnen in der zus\"atzlichen
Anforderung ``Es darf nie zu Unf\"allen kommen'' zusammengefa\3t
werden, wobei eine notwendige Bedingung f\"ur einen Unfall durch
eine entsprechende Aufz\"ahlung aller kritischen
Maschinenkombinationen (z.B.\ Roboter/Presse) beschrieben wird. 

Ein Nachteil dieses Vorgehens besteht in dem Risiko, beim
Schreiben der Spezifikation bestimmte m\"ogliche
Konfliktsituationen zu \"ubersehen. So war z.B.\ in der ersten
Version der informellen Beschreibung der
Sicherheitsanforderungen nicht verlangt, da\3 das
Zuf\"uhrf\"orderband nur dann Me\-tall\-pl\"att\-chen
transportieren darf, wenn der Hubdrehtisch leer ist. 

Jede der informellen Sicherheitsanforderungen aus
Abschnitt~\ref{Sicherheitsanforderungen} ergibt sich aus einem
der folgenden Prinzipien: 

\bi
\item Maschinenkollisionen m\"ussen vermieden werden (1, 2, 5, 6),
\item die Beweglichkeitsbeschr\"ankungen der Maschinen m\"ussen
	respektiert werden (3, 4, 5),
\item Metallpl\"attchen d\"urfen nicht aus gro\3er H\"ohe herabfallen
	(7, 9),
\item die Metallpl\"attchen m\"ussen gen\"ugend gut separiert bleiben
	(8, 10).
\ei

Es ist grunds\"atzlich m\"oglich, die Sicherheitsanforderungen
an die Fertigungszelle auf diese vier Prinzipien zu gr\"unden.
Eine Formalisierung des ersten Prinzips verlangt jedoch eine
vollst\"andige Beschreibung aller Maschinenabmessungen
einschlie\3lich ihrer Bewegungsbahnen und dar\"uber hinaus f\"ur
jedes der $n \cdot (n-1)$ Maschinenpaare einen Beweis, da\3 sie
nicht zusammensto\3en k\"onnen, unabh\"angig davon, wie weit sie
tats\"achlich voneinander entfernt stehen. Da beides sehr
aufwendig ist, haben wir uns stattdessen dazu entschlossen, die
m\"oglichen Kollisionssituationen explizit anzugeben.

\subsection{Explizitheitsgrad}
\label{Explizitheitsgrad}

Die Voraussetzungen \"uber das Maschinenverhalten und \"uber die
Gesamtkonfiguration der Fertigungszelle sind in den
entsprechenden Modulen der Spezifikation explizit aufgef\"uhrt.
Dar\"uber hinaus ist es m\"oglich, w\"ahrend des Beweises
weitere ben\"otigte Anforderungen an Verhalten oder
Konfiguration abzuleiten, siehe Abschnitt~\ref{Einbeziehung der
Mechanik}.

\subsection{Statistik}
\label{Statistik}

Es ist schwierig, den Zeitaufwand f\"ur die Durchf\"uhrung des
Beweises abzusch\"atzen, da parallel dazu das
Unterst\"utzungswerkzeug {\sysyfos} weiterentwickelt werden
mu\3te, um den Beweis \"uberhaupt in den Griff zu bekommen. Als
Nebenergebnis der Fallstudie wurde eine halbgraphische
Be\-nut\-zer\-ober\-fl\"a\-che und ein
Beweiswiederholungs--Mechanismus in das Werkzeug eingebaut; in
der zweiten Phase erforderte die eingeschr\"ankte Unifikation
h\"oherer Ordnung f\"ur $unt$ und $ldt$ einige
Implementierungsarbeit. Unter diesen Vorbehalten kann der
Aufwand f\"ur das Finden bzw.\ Verifizieren der Teilschaltung
zum Transport eines Metallpl\"attchens vom Hubdrehtisch in die
Presse mit etwa 1 bis 2 Mannwochen angegeben werden. Dieser
Teilbeweis besteht aus 210 Schritten ohne die Verwendung von
$unt$ und $ldt$ und war der erste Teilbeweis innerhalb der
Fallstudie. Ein sp\"ater durchgef\"uhrter vergleichbar gro\3er
Teilbeweis ben\"otigte gr\"o\3enordnungsm\"a\3ig nur noch 1 bis
2 Manntage, haupts\"achlich aufgrund der vorhandenen Erfahrungen
speziell bzgl.\ der in Abschnitt~\ref{Zeitmodellierung}
diskutierten Stetigkeitsaspekte.

\subsection{Wartung}
\label{Wartung}

Der Hauptaufwand f\"ur die Entwicklung einer Steuerung f\"ur
eine ge\"anderte, vergleichbare Fertigungszelle besteht im
F\"uhren eines neuen Beweises. Aufbauend auf der vorhandenen
Terminologie sollte es leicht fallen, eine neue formale
Spezifikation zu erstellen. Sofern der auf reiner
Pr\"adikatenlogik erster Ordnung basierende Ansatz verwendet
wird, d\"urften nur wenige Teile des Originalbeweises
wiederverwendbar sein, je nach dem Grad der \"Ahnlichkeit der
beiden Spezifikationen. Im $unt$/$ldt$--Ansatz jedoch wurde ein
gro\3er Teil des Aufwandes in die Schematisierung von
Steuerungswissen als Hintergrundtheoreme gesteckt, der beim
zweiten Mal nicht wieder anf\"allt; siehe etwa Satz {\bf 61} in
Abb.~\ref{Synthesebeweis der Schaltung aus Abb. Einfache
Steuerungsschaltung unter Verwendung hoherer logischer
Operatoren}, der dort die zentrale Rolle im Beweis spielt. Wir
w\"urden erwarten, da\3 der verbleibende Beweisaufwand, um die
Hintergrundtheoreme f\"ur die neue Situation geeignet zu
instanziieren, eher gering ist. In jedem Fall ist jedoch der
Aufwand, eine neue verifizierte Steuerungsschaltung zu erhalten,
sehr viel gr\"o\3er als etwa der f\"ur die Rekonfiguration eines
objektorientierten Steuerungsprogramms.

\subsection{Effizienz}
\label{Effizienz}

Das Paradigma der deduktiven Programmsynthese macht keine
Aussagen \"uber die Effizienz der konstruierten Programme.
Dar\"uber hinaus bedeutet Effizienz im Fall der Fertigungszelle
nicht, kurze Software--Reaktionszeiten zu erreichen, sondern
eine hohe Gesamtdurchsatzrate. Gem\"a\3 dem Ansatz,
Software--Entwicklungsmethoden auch auf die Anwendung auf
mechanische Probleme auszuweiten, k\"onnte man die
``algorithmische Komplexit\"at'' der gesamten Fertigungszelle
absch\"atzen. Dazu w\"are eine entsprechende Verallgemeinerung
eines Komplexit\"atskalk\"uls f\"ur reaktive Systeme notwendig.
Da in der Fertigungszelle keine Rekursion auftritt, k\"onnte die
maximale Bearbeitungszeit f\"ur ein Me\-tall\-pl\"att\-chen
exakt berechnet werden. Etwa aus dem Teilbeweis in Anhang~\refC\
ergibt sich die Zeit zum Zu\-r\"uck\-fah\-ren des ersten
Roboteramrs von der Presse zum Hubdrehtisch als
$trv_1(r(\ldots),winkel\_xy(d_4,d_5),t_0) - t_0$, vgl.\ Formel
$325$. Ein Beweis daf\"ur, da\3 die gefundene Konfiguration und
Steuerung der Zelle einen {\em maximalen\/} Durchsatz
garantiert, scheint jedoch ebenso schwierig wie der
Nachweis unterer Kom\-ple\-xi\-t\"ats\-schran\-ken f\"ur
gew\"ohnliche algorithmische Probleme.

\subsection{Mechanische Anforderungen}
\label{Mechanische Anforderungen}

Wie bereits in Abschnitt~\ref{Einbeziehung der Mechanik}
erw\"ahnt, wurden w\"ahrend der Synthese eine Reihe
zus\"atzlicher Anforderungen an die Konfiguration der
Fertigungszelle hergeleitet. Sie verlangen meist, da\3 die
Beweglichkeitsbeschr\"ankungen der Maschinen es ihnen erlauben,
die ben\"otigten Punkte tats\"achlich zu erreichen, etwa, da\3
der erste Roboterarm den Hubdrehtisch erreichen kann, vgl.\
Axiom $u22$ in Anhang~\ref{Formale Spezifikation der
Fertigungszelle}. Eine zweite Gruppe von Anforderungen betrifft
die Kompatibilit\"at von Abmessungen und Winkeln, etwa, da\3 die
obere Position des Hubdrehtisches, der erste Roboterarm und die
mittlere Position der Presse alle in derselben H\"ohe liegen
m\"ussen, vgl.\ Axiome $u15$ und $u46$. 

Einige weitere Bedingungen sind nicht unbedingt notwendig,
f\"uhren aber zu einer vereinfachten Steuerungsschaltung
Wenn z.B.\ bekannt ist, da\3 der Abstand vom.
Drehzentrum des Roboters zum Hubdrehtisch gr\"o\3er ist als zur
Presse, so gen\"ugt es, den ersten Arm auf dem Weg zur Presse
einzufahren, w\"ahrend sonst die Schaltung zus\"atzlich die
M\"oglichkeit des Ausfahrens vorsehen m\"u\3te; vgl.\ auch Axiom
$r6$ in Anhang~\ref{Formale Spezifikation der Fertigungszelle}. 

Wenn die Fertigungszelle offen, d.h.\ ohne das
Handhabungsger\"at, betrieben wird, fallen weitere Anforderungen
\"uber das Be- und Entladeverhalten an. Zum Beispiel darf das
Zuf\"uhrband nur beladen werden, wenn an seinem Anfang
gen\"ugend Platz daf\"ur frei ist. Diese Bedingung macht einen
zus\"atzlichen Sensor notwendig, entweder am Anfang des
Zuf\"uhrbandes oder ---~was zu einer einfacheren und robusteren
Steuerung f\"uhrt~--- an dessen Ende. 

Unsere Modellierung basiert auf der idealisierenden Annahme,
da\3 alle geometrischen Abmessungen exakt sind. In der Praxis
wird dies jedoch nicht der Fall sein, etwa das Zuf\"uhrband wird
nicht jedes Metallpl\"attchen genau bis zur Mitte des
Hubdrehtischs (Punkt $d_3$ in Abb.~\ref{Schematische Darstellung der
Fertigungszelle}) transportieren. Ein Modell der
Fertigungszelle, das diesen Ungenauigkeiten Rechnung tr\"agt,
m\"usste mit zul\"assigen Toleranzintervallen umgehen k\"onnen.
So w\"urde man z.B.\ in der Spezifikation fordern, da\3 der
Roboter jedes Metallpl\"attchen, das im Bereich $d_3+x$ mit
${\mid\!\mid\!\! x \!\!\mid\!\mid} < \varepsilon_3$ liegt, noch
sicher aufnimmt und zur Presse transportiert. Jede Maschine
w\"urde ihre eigene Ungenauigkeit zum Toleranzintervall
hinzuf\"ugen, in manchen F\"allen dieses Intervall aber auch
durch gewisse Ausrichtungseffekte wieder verkleinern, etwa bei
Lichtschranken. Es m\"u\3te dann zus\"atzlich verlangt werden,
da\3 die Toleranzintervalle klein genug bleiben, um die
problemlose Weiterverarbeitung zu erm\"oglichen. Das
Toleranzintervall eines Metallpl\"attchens in der Presse zum
Beispiel enth\"alt die Toleranzen des ersten Roboterarms, des
Hubdrehtischs, des Zuf\"uhrf\"orderbands und dessen (externen)
Beladungsger\"ats, es mu\3 sichergestellt sein, da\3 diese
Abweichung klein genug bleibt, um das Metallpl\"attchen
zuverl\"assig pressen zu k\"onnen.

\subsection{Thesen}
\label{Thesen}

Unsere Erfahrungen mit der Fertigungszelle scheinen folgende Thesen zu
best\"atigen:

\bi
\item {\em Eine gute Spezifikation sollte aus einer Sammlung beinahe
	offensichtlicher Fakten in formaler Notation bestehen.}

	Der Verzicht auf Ausf\"uhrbarkeit garantiert die Freiheit, die
	formalen Anforderungen als eine m\"og\-lichst direkte
	\"Ubersetzung der nat\"urlichsprachlichen Beschreibung
	aufzustellen. Erstere k\"onnen lokal (d.h.\ Axiom f\"ur Axiom,
	ohne Ber\"ucksichtigung von Querverweisen) gegen letztere
	validiert werden.

\item[]
\item {\em Module der Anforderungsspezifikation beschreiben
	verschiedene Aspekte der modellierten Realit\"at, nicht der
	Implementierung.}

	Im Unterschied zu Entwurfsspezifikationen k\"onnen sie nicht in
	Implementierungsmodule verfeinert werden, stattdessen sind sie
	zu ihnen orthogonal.

\item[]
\item {\em Pr\"adikatenlogik kann als eine ``Assemblersprache'' f\"ur
	Spezifikationen angesehen werden.}

	Es ist w\"unschenswert, h\"ohere, auch anwendungsabh\"angige,
	Sprachkonstrukte darauf
	aufzubauen, um k\"urzere Spezifikationen und Beweise zu
	erhalten.

\item[]
\item {\em Formale Beschreibungen k\"onnen bereits auf der obersten
	Ebene eingesetzt werden, auf der nur rein technische Aspekte
	auftreten.}

	Es scheint keinen Grund zu geben, sie erst ab der Ebene der
	Software--Entwicklung einzusetzen, logik--basierte Methoden
	k\"onnen im Gegenteil als ein Integrationsrahmen f\"ur eine
	verifizierte Entwicklung des Gesamtsystems, einschlie\3lich
	klassischer Mechanik, dienen. Dies wurde durch unsere Behandlung
	der Fallstudie Fertigungszelle demonstriert, die g\"anzlich im
	rein technischen Bereich liegt und deren Spezifikation das
	Hauptziel (Fertigung bearbeiteter Metallpl\"attchen) beinhaltet.
	Wenn das Hauptziel andererseits nichttechnischer Natur ist, wie
	z.B.\ in einem medizinischen Informationssystem, ist dieser
	Ansatz nicht voll anwendbar.

\item[]
\item {\em Es gibt nur wenige ad\"aquate Beschreibungsebenen.}

	Unsere Erfahrung hat gezeigt, da\3 die Entscheidung, die Zeit
	nicht--diskret zu modellieren, notwendigerweise eine auf
	reellwertiger Zeit und stetigen Funktionen basierende
	Modellierung zur Folge hat, dazwischen scheint es keine
	ad\"aquate Ebene mehr zu geben (etwa rationalwertige Zeit und
	beliebige Funktionen). Ein realistischeres Vorgehen w\"are die
	Benutzung differenzierbarer Funktionen, um
	Aussagen \"uber Beschleunigungen und Startgeschwindigkeiten
	machen zu k\"onnen.
	W\"ahrend eine solche
	Beschreibungsebene f\"ur die Fertigungszelle nicht unbedingt
	notwendig war, ist sie f\"ur zeitkritische Anwendungen
	unvermeidbar, etwa im Bereich von Fahrzeugsteuerungssystemen, wo
	Aussagen \"uber Beschleunigungs- und Bremszeiten lebenswichtig
	sind.

\ei

%\clearpage

\nocite{Bledsoe.1977}
\nocite{Murray.1978}
\nocite{Murray.1982}
\nocite{Traugott.1986}
\nocite{Traugott.1986b}
\nocite{Wilkins.1973}

\bibliographystyle{alpha}
\bibliography{lit}

\clearpage

{\LARGE\bf Anhang}

\vspace*{1.0cm}

\appendix

Anhang~\ref{Informelle Beschreibung der verwendeten Bezeichner}
zeigt die informelle Beschreibung der in der Spezifikation der
Fertigungszelle verwendeten Bezeichner.
Abbildung~\ref{Informelle Beschreibung der verwendeten Variablenkonventionen}
zeigt die Variablenkonventionen,
Abb.~\ref{Informelle Beschreibung der verwendeten Pradikate}
die Pr\"adikate,
Abb.~\ref{Informelle Beschreibung der verwendeten Funktionen}
die Funktionen,
Abb.~\ref{Informelle Beschreibung der verwendeten Konstanten}
die Konstanten
und Abb.~\ref{Informelle Beschreibung der verwendeten expliziten
Skolemfunktionen}
die expliziten Skolemfunktionen.
In der rechten Spalte sind jeweils die in Anhang~\refC\ verwendeten
Kurzbezeichner aufgef\"uhrt.

Anhang~\ref{Formale Spezifikation der Fertigungszelle}
zeigt die formale Anforderungsspezifikation der
Fertigungszelle unter Verwendung der in Anhang~\ref{Informelle
Beschreibung der verwendeten Bezeichner} definierten Sprachebene
(Axiome $u1$ bis $u77$). Die Axiome f\"ur Sicherheitsanforderungen sind
weggelassen.
Der Text wurde aus einer {\web}--artigen
Darstellung \cite{Knuth.1984} gewonnen, die sich andererseits
auch direkt in die {\prolog}--Eingabe f\"ur {\sysyfos}
transformieren l\"a\3t. 
Im Anschlu\3 daran sind die f\"ur die Synthese von Modul 1
ben\"otigten Axiome und das Beweisziel aufgef\"uhrt (Axiome $r1$ bis
$r11$, Beweisziel $r20$).

Anhang~\refC\ zeigt den Beweisbaum f\"ur den Hauptteil der
Verifikation des Moduls 1 aus Abschnitt~\ref{Entwicklung der
Robotersteuerung}.

\newpage

\section{Informelle Beschreibung der verwendeten Bezeichner}
\label{Informelle Beschreibung der verwendeten Bezeichner}

\begin{figure}
\begin{center}
\begin{tabular}[t]{@{}lll@{}}
$p$ & :Presse	\\
$x$ & :Raumkoordinaten	\\
$t$ & :Zeitpunkt	\\
$s$ & :Schiene (Metallpl\"attchen)	\\
$r$ & :Zweiarmiger Roboter	\\
$h$ & :Hubdrehtisch	\\
$f$ & :F\"orderband	\\
$\alpha$ & :Winkel	\\
\end{tabular}
\end{center}
\caption{Informelle Beschreibung der verwendeten Variablenkonventionen}
\label{Informelle Beschreibung der verwendeten Variablenkonventionen}
\end{figure}

\begin{figure}
\begin{center}
\begin{tabular}[t]{@{}ll@{\hspace*{0.5cm}}l@{\hspace*{0.5cm}}l@{}}
$presse(p,x)$ & $\Lra$
	& $p$ ist eine Presse und steht am Ort $x$
& $prs$	\\
$roboter(r,x)$ & $\Lra$
	& $r$ ist ein zweiarmiger Roboter und steht am Ort $x$
& $rob$	\\
$greift1(r,t)$ & $\Lra$
        & Der erste Arm des Roboters $r$ greift zum Zeitpunkt $t$ \\
	&& (d.h.\ Magnet ein)
& $gr1$	\\
$greift2(r,t)$ & $\Lra$
        & Der zweite Arm des Roboters $r$ greift zum Zeitpunkt $t$
& $gr2$	\\
$haelt1(r,s,t)$ & $\Lra$
	& Der erste Arm des Roboters $r$ h\"alt zum Zeitpunkt $t$
	die	\\
	&& Schiene $s$ fest (d.h.\ magnetisch angezogen)
& $ha1$	\\
$haelt2(r,s,t)$ & $\Lra$
	& Der zweite Arm von $r$ h\"alt zum Zeitpunkt $t$
	die Schiene $s$ fest
& $ha2$	\\
\end{tabular}
\end{center}
\caption{Informelle Beschreibung der verwendeten Pr\"adikate}
\label{Informelle Beschreibung der verwendeten Pradikate}
\end{figure}

\begin{figure}
\begin{center}
\begin{tabular}[t]{@{}ll@{\hspace*{0.5cm}}l@{\hspace*{0.5cm}}l@{}}
$ort(s,t)$ & $=$ &
Raumkoordinaten des Ortes, an dem sich die Schiene $s$	\\
&& zum Zeitpunkt $t$ befindet
& $ort$	\\
$bearbeitungszustand(s,t)$ & $=$ &
der Bearbeitungszustand der Schiene $s$ zum Zeitpunkt $t$	\\
&& ($\in \{bearbeitet,unbearbeitet\}$)
& $btz$	\\
$pos1(r,t)$ & $=$ &
Koordinaten des Greifers des ersten Arms des Roboters $r$ 	\\
&& zum Zeitpunkt $t$
& $ps1$	\\
$proj\_xy(x)$ & $=$ &
Projektion der Raumkoordinaten $x$ auf die Bodenebene
& $pxy$	\\
$hoehe(x,t)$ & $=$ &
H\"ohe der Arbeitsfl\"ache \"uber dem Boden
am Punkt	\\
\end{tabular}
\end{center}
\caption{Informelle Beschreibung der verwendeten Funktionen}
\label{Informelle Beschreibung der verwendeten Funktionen}
\end{figure}

\begin{figure}
\begin{center}
\begin{tabular}[t]{@{}ll@{\hspace*{0.5cm}}l@{\hspace*{0.5cm}}l@{}}
$bearbeitet$ & $=$ &
	m\"oglicher Zustand eines Metallpl\"attchens
	& $bbt$	\\
$unbearbeitet$ & $=$ &
	m\"oglicher Zustand eines Metallpl\"attchens
	& $ubt$	\\
$d_1$ & $=$ &
	Anfangspunkt des Zuf\"uhrf\"orderbands
	(kartesischer Koordinatenvektor)
	& $d_1$	\\
$d_2$ & $=$ &
	Endpunkt des Zuf\"uhrf\"orderbands
	(kartesischer Koordinatenvektor)
	& $d_2$	\\
$d_3$ & $=$ &
	Mittelpunkt des Hubdrehtischs
	(kartesischer Koordinatenvektor)
	& $d_3$	\\
$d_4$ & $=$ &
	Drehzentrum des Roboters
	(kartesischer Koordinatenvektor)
	& $d_4$	\\
$d_5$ & $=$ &
	Mittelpunkt der Presse
	(kartesischer Koordinatenvektor)
	& $d_5$	\\
$d_6$ & $=$ &
	Anfangspunkt des Ablagef\"orderbands
	(kartesischer Koordinatenvektor)
	& $d_6$	\\
\end{tabular}
\end{center}
\caption{Informelle Beschreibung der verwendeten Konstanten}
\label{Informelle Beschreibung der verwendeten Konstanten}
\end{figure}

\begin{figure}
\begin{center}
\begin{tabular}[t]{@{}ll@{\hspace*{0.5cm}}l@{\hspace*{0.5cm}}l@{}}
$trv1(r,\alpha,t)$ & $=$ &
Zeitpunkt, zu dem sich der Roboter $r$ so weit vor gedreht hat, \\
&& da\3
Arm1 den Winkel $\alpha$ hat, wenn er sich seit Zeitpunkt $t$ vor dreht
& $trv1$	\\
$trv2(r,\alpha,t)$ & $=$ &
Zeitpunkt, zu dem sich der Roboter $r$ so weit vor gedreht hat, \\
&& da\3
Arm2 den Winkel $\alpha$ hat, wenn er sich seit Zeitpunkt $t$ vor dreht
& $trv2$	\\
$trz1(r,\alpha,t)$ & $=$ &
Zeitpunkt, zu dem sich der Roboter $r$ so weit zur\"uck gedreht hat, \\
&& da\3 Arm1 den Winkel $\alpha$ hat, wenn er sich seit Zeitpunkt $t$
zur\"uck dreht
& $trz1$	\\
$trz2(r,\alpha,t)$ & $=$ &
Zeitpunkt, zu dem sich der Roboter $r$ so weit zur\"uck gedreht hat, \\
&& da\3 Arm2 den Winkel $\alpha$ hat, wenn er sich seit Zeitpunkt $t$
zur\"uck dreht
& $trz2$	\\
$tra1(r,d,t)$ & $=$ &
Zeitpunkt, zu dem der Roboter $r$ Arm1 bis zur L\"ange $d$ ausgefahren
hat,	\\
&& wenn er ihn seit Zeitpunkt $t$ ausf\"ahrt
& $tra1$	\\
$tra2(r,d,t)$ & $=$ &
Zeitpunkt, zu dem der Roboter $r$ Arm2 bis zur L\"ange $d$ ausgefahren
hat,	\\
&& wenn er ihn seit Zeitpunkt $t$ ausf\"ahrt
& $tra2$	\\
$tre1(r,d,t)$ & $=$ &
Zeitpunkt, zu dem der Roboter $r$ Arm1 bis zur L\"ange $d$ eingefahren
hat,	\\
&& wenn er ihn seit Zeitpunkt $t$ einf\"ahrt
& $tre1$	\\
$tre2(r,d,t)$ & $=$ &
Zeitpunkt, zu dem der Roboter $r$ Arm2 bis zur L\"ange $d$ eingefahren
hat,	\\
&& wenn er ihn seit Zeitpunkt $t$ einf\"ahrt
& $tre2$	\\
\end{tabular}
\end{center}
\caption{Informelle Beschreibung der verwendeten
	expliziten Skolemfunktionen}
\label{Informelle Beschreibung der verwendeten
	expliziten Skolemfunktionen}
\end{figure}

\clearpage

\section{Formale Spezifikation der Fertigungszelle}
\label{Formale Spezifikation der Fertigungszelle}

\newcommand{\eqd}[1]{\item[{\bf #1.}]}

\begin{itemize}
\item[]
{\small $\begin{array}[t]{@{}r@{\;}l@{}}&
\end{array}$ } \vspace{1.0cm} \par {\bf Presse
} \vspace{-0.3cm} \par {\small $\begin{array}[t]{@{}r@{\;}l@{}} &
		\end{array}$ } \vspace{3.5\parskip} \par  \eqd{u1}
		  Die Presse pre\3t in oberer Position
		 eine in ihr liegende unbearbeitete Schiene
\par\vspace{-0.5\parskip}{\small$\begin{array}[t]{@{}r@{\;}l@{}}&
 \forall [ p,x,s,t,t_{2}]\; ( \\ &
   presse(p,x)
\\  \wedge  & bearbeitungszustand(s,t)=unbearbeitet
\\  \wedge  & proj\_xy(ort(s,t))=proj\_xy(x)
\\  \wedge  & winkel(s,t)=180
\\  \wedge  &  \exists [ t_{1}]\; (t \.\leq  t_{1}  \.\leq t_{2} \wedge  hoehe(proj\_xy(x),t_{1})=zp\_oben)
\\  \ra  & bearbeitungszustand(s,t_{2})=bearbeitet
\\  \wedge  & proj\_xy(ort(s,t_{2}))=proj\_xy(x)
\\  \wedge  & winkel(s,t_{2})=180 )
                \end{array}$ } \vspace{3.5\parskip} \par  \eqd{u2}
		  Wenn sich die Presse nur lange genug hebt,
                 erreicht sie schlie\3lich die obere Position
\par\vspace{-0.5\parskip}{\small$\begin{array}[t]{@{}r@{\;}l@{}}&
 \forall [ t]\;  \exists [ t_{2}]\;  \forall [ p,x]\; ( \\ &
   presse(p,x)
\\  \ra  & (    \forall [ t_{1}]\; (t \.\leq  t_{1} <t_{2} \ra  hebt(p,t_{1}))
    \ra  hoehe(proj\_xy(x),t_{2})=zp\_oben )
\\  \wedge  &  \forall [ t_{3}]\; (t \.\leq  t_{3} <t_{2} \ra  \neg hoehe(proj\_xy(x),t_{3})=zp\_oben) )
                \end{array}$ } \vspace{3.5\parskip} \par  \eqd{u3}
		  Wenn sich die Presse aus der unteren Position
		  nur lange genug hebt, erreicht sie schlie\3lich
		 die mittlere Position
\par\vspace{-0.5\parskip}{\small$\begin{array}[t]{@{}r@{\;}l@{}}&
 \forall [ t]\;  \exists [ t_{2}]\;  \forall [ p,x]\; ( \\ &
   presse(p,x)
\\  \wedge  & hoehe(proj\_xy(x),t)=zp\_unten
\\  \ra  & (    \forall [ t_{1}]\; (t \.\leq  t_{1}  \.\leq t_{2} \ra  hebt(p,t_{1}))
    \ra  hoehe(proj\_xy(x),t_{2})=zp\_mitte )
\\  \wedge  &  \forall [ t_{3}]\; (t \.\leq  t_{3} <t_{2} \ra  \neg hoehe(proj\_xy(x),t_{3})=zp\_mitte) )
                \end{array}$ } \vspace{3.5\parskip} \par  \eqd{u4}
		  Wenn sich die Presse nur lange genug senkt,
                 erreicht sie schlie\3lich die untere Position
\par\vspace{-0.5\parskip}{\small$\begin{array}[t]{@{}r@{\;}l@{}}&
 \forall [ t]\;  \exists [ t_{2}]\;  \forall [ p,x]\; ( \\ &
   presse(p,x)
\\  \ra  & (    \forall [ t_{1}]\; (t \.\leq  t_{1}  \.\leq t_{2} \ra  senkt(p,t_{1}))
    \ra  hoehe(proj\_xy(x),t_{2})=zp\_unten )
\\  \wedge  &  \forall [ t_{3}]\; (t \.\leq  t_{3} <t_{2} \ra  \neg hoehe(proj\_xy(x),t_{3})=zp\_unten) )
		\end{array}$ } \vspace{3.5\parskip} \par  \eqd{u5}
\par\vspace{-0.5\parskip}{\small$\begin{array}[t]{@{}r@{\;}l@{}}&
 \forall [ p,x,s,t]\; ( \\ &
   presse(p,x)
\\  \ra  & (    proj\_xy(ort(s,t))=proj\_xy(x) \wedge  (hebt(p,t) \vee  senkt(p,t))
    \lra  bewegt(p,s,t)) )
		\end{array}$ } \vspace{3.5\parskip} \par  \eqd{u6}
		  Motorsteuerung und Sensoren	\\
		  $c_{1}$: $+1$ hebt,  $0$ steht	\\
		  $c_{2}$: $+1$ senkt, $0$ steht	\\
		  $s_{1}$: $+1$ unten		\\
		  $s_{2}$: $+1$ mitte		\\
		 $s_{3}$: $+1$ oben
\par\vspace{-0.5\parskip}{\small$\begin{array}[t]{@{}r@{\;}l@{}}&
 \forall [ c_{1},c_{2},s_{1},s_{2},s_{3},x,t]\; ( \\ &
   presse(p(c_{1},c_{2},s_{1},s_{2},s_{3}),x)
\\  \ra  & (hebt(p(c_{1},c_{2},s_{1},s_{2},s_{3}),t)    \lra  val(c_{1},t) = 1)
\\  \wedge  & (senkt(p(c_{1},c_{2},s_{1},s_{2},s_{3}),t)   \lra  val(c_{2},t) = 1)
\\  \wedge  & (hoehe(proj\_xy(x),t)=zp\_unten \lra  val(s_{1},t) = 1)
\\  \wedge  & (hoehe(proj\_xy(x),t)=zp\_mitte \lra  val(s_{2},t) = 1)
\\  \wedge  & (hoehe(proj\_xy(x),t)=zp\_oben  \lra  val(s_{3},t) = 1) )
\end{array}$ } \vspace{1.0cm} \par {\bf Zweiarmiger Roboter
} \vspace{-0.3cm} \par {\small $\begin{array}[t]{@{}r@{\;}l@{}} &
		\end{array}$ } \vspace{3.5\parskip} \par  \eqd{u7}
		  Wenn der Roboter mit Arm 1 zugreift,
		  h\"alt er eine Schiene, die sich im richtigen
		 Winkel unter Arm 1 befindet
\par\vspace{-0.5\parskip}{\small$\begin{array}[t]{@{}r@{\;}l@{}}&
 \forall [ r,x,t,s,x_{1}]\; ( \\ &
   roboter(r,x)
\\  \wedge  & pos_{1}(r,t)=x_{1}
\\  \wedge  & ort(s,t)=x_{1}
\\  \wedge  & winkel(s,t)=winkel\_xy(x,x_{1})-90
\\  \wedge  & greift_{1}(r,t)
\\  \ra  & haelt_{1}(r,s,t) )
		\end{array}$ } \vspace{3.5\parskip} \par  \eqd{u8}
		  Wenn der Roboter mit Arm 2 zugreift,
		  h\"alt er eine Schiene, die sich im richtigen
		 Winkel unter Arm 2 befindet
\par\vspace{-0.5\parskip}{\small$\begin{array}[t]{@{}r@{\;}l@{}}&
 \forall [ r,x,t,s,x_{1}]\; ( \\ &
   roboter(r,x)
\\  \wedge  & pos_{2}(r,t)=x_{1}
\\  \wedge  & ort(s,t)=x_{1}
\\  \wedge  & winkel(s,t)=winkel\_xy(x,x_{1})-90
\\  \wedge  & greift_{2}(r,t)
\\  \ra  & haelt_{2}(r,s,t) )
		\end{array}$ } \vspace{3.5\parskip} \par  \eqd{u9}
		  Der Roboter h\"alt eine einmal gegriffene Schiene
		 mit Arm 1 fest, bis er sie wieder losl\"a\3t
\par\vspace{-0.5\parskip}{\small$\begin{array}[t]{@{}r@{\;}l@{}}&
 \forall [ r,x,s,t,t_{2}]\; ( \\ &
   roboter(r,x)
\\  \wedge  & haelt_{1}(r,s,t)
\\  \wedge  &  \forall [ t_{1}]\; (t \.\leq  t_{1}  \.\leq t_{2} \ra  greift_{1}(r,t_{1}))
\\  \ra  & haelt_{1}(r,s,t_{2})
\\  \wedge  & bearbeitungszustand(s,t_{2})=bearbeitungszustand(s,t_{0}) )
		\end{array}$ } \vspace{3.5\parskip} \par  \eqd{u10}
		  Der Roboter h\"alt eine einmal gegriffene Schiene
		 mit Arm 2 fest, bis er sie wieder losl\"a\3t
\par\vspace{-0.5\parskip}{\small$\begin{array}[t]{@{}r@{\;}l@{}}&
 \forall [ r,x,s,t,t_{2}]\; ( \\ &
   roboter(r,x)
\\  \wedge  & haelt_{2}(r,s,t)
\\  \wedge  &  \forall [ t_{1}]\; (t \.\leq  t_{1}  \.\leq t_{2} \ra  greift_{2}(r,t_{1}))
\\  \ra  & haelt_{2}(r,s,t_{2})
\\  \wedge  & bearbeitungszustand(s,t_{2})=bearbeitungszustand(s,t_{0}) )
		\end{array}$ } \vspace{3.5\parskip} \par  \eqd{u11}
		  Der Roboter bewegt eine gehaltene Schiene
		 mit seinem Arm 1 mit
\par\vspace{-0.5\parskip}{\small$\begin{array}[t]{@{}r@{\;}l@{}}&
 \forall [ r,x,s,t]\; ( \\ &
   roboter(r,x)
\\  \wedge  & haelt_{1}(r,s,t)
\\  \ra  & ort(s,t)=pos_{1}(r,t)
\\  \wedge  & winkel(s,t)=winkel\_xy(x,pos_{1}(r,t))-90 )
		\end{array}$ } \vspace{3.5\parskip} \par  \eqd{u12}
		  Der Roboter bewegt eine gehaltene Schiene
		 mit seinem Arm 2 mit
\par\vspace{-0.5\parskip}{\small$\begin{array}[t]{@{}r@{\;}l@{}}&
 \forall [ r,x,s,t]\; ( \\ &
   roboter(r,x)
\\  \wedge  & haelt_{2}(r,s,t)
\\  \ra  & ort(s,t)=pos_{2}(r,t)
\\  \wedge  & winkel(s,t)=winkel\_xy(x,pos_{2}(r,t)-90) )
		\end{array}$ } \vspace{3.5\parskip} \par  \eqd{u13}
		  Wenn der Roboter mit Arm 1 eine Schiene h\"alt
		  und dann direkt \"uber einer freien Arbeitsfl\"ache
		 losl\"a\3t, bleibt sie darauf liegen
\par\vspace{-0.5\parskip}{\small$\begin{array}[t]{@{}r@{\;}l@{}}&
 \forall [ r,x,t_{2},s,x_{1}]\; ( \\ &
   roboter(r,x)
\\  \wedge  &  \exists [ t]\; (t<t_{2} \wedge   \forall [ t_{1}]\; (t \.\leq  t_{1} <t_{2} \ra  haelt_{1}(r,s,t_{1})))
\\  \wedge  &  \forall [ s_{1}]\; (ort(s_{1},t_{2})=x_{1} \ra  s_{1}=s)
\\  \wedge  & pos_{1}(r,t_{2})=x_{1}
\\  \wedge  & \neg greift_{1}(r,t_{2})
\\  \wedge  & hoehe(proj\_xy(x_{1}),t_{2})=proj\_z(x_{1})
\\  \ra  & ort(s,t_{2})=x_{1} )
		\end{array}$ } \vspace{3.5\parskip} \par  \eqd{u14}
		  Wenn der Roboter mit Arm 2 eine Schiene h\"alt
		  und dann direkt \"uber einer freien Arbeitsfl\"ache
		 losl\"a\3t, bleibt sie darauf liegen
\par\vspace{-0.5\parskip}{\small$\begin{array}[t]{@{}r@{\;}l@{}}&
 \forall [ r,x,t_{2},s,x_{1}]\; ( \\ &
   roboter(r,x)
\\  \wedge  &  \exists [ t]\; (  t<t_{2}
	   \wedge   \forall [ t_{1}]\; (t \.\leq  t_{1} <t_{2} \ra  haelt_{2}(r,s,t_{1})))
\\  \wedge  & pos_{2}(r,t_{2})=x_{1}
\\  \wedge  & \neg greift_{2}(r,t_{2})
\\  \wedge  & hoehe(proj\_xy(x_{1}),t_{2})=proj\_z(x_{1})
\\  \ra  & ort(s,t_{2})=x_{1} )
		\end{array}$ } \vspace{3.5\parskip} \par  \eqd{u15}
		 Die Roboter--Arme haben feste H\"ohe
\par\vspace{-0.5\parskip}{\small$\begin{array}[t]{@{}r@{\;}l@{}}&
 \forall [ r,x,t]\; ( \\ &
   roboter(r,x)
\\  \ra  & proj\_z(pos_{1}(r,t))=zp\_mitte
\\  \wedge  & proj\_z(pos_{2}(r,t))=zp\_unten )
		\end{array}$ } \vspace{3.5\parskip} \par  \eqd{u16}
		  Wenn sich der Roboter nur lange genug vordreht,
		  kann er f\"ur Arm 1 jeden Winkel zwischen dem
		 jetzigen und 270$^\circ$ erreichen
\par\vspace{-0.5\parskip}{\small$\begin{array}[t]{@{}r@{\;}l@{}}&
 \forall [ r,x, \alpha ,t]\; ( \\ &
   roboter(r,x)
\\  \wedge  & winkel\_xy(x,pos_{1}(r,t)) \.\leq   \alpha   \.\leq 270
\\  \ra  & (    \forall [ t_{1}]\; (t \.\leq  t_{1} <trv_{1}(r, \alpha ,t) \ra  dreht\_vor(r,t_{1}))
    \ra  winkel\_xy(x,pos_{1}(r,trv_{1}(r, \alpha ,t)))= \alpha  )
\\  \wedge  &  \forall [ t_{3}]\; (   t \.\leq  t_{3} <trv_{1}(r, \alpha ,t)
	     \ra  winkel\_xy(x,pos_{1}(r,t_{3}))< \alpha ) )
		\end{array}$ } \vspace{3.5\parskip} \par  \eqd{u17}
		  Wenn sich der Roboter nur lange genug zur\"uckdreht,
		  kann er f\"ur Arm 1 jeden Winkel zwischen dem
		 jetzigen und 90$^\circ$ erreichen
\par\vspace{-0.5\parskip}{\small$\begin{array}[t]{@{}r@{\;}l@{}}&
 \forall [ r,x, \alpha ,t]\; ( \\ &
   roboter(r,x)
\\  \wedge  & 90 \.\leq   \alpha   \.\leq winkel\_xy(x,pos_{1}(r,t))
\\  \ra  & (    \forall [ t_{1}]\; (t \.\leq  t_{1} <trz_{1}(r, \alpha ,t) \ra  dreht\_zurueck(r,t_{1}))
							        \\&\tab
    \ra  winkel\_xy(x,pos_{1}(r,trz_{1}(r, \alpha ,t)))= \alpha  )
\\  \wedge  &  \forall [ t_{3}]\; (   t \.\leq  t_{3} <trz_{1}(r, \alpha ,t)
	     \ra   \alpha <winkel\_xy(x,pos_{1}(r,t_{3}))) )
		\end{array}$ } \vspace{3.5\parskip} \par  \eqd{u18}
		  Wenn sich der Roboter nur lange genug vordreht,
		  kann er f\"ur Arm 2 jeden Winkel zwischen dem
		 jetzigen und 360$^\circ$ erreichen
\par\vspace{-0.5\parskip}{\small$\begin{array}[t]{@{}r@{\;}l@{}}&
 \forall [ r,x, \alpha ,t]\; ( \\ &
   roboter(r,x)
\\  \wedge  & winkel\_xy(x,pos_{2}(r,t)) \.\leq   \alpha   \.\leq 360
\\  \ra  & (    \forall [ t_{1}]\; (t \.\leq  t_{1} <trv_{2}(r, \alpha ,t) \ra  dreht\_vor(r,t_{1}))
    \ra  winkel\_xy(x,pos_{2}(r,trv_{2}(r, \alpha ,t)))= \alpha  )
\\  \wedge  &  \forall [ t_{3}]\; (   t \.\leq  t_{3} <trv_{2}(r, \alpha ,t)
	     \ra  winkel\_xy(x,pos_{2}(r,t_{3}))< \alpha ) )
		\end{array}$ } \vspace{3.5\parskip} \par  \eqd{u19}
		  Wenn sich der Roboter nur lange genug zur\"uckdreht,
		  kann er f\"ur Arm 2 jeden Winkel zwischen dem
		 jetzigen und 180$^\circ$ erreichen
\par\vspace{-0.5\parskip}{\small$\begin{array}[t]{@{}r@{\;}l@{}}&
 \forall [ r,x, \alpha ,t]\; ( \\ &
   roboter(r,x)
\\  \wedge  & 180 \.\leq   \alpha   \.\leq winkel\_xy(x,pos_{2}(r,t))
\\  \ra  & (    \forall [ t_{1}]\; (t \.\leq  t_{1} <trz_{2}(r, \alpha ,t) \ra  dreht\_zurueck(r,t_{1}))
							        \\&\tab
    \ra  winkel\_xy(x,pos_{2}(r,trz_{2}(r, \alpha ,t)))= \alpha  )
\\  \wedge  &  \forall [ t_{3}]\; (   t \.\leq  t_{3} <trz_{2}(r, \alpha ,t)
	     \ra   \alpha <winkel\_xy(x,pos_{2}(r,t_{3}))) )
		\end{array}$ } \vspace{3.5\parskip} \par  \eqd{u20}
		  Wenn Arm 1 nur lange genug ausf\"ahrt, kann er jede
		 L\"ange zwischen der jetzigen und $maxlg_{1}$ erreichen
\par\vspace{-0.5\parskip}{\small$\begin{array}[t]{@{}r@{\;}l@{}}&
 \forall [ r,x,t,d]\; ( \\ &
   roboter(r,x)
\\  \wedge  & dist\_xy(x,pos_{1}(r,t)) \.\leq  d  \.\leq maxlg_{1}
\\  \ra  & (    \forall [ t_{1}]\; (t \.\leq  t_{1} <tra_{1}(r,d,t) \ra  faehrt\_aus_{1}(r,t_{1}))
    \ra  dist\_xy(x,pos_{1}(r,tra_{1}(r,d,t)))=d )
\\  \wedge  &  \forall [ t_{3}]\; (t \.\leq  t_{3} <tra_{1}(r,d,t) \ra  dist\_xy(x,pos_{1}(r,t_{3}))<d) )
		\end{array}$ } \vspace{3.5\parskip} \par  \eqd{u21}
		  Nur wenn Arm 1 ausf\"ahrt,
		 kann seine L\"ange gr\"o\3er werden
\par\vspace{-0.5\parskip}{\small$\begin{array}[t]{@{}r@{\;}l@{}}&
 \forall [ r,x,t,t_{2}]\; ( \\ &
   roboter(r,x)
\\  \wedge  & t \.\leq t_{2}
\\  \wedge  & dist\_xy(x,pos_{1}(r,t))<dist\_xy(x,pos_{1}(r,t_{2}))
\\  \ra  &  \exists [ t_{1}]\; (t< t_{1} <t_{2} \wedge  faehrt\_aus_{1}(r,t_{1})) )
		\end{array}$ } \vspace{3.5\parskip} \par  \eqd{u22}
		  Wenn Arm 1 nur lange genug einf\"ahrt, kann er jede
		 L\"ange zwischen der jetzigen und $minlg_{1}$ erreichen
\par\vspace{-0.5\parskip}{\small$\begin{array}[t]{@{}r@{\;}l@{}}&
 \forall [ r,x,t,d]\; ( \\ &
   roboter(r,x)
\\  \wedge  & minlg_{1} \.\leq  d  \.\leq dist\_xy(x,pos_{1}(r,t))
\\  \ra  & (    \forall [ t_{1}]\; (t \.\leq  t_{1} <tre_{1}(r,d,t) \ra  faehrt\_ein_{1}(r,t_{1}))
    \ra  dist\_xy(x,pos_{1}(r,tre_{1}(r,d,t)))=d )
\\  \wedge  &  \forall [ t_{3}]\; (t \.\leq  t_{3} <tre_{1}(r,d,t) \ra  d<dist\_xy(x,pos_{1}(r,t_{3}))) )
		\end{array}$ } \vspace{3.5\parskip} \par  \eqd{u23}
		  Nur wenn Arm 1 einf\"ahrt,
		 kann seine L\"ange kleiner werden
\par\vspace{-0.5\parskip}{\small$\begin{array}[t]{@{}r@{\;}l@{}}&
 \forall [ r,x,t,t_{2}]\; ( \\ &
   roboter(r,x)
\\  \wedge  & t \.\leq t_{2}
\\  \wedge  & dist\_xy(x,pos_{1}(r,t_{2}))<dist\_xy(x,pos_{1}(r,t))
\\  \ra  &  \exists [ t_{1}]\; (t< t_{1} <t_{2} \wedge  faehrt\_ein_{1}(r,t_{1})) )
		\end{array}$ } \vspace{3.5\parskip} \par  \eqd{u24}
		  Wenn Arm 2 nur lange genug ausf\"ahrt, kann er jede
		 L\"ange zwischen der jetzigen und $maxlg_{2}$ erreichen
\par\vspace{-0.5\parskip}{\small$\begin{array}[t]{@{}r@{\;}l@{}}&
 \forall [ r,x,t,d]\; ( \\ &
   roboter(r,x)
\\  \wedge  & dist\_xy(x,pos_{2}(r,t)) \.\leq  d  \.\leq maxlg_{2}
\\  \ra  & (    \forall [ t_{1}]\; (t \.\leq  t_{1} <tra_{2}(r,d,t) \ra  faehrt\_aus_{2}(r,t_{1}))
    \ra  dist\_xy(x,pos_{2}(r,tra_{2}(r,d,t)))=d )
\\  \wedge  &  \forall [ t_{3}]\; (t \.\leq  t_{3} <tra_{2}(r,d,t) \ra  dist\_xy(x,pos_{2}(r,t_{3}))<d) )
		\end{array}$ } \vspace{3.5\parskip} \par  \eqd{u25}
		  Wenn Arm 2 nur lange genug einf\"ahrt, kann er jede
		 L\"ange zwischen der jetzigen und $minlg_{2}$ erreichen
\par\vspace{-0.5\parskip}{\small$\begin{array}[t]{@{}r@{\;}l@{}}&
 \forall [ r,x,t,d]\; ( \\ &
   roboter(r,x)
\\  \wedge  & minlg_{2} \.\leq  d  \.\leq dist\_xy(x,pos_{2}(r,t))
\\  \ra  & (    \forall [ t_{1}]\; (t \.\leq  t_{1} <tre_{2}(r,d,t) \ra  faehrt\_ein_{2}(r,t_{1}))
    \ra  dist\_xy(x,pos_{2}(r,tre_{2}(r,d,t)))=d )
\\  \wedge  &  \forall [ t_{3}]\; (t \.\leq  t_{3} <tre_{2}(r,d,t) \ra  d<dist\_xy(x,pos_{2}(r,t_{3}))) )
		\end{array}$ } \vspace{3.5\parskip} \par  \eqd{u26}
		  Die Winkel der Arme bleiben unver\"andert, wenn sich
		 der Roboter nicht dreht
\par\vspace{-0.5\parskip}{\small$\begin{array}[t]{@{}r@{\;}l@{}}&
 \forall [ r,x,t,t_{2}]\; ( \\ &
   roboter(r,x)
\\  \wedge  &  \forall [ t_{1}]\; (t \.\leq  t_{1}  \.\leq t_{2} \ra  \neg dreht\_vor(r,t_{1}) \wedge  \neg dreht\_zurueck(r,t_{1}))
\\  \ra  & winkel\_xy(x,pos_{1}(r,t))=winkel\_xy(x,pos_{1}(r,t_{2}))
\\  \wedge  & winkel\_xy(x,pos_{2}(r,t))=winkel\_xy(x,pos_{2}(r,t_{2})) )
		\end{array}$ } \vspace{3.5\parskip} \par  \eqd{u27}
		  Die L\"ange von Arm 1 bleibt unver\"andert, wenn ihn
		 der Roboter weder ein- noch ausf\"ahrt
\par\vspace{-0.5\parskip}{\small$\begin{array}[t]{@{}r@{\;}l@{}}&
 \forall [ r,x,t,t_{2}]\; ( \\ &
   roboter(r,x)
\\  \wedge  &  \forall [ t_{1}]\; (t \.\leq  t_{1}  \.\leq t_{2} \ra  \neg faehrt\_aus_{1}(r,t_{1}) \wedge  \neg faehrt\_ein_{1}(r,t_{1}))
\\  \ra  & dist\_xy(x,pos_{1}(r,t))=dist\_xy(x,pos_{1}(r,t_{2})) )
		\end{array}$ } \vspace{3.5\parskip} \par  \eqd{u28}
		  Die L\"ange von Arm 2 bleibt unver\"andert, wenn ihn
		 der Roboter weder ein- noch ausf\"ahrt
\par\vspace{-0.5\parskip}{\small$\begin{array}[t]{@{}r@{\;}l@{}}&
 \forall [ r,x,t,t_{2}]\; ( \\ &
   roboter(r,x)
\\  \wedge  &  \forall [ t_{1}]\; (t \.\leq  t_{1}  \.\leq t_{2} \ra  \neg faehrt\_aus_{2}(r,t_{1}) \wedge  \neg faehrt\_ein_{2}(r,t_{1}))
\\  \ra  & dist\_xy(x,pos_{2}(r,t))=dist\_xy(x,pos_{2}(r,t_{2})) )
		\end{array}$ } \vspace{3.5\parskip} \par  \eqd{u29}
\par\vspace{-0.5\parskip}{\small$\begin{array}[t]{@{}r@{\;}l@{}}&
 \forall [ r,x,t,s]\; ( \\ &
   roboter(r,x)
\\  \ra  & (    haelt_{1}(r,s,t) \wedge  (  faehrt\_aus_{1}(r,t)		 \\ &
		         \vee  faehrt\_ein_{1}(r,t)		 \\ &
		         \vee  dreht\_vor(r,t)		 \\ &
		         \vee  dreht\_zurueck(r,t))		 \\ &
      \vee  haelt_{2}(r,s,t) \wedge  (  faehrt\_aus_{2}(r,t)		 \\ &
		         \vee  faehrt\_ein_{2}(r,t)		 \\ &
		         \vee  dreht\_vor(r,t)		 \\ &
		         \vee  dreht\_zurueck(r,t))		 \\ &
    \lra  bewegt(r,s,t)) )
		\end{array}$ } \vspace{3.5\parskip} \par  \eqd{u29a}
		 explizite Skolemfunktionen
\par\vspace{-0.5\parskip}{\small$\begin{array}[t]{@{}r@{\;}l@{}}&
 \forall [ r,d, \alpha ,t]\; ( \\ &
   t \.\leq trv_{1}(r, \alpha ,t)
\\  \wedge  & t \.\leq trv_{2}(r, \alpha ,t)
\\  \wedge  & t \.\leq trz_{1}(r, \alpha ,t)
\\  \wedge  & t \.\leq trz_{2}(r, \alpha ,t)
\\  \wedge  & t \.\leq tre_{1}(r,d,t)
\\  \wedge  & t \.\leq tre_{2}(r,d,t)
\\  \wedge  & t \.\leq tra_{1}(r,d,t)
\\  \wedge  & t \.\leq tra_{2}(r,d,t) )
		\end{array}$ } \vspace{3.5\parskip} \par  \eqd{u30}
		  Motorsteuerung und Sensoren			\\
		  $c_{1}$: $+1$ Arm 1 aus,       $0$ Arm 1 stop	\\
		  $c_{2}$: $+1$ Arm 1 ein,       $0$ Arm 1 stop	\\
		  $c_{3}$: $+1$ Arm 2 aus,       $0$ Arm 2 stop	\\
		  $c_{4}$: $+1$ Arm 2 ein,       $0$ Arm 2 stop	\\
		  $c_{5}$: $+1$ Arm 1 greift			\\
		  $c_{6}$: $+1$ Arm 2 greift			\\
		  $c_{7}$: $+1$ dreht vor,      $0$ stop		\\
		  $c_{8}$: $+1$ dreht zur\"uck, $0$ stop		\\
		  $s_{1}$: L\"ange von Arm 1			\\
		  $s_{2}$: L\"ange von Arm 2			\\
		 $s_{3}$: Winkel von Arm 1
\par\vspace{-0.5\parskip}{\small$\begin{array}[t]{@{}r@{\;}l@{}}&
 \forall [ c_{1},c_{2},c_{3},c_{4},c_{5},c_{6},c_{7},c_{8},s_{1},s_{2},s_{3},x,t]\; ( \\ &
   roboter(r(c_{1},c_{2},c_{3},c_{4},c_{5},c_{6},c_{7},c_{8},s_{1},s_{2},s_{3}),x)
\\  \ra  & (faehrt\_aus_{1}(r(c_{1},c_{2},c_{3},c_{4},c_{5},c_{6},c_{7},c_{8},s_{1},s_{2},s_{3}),t)  \lra val(c_{1},t)=1)
\\  \wedge  & (faehrt\_ein_{1}(r(c_{1},c_{2},c_{3},c_{4},c_{5},c_{6},c_{7},c_{8},s_{1},s_{2},s_{3}),t)  \lra val(c_{2},t)=1)
\\  \wedge  & (faehrt\_aus_{2}(r(c_{1},c_{2},c_{3},c_{4},c_{5},c_{6},c_{7},c_{8},s_{1},s_{2},s_{3}),t)  \lra val(c_{3},t)=1)
\\  \wedge  & (faehrt\_ein_{2}(r(c_{1},c_{2},c_{3},c_{4},c_{5},c_{6},c_{7},c_{8},s_{1},s_{2},s_{3}),t)  \lra val(c_{4},t)=1)
\\  \wedge  & (greift_{1}(r(c_{1},c_{2},c_{3},c_{4},c_{5},c_{6},c_{7},c_{8},s_{1},s_{2},s_{3}),t)       \lra val(c_{5},t)=1)
\\  \wedge  & (greift_{2}(r(c_{1},c_{2},c_{3},c_{4},c_{5},c_{6},c_{7},c_{8},s_{1},s_{2},s_{3}),t)       \lra val(c_{6},t)=1)
\\  \wedge  & (dreht\_vor(r(c_{1},c_{2},c_{3},c_{4},c_{5},c_{6},c_{7},c_{8},s_{1},s_{2},s_{3}),t)     \lra val(c_{7},t)=1)
\\  \wedge  & (dreht\_zurueck(r(c_{1},c_{2},c_{3},c_{4},c_{5},c_{6},c_{7},c_{8},s_{1},s_{2},s_{3}),t)\lra val(c_{8},t)=1)
\\  \wedge  & (dist\_xy(x,pos_{1}(r(c_{1},c_{2},c_{3},c_{4},c_{5},c_{6},c_{7},c_{8},s_{1},s_{2},s_{3}),t))  =val(s_{1},t))
\\  \wedge  & (dist\_xy(x,pos_{2}(r(c_{1},c_{2},c_{3},c_{4},c_{5},c_{6},c_{7},c_{8},s_{1},s_{2},s_{3}),t))  =val(s_{2},t))
\\  \wedge  & (winkel\_xy(x,pos_{1}(r(c_{1},c_{2},c_{3},c_{4},c_{5},c_{6},c_{7},c_{8},s_{1},s_{2},s_{3}),t))=val(s_{3},t)))
\end{array}$ } \vspace{1.0cm} \par {\bf Hubdrehtisch
} \vspace{-0.3cm} \par {\small $\begin{array}[t]{@{}r@{\;}l@{}} &
		\end{array}$ } \vspace{3.5\parskip} \par  \eqd{u32}
		  Die H\"ohe einer auf dem Hubdrehtisch liegenden
		  Schiene bleibt unver\"andert, solange er weder
		 hebt noch senkt
\par\vspace{-0.5\parskip}{\small$\begin{array}[t]{@{}r@{\;}l@{}}&
 \forall [ h,t,x,s,t_{2}]\; ( \\ &
   hubdrehtisch(h,x)
\\  \wedge  & proj\_xy(ort(s,t))=proj\_xy(x)
\\  \wedge  & t \.\leq t_{2}
\\  \wedge  &  \forall [ m,t_{1}]\; (t \.\leq  t_{1}  \.\leq t_{2} \wedge  bewegt(m,s,t_{1}) \ra  m=h)
\\  \wedge  &  \forall [ t_{1}]\; (t \.\leq  t_{1}  \.\leq t_{2} \ra  \neg hebt(h,t_{1}) \wedge  \neg senkt(h,t_{1}))
\\  \ra  & proj\_z(ort(s,t))=proj\_z(ort(s,t_{2})) )
		\end{array}$ } \vspace{3.5\parskip} \par  \eqd{u33}
		  Der Winkel einer auf dem Hubdrehtisch liegenden
		  Schiene bleibt unver\"andert, solange er sich
		 nicht dreht
\par\vspace{-0.5\parskip}{\small$\begin{array}[t]{@{}r@{\;}l@{}}&
 \forall [ h,t,x,s,t_{2}]\; ( \\ &
   hubdrehtisch(h,x)
\\  \wedge  & proj\_xy(ort(s,t))=proj\_xy(x)
\\  \wedge  & t \.\leq t_{2}
\\  \wedge  &  \forall [ m,t_{1}]\; (t \.\leq  t_{1}  \.\leq t_{2} \wedge  bewegt(m,s,t_{1}) \ra  m=h)
\\  \wedge  &  \forall [ t_{1}]\; (   t \.\leq  t_{1}  \.\leq t_{2}
	     \ra  \neg dreht\_vor(h,t_{1})
	      \wedge  \neg dreht\_zurueck(h,t))
\\  \ra  & winkel(s,t)=winkel(s,t_{2}) )
		\end{array}$ } \vspace{3.5\parskip} \par  \eqd{u34}
		  Der Hubdrehtisch dreht eine auf ihm liegende
		 Schiene entsprechend mit
\par\vspace{-0.5\parskip}{\small$\begin{array}[t]{@{}r@{\;}l@{}}&
 \forall [ h,t,x,s,t_{2}]\; ( \\ &
   hubdrehtisch(h,x)
\\  \wedge  & proj\_xy(ort(s,t))=proj\_xy(x)
\\  \wedge  & t \.\leq t_{2}
\\  \wedge  &  \forall [ m,t_{1}]\; (t \.\leq  t_{1}  \.\leq t_{2} \wedge  bewegt(m,s,t_{1}) \ra  m=h)
\\  \ra  & winkel(s,t_{2})-winkel(s,t) = winkel(h,t_{2})-winkel(h,t) )
		\end{array}$ } \vspace{3.5\parskip} \par  \eqd{u35}
		  Wenn sich der Hubdrehtisch nur lange genug vordreht,
		  kann er jeden Winkel zwischen dem jetzigen und
		 360$^\circ$ erreichen
\par\vspace{-0.5\parskip}{\small$\begin{array}[t]{@{}r@{\;}l@{}}&
 \forall [ h,x,t, \alpha ]\;  \exists [ t_{2}]\; ( \\ &
   hubdrehtisch(h,x)
\\  \wedge  & winkel(h,t) \.\leq   \alpha   \.\leq 360
\\  \ra  & (    \forall [ t_{1}]\; (t \.\leq  t_{1} <t_{2} \ra  dreht\_vor(h,t_{1}))
    \ra  winkel(h,t_{2})= \alpha  )
\\  \wedge  &  \forall [ t_{3}]\; (t \.\leq  t_{3} <t_{2} \ra  \neg winkel(h,t_{3})= \alpha ) )
		\end{array}$ } \vspace{3.5\parskip} \par  \eqd{u36}
		  Wenn sich der Hubdrehtisch nur lange genug
		  zur\"uckdreht, kann er jeden Winkel zwischen dem
		 jetzigen und 0$^\circ$ erreichen
\par\vspace{-0.5\parskip}{\small$\begin{array}[t]{@{}r@{\;}l@{}}&
 \forall [ h,x,t, \alpha ]\;  \exists [ t_{2}]\; ( \\ &
   hubdrehtisch(h,x)
\\  \wedge  & 0 \.\leq   \alpha   \.\leq winkel(h,t)
\\  \ra  & (    \forall [ t_{1}]\; (t \.\leq  t_{1} <t_{2} \ra  dreht\_zurueck(h,t_{1}))
    \ra  winkel(h,t_{2})= \alpha  )
\\  \wedge  &  \forall [ t_{3}]\; (t \.\leq  t_{3} <t_{2} \ra  \neg winkel(h,t_{3})= \alpha ) )
		\end{array}$ } \vspace{3.5\parskip} \par  \eqd{u37}
		  Wenn sich der Hubdrehtisch nur lange genug hebt,
		 erreicht er schlie\3lich die obere Position
\par\vspace{-0.5\parskip}{\small$\begin{array}[t]{@{}r@{\;}l@{}}&
 \forall [ t]\;  \exists [ t_{2}]\;  \forall [ h,x]\; ( \\ &
   hubdrehtisch(h,x)
\\  \ra  & (    \forall [ t_{1}]\; (t \.\leq  t_{1}  \.\leq t_{2} \ra  hebt(h,t_{1}))
    \ra  hoehe(proj\_xy(x),t_{2})=zh\_oben )
\\  \wedge  &  \forall [ t_{3}]\; (t \.\leq  t_{3} <t_{2} \ra  \neg hoehe(proj\_xy(x),t_{3})=zh\_oben) )
		\end{array}$ } \vspace{3.5\parskip} \par  \eqd{u38}
		  Wenn sich der Hubdrehtisch nur lange genug senkt,
		 erreicht er schlie\3lich die untere Position
\par\vspace{-0.5\parskip}{\small$\begin{array}[t]{@{}r@{\;}l@{}}&
 \forall [ t]\;  \exists [ t_{2}]\;  \forall [ h,x]\; ( \\ &
   hubdrehtisch(h,x)
\\  \ra  & (    \forall [ t_{1}]\; (t \.\leq  t_{1}  \.\leq t_{2} \ra  senkt(h,t_{1}))
    \ra  hoehe(proj\_xy(x),t_{2})=zh\_unten )
\\  \wedge  &  \forall [ t_{3}]\; (t \.\leq  t_{3} <t_{2} \ra  \neg hoehe(proj\_xy(x),t_{3})=zh\_unten) )
		\end{array}$ } \vspace{3.5\parskip} \par  \eqd{u39}
		  Motorsteuerung und Sensoren			\\
		  $c_{1}$: $+1$ hebt,           $0$ steht		\\
		  $c_{2}$: $+1$ senkt,          $0$ steht		\\
		  $c_{3}$: $+1$ dreht vor,      $0$ steht		\\
		  $c_{4}$: $+1$ dreht zur\"uck, $0$ steht		\\
		  $s_{1}$: $+1$ unten				\\
		  $s_{2}$: $+1$ oben				\\
		 $s_{3}$: Winkel
\par\vspace{-0.5\parskip}{\small$\begin{array}[t]{@{}r@{\;}l@{}}&
 \forall [ c_{1},c_{2},c_{3},c_{4},s_{1},s_{2},s_{3},t,x]\; ( \\ &
   hubdrehtisch(h(c_{1},c_{2},c_{3},c_{4},s_{1},s_{2},s_{3}),x)
\\  \ra  & (hebt(h(c_{1},c_{2},c_{3},c_{4},s_{1},s_{2},s_{3}),t)           \lra  val(c_{1},t) = 1)
\\  \wedge  & (senkt(h(c_{1},c_{2},c_{3},c_{4},s_{1},s_{2},s_{3}),t)          \lra  val(c_{2},t) = 1)
\\  \wedge  & (dreht\_vor(h(c_{1},c_{2},c_{3},c_{4},s_{1},s_{2},s_{3}),t)      \lra  val(c_{3},t) = 1)
\\  \wedge  & (dreht\_zurueck(h(c_{1},c_{2},c_{3},c_{4},s_{1},s_{2},s_{3}),t) \lra  val(c_{4},t) = 1)
\\  \wedge  & (hoehe(proj\_xy(x),t)=zh\_unten             \lra  val(s_{1},t) =  1)
\\  \wedge  & (hoehe(proj\_xy(x),t)=zh\_oben              \lra  val(s_{2},t) =  1)
\\  \wedge  & (winkel(h(c_{1},c_{2},c_{3},c_{4},s_{1},s_{2},s_{3}),t) = val(s_{3},t)) )
\end{array}$ } \vspace{1.0cm} \par {\bf F\"orderband
} \vspace{-0.3cm} \par {\small $\begin{array}[t]{@{}r@{\;}l@{}} &
		\end{array}$ } \vspace{3.5\parskip} \par  \eqd{u40}
		  Wenn das F\"orderband nur lange genug l\"auft,
		  transportiert es eine auf ihm liegende Schiene
		 vom Anfang zum Ende
\par\vspace{-0.5\parskip}{\small$\begin{array}[t]{@{}r@{\;}l@{}}&
 \forall [ f,s,t,x_{1},x_{2}]\;  \exists [ t_{2}]\; ( \\ &
   foerderband(f,x_{1},x_{2})
\\  \wedge  & ort(s,t)=x_{1}
\\  \ra  & (    \forall [ t_{1}]\; (t \.\leq  t_{1} <t_{2} \ra  laeuft(f,t_{1}))
    \ra  ort(s,t_{2})=x_{2} )
\\  \wedge  &  \forall [ t_{3}]\; (t \.\leq  t_{3} <t_{2} \ra  \neg ort(s,t_{3})=x_{2}) )
		\end{array}$ } \vspace{3.5\parskip} \par  \eqd{u41}
		  Motorsteuerung und Sensoren	\\
		  $c_{1}$: $0$ steht, $+1$ l\"auft	\\
		 $s_{1}$:            $+1$ Ende
\par\vspace{-0.5\parskip}{\small$\begin{array}[t]{@{}r@{\;}l@{}}&
 \forall [ c_{1},s_{1},x_{1},x_{2},t]\; ( \\ &
   foerderband(f(c_{1},s_{1}),x_{1},x_{2})
\\  \ra  & (laeuft(f(c_{1},s_{1}),t)   \lra  val(c_{1},t) =  1)
\\  \wedge  & (( \exists [ s]\; ort(s,t)=x_{2}) \lra  val(s_{1},t) =  1) )
\end{array}$ } \vspace{1.0cm} \par {\bf Gesamte Fabrik
} \vspace{-0.3cm} \par {\small $\begin{array}[t]{@{}r@{\;}l@{}} &
		\end{array}$ } \vspace{3.5\parskip} \par  \eqd{u42}
		  Konfiguration der Fabrik			\\
		  $c_{1}$  Zuf\"uhrf\"orderband laufen		\\
		  $c_{2}$  Hubdrehtisch heben			\\
		  $c_{3}$  Hubdrehtisch senken			\\
		  $c_{4}$  Hubdrehtisch vor drehen		\\
		  $c_{5}$  Hubdrehtisch zur\"uck drehen		\\
		  $c_{6}$  Roboter Arm 1 aus			\\
		  $c_{7}$  Roboter Arm 1 ein			\\
		  $c_{8}$  Roboter Arm 2 aus			\\
		  $c_{9}$  Roboter Arm 2 ein			\\
		  $c_{10}$ Roboter Arm 1 greifen			\\
		  $c_{11}$ Roboter Arm 2 greifen			\\
		  $c_{12}$ Roboter vor drehen			\\
		  $c_{13}$ Roboter zur\"uck drehen		\\
		  $c_{14}$ Presse heben				\\
		  $c_{15}$ Presse senken				\\
		  $c_{16}$ Ablagef\"orderband laufen		\\
		  $s_{1}$  Zuf\"uhrf\"orderband (1:Ende)		\\
		  $s_{2}$  Hubdrehtisch Vertikal (1:unten)	\\
		  $s_{3}$  Hubdrehtisch Vertikal (1:oben)		\\
		  $s_{4}$  Hubdrehtisch Drehung (Winkel)		\\
		  $s_{5}$  Roboter Arm 1 (L\"ange)		\\
		  $s_{6}$  Roboter Arm 2 (L\"ange)		\\
		  $s_{7}$  Roboter Drehung (Winkel Arm 1)		\\
		  $s_{8}$  Presse (1:unten)			\\
		  $s_{9}$  Presse (1:mitte)			\\
		  $s_{10}$ Presse (1:oben)			\\
		  $s_{11}$ Ablagef\"orderband (1:Ende)		\\
		  $d_{1}$  Standort Zuf\"uhrf\"orderband Anfang	\\
		  $d_{2}$  Standort Zuf\"uhrf\"orderband Ende	\\
		  $d_{3}$  Standort Hubdrehtisch			\\
		  $d_{4}$  Standort Roboter			\\
		  $d_{5}$  Standort Presse			\\
		  $d_{6}$  Standort Ablagef\"orderband Anfang	\\
		 $d_{7}$  Standort Ablagef\"orderband Ende
\par\vspace{-0.5\parskip}{\small$\begin{array}[t]{@{}r@{\;}l@{}}&
 \forall [ c_{1},c_{2},c_{3},c_{4},c_{5},c_{6},c_{7},c_{8},c_{9},c_{10},c_{11},c_{12},c_{13},c_{14},c_{15},c_{16},
							        \\&\tab
    s_{1},s_{2},s_{3},s_{4},s_{5},s_{6},s_{7},s_{8},s_{9},s_{10},s_{11},x]\; ( \\ &
fabrik(fa(c_{1},c_{2},c_{3},c_{4},c_{5},c_{6},c_{7},c_{8},c_{9},c_{10},c_{11},c_{12},c_{13},c_{14},c_{15},c_{16},
							        \\&\tab
	  s_{1},s_{2},s_{3},s_{4},s_{5},s_{6},s_{7},s_{8},s_{9},s_{10},s_{11}),x)
\\  \lra  & foerderband(f(c_{1},s_{1}),x+d_{1},x+d_{2})
\\  \wedge  &  hubdrehtisch(h(c_{2},c_{3},c_{4},c_{5},s_{2},s_{3},s_{4}),x+d_{3})
\\  \wedge  &  roboter(r(c_{6},c_{7},c_{8},c_{9},c_{10},c_{11},c_{12},c_{13},s_{5},s_{6},s_{7}),x+d_{4})
\\  \wedge  &  presse(p(c_{14},c_{15},s_{8},s_{9},s_{10}),x+d_{5})
\\  \wedge  &  foerderband(f(c_{16},s_{11}),x+d_{6},x+d_{7})
\\  \wedge  &   \forall [ s,t]\;  \exists [ t_{2}]\; (   ort(s,t)=x+d_{1}
		        \wedge  bearbeitungszustand(s,t)=unbearbeitet	    \\&
		       \ra  ort(s,t_{2})=x+d_{7}
			\wedge  bearbeitungszustand(s,t_{2})=bearbeitet) )
		\end{array}$ } \vspace{3.5\parskip} \par  \eqd{u43}
		  Das Handhabungsger\"at ``verbraucht'' die Schienen
		  beim Zur\"uckf\"uhren, um einen geschlossenen
		  Kreislauf zu erm\"oglichen			\\
		  $c_{17}$ Handhabungsger\"at links		\\
		  $c_{18}$ Handhabungsger\"at rechts		\\
		  $c_{19}$ Handhabungsger\"at heben		\\
		  $c_{20}$ Handhabungsger\"at senken		\\
		  $c_{21}$ Handhabungsger\"at greifen		\\
		  $s_{12}$ Handhabungsger\"at Horizontal (1:links)\\
		  $s_{13}$ Handhabungsger\"at Horizontal (1:rechts)\\
		  $s_{14}$ Handhabungsger\"at Vertikal (H\"ohe)	\\
		  $d_{8}$  Standort Handhabungsger\"at Ende (links)\\
		 $d_{9}$  Standort Handhabungsger\"at Anfang (rechts)
\par\vspace{-0.5\parskip}{\small$
\begin{array}[t]{@{}r@{\;}l@{}}&
 \forall [ c_{17},c_{18},c_{19},c_{20},c_{21},s_{12},s_{13},s_{14},x]\; ( \\ &
    verbraucher(v(c_{17},c_{18},c_{19},c_{20},c_{21},s_{12},s_{13},s_{14}),x)
\\  \lra  & handhabungsgeraet(ha(c_{17},c_{18},c_{19},c_{20},c_{21},s_{12},s_{13},s_{14}),x+d_{8},x+d_{9})
\\  \wedge  &   \forall [ s,t]\;  \exists [ t_{2}]\; (   ort(s,t)=x+d_{9}
			\wedge  bearbeitungszustand(s,t)=bearbeitet	    \\&
		       \ra  ort(s,t_{2})=x+d_{8}
			\wedge  bearbeitungszustand(s,t_{2})=unbearbeitet) )
		\end{array}$ } \vspace{3.5\parskip} \par  \eqd{u44}
		 Nur die Presse bearbeitet Schienen,
\par\vspace{-0.5\parskip}{\small$\begin{array}[t]{@{}r@{\;}l@{}}&
 \forall [ s,t,t_{2}]\; ( \\ &
   bearbeitungszustand(s,t)=unbearbeitet
\\  \wedge  & bearbeitungszustand(s,t_{2})=bearbeitet
\\  \ra  &  \exists [ t_{1},p,x]\; (  t \.\leq  t_{1}  \.\leq t_{2}
		\wedge  presse(p,x)
		\wedge  bewegt(p,s,t)) )
		\end{array}$ } \vspace{3.5\parskip} \par  \eqd{u45}
		 Nur das Handhabungsger\"at ``verbraucht'' Schienen
\par\vspace{-0.5\parskip}{\small$\begin{array}[t]{@{}r@{\;}l@{}}&
 \forall [ s,t,t_{2}]\; ( \\ &
   bearbeitungszustand(s,t)=bearbeitet
\\  \wedge  & bearbeitungszustand(s,t_{2})=unbearbeitet
\\  \ra  &  \exists [ t_{1},h,x_{1},x_{2}]\; (  t \.\leq  t_{1}  \.\leq t_{2}
		    \wedge  handhabungsgeraet(h,x_{1},x_{2})
		    \wedge  bewegt(h,s,t)) )
		\end{array}$ } \vspace{3.5\parskip} \par  \eqd{u46}
		  Hubdrehtisch in oberer Position
		  und Presse in unterer Position
		 haben die gleiche H\"ohe
\par\vspace{-0.5\parskip}{\small$\begin{array}[t]{@{}r@{\;}l@{}}&
zh\_oben=zp\_mitte
		\end{array}$ } \vspace{3.5\parskip} \par  \eqd{u47}
		  Roboter und Presse stehen im richtigen Winkel
		 zueinander
\par\vspace{-0.5\parskip}{\small$\begin{array}[t]{@{}r@{\;}l@{}}&
winkel\_xy(d_{4},d_{5})=270
		\end{array}$ } \vspace{3.5\parskip} \par  \eqd{u47a}
		  Roboterarm 1 kann schneller eingezogen werden als
		 vom Hubdrehtisch zur Presse vorgedreht werden
\par\vspace{-0.5\parskip}{\small$\begin{array}[t]{@{}r@{\;}l@{}}&
 \forall [ r,t]\; ( \\ &
   roboter(r,d_{4})
\\  \wedge  & pos_{1}(r,t)=d_{3}
\\  \ra  & tre_{1}(r,dist\_xy(d_{4},d_{5}),t)<trv_{1}(r,winkel\_xy(d_{4},d_{5}),t) )
		\end{array}$ } \vspace{3.5\parskip} \par  \eqd{u47b}
		  Roboter und Hubdrehtisch stehen nahe genug
		 beieinander
\par\vspace{-0.5\parskip}{\small$\begin{array}[t]{@{}r@{\;}l@{}}&
dist\_xy(d_4,d_3) \leq maxlg_1
\end{array}$ } \vspace{1.0cm} \par {\bf Physik
} \vspace{-0.3cm} \par {\small $\begin{array}[t]{@{}r@{\;}l@{}} &
		\end{array}$ } \vspace{3.5\parskip} \par  \eqd{u48}
		 Schwerkraftgesetz
\par\vspace{-0.5\parskip}{\small$\begin{array}[t]{@{}r@{\;}l@{}}&
 \forall [ s,t]\; ( \\ &
proj\_z(ort(s,t))=hoehe(proj\_xy(ort(s,t)),t) )
		\end{array}$ } \vspace{3.5\parskip} \par  \eqd{u49}
		 Zwei Dinge sind nicht gleichzeitig am selben Ort
\par\vspace{-0.5\parskip}{\small$\begin{array}[t]{@{}r@{\;}l@{}}&
 \forall [ s_{1},s_{2},t]\; ( \\ &
   ort(s_{1},t)=ort(s_{2},t)
\\  \ra  & s_{1}=s_{2} )
		\end{array}$ } \vspace{3.5\parskip} \par  \eqd{u50}
		 Kein Ding bewegt sich von selbst
\par\vspace{-0.5\parskip}{\small$\begin{array}[t]{@{}r@{\;}l@{}}&
 \forall [ s,t,t_{2}]\; ( \\ &
   t \.\leq t_{2}
\\  \wedge  & (\neg ort(s,t)=ort(s,t_{2}) \vee  \neg winkel(s,t)=winkel(s,t_{2}))
\\  \ra  &  \exists [ m,t_{1}]\; bewegt(m,s,t_{1}) )
\end{array}$ } \vspace{1.0cm} \par {\bf Mathematik
} \vspace{-0.3cm} \par {\small $\begin{array}[t]{@{}r@{\;}l@{}} &
		\end{array}$ } \vspace{3.5\parskip} \par  \eqd{u51}
		 Dreidimensionaler Raum
\par\vspace{-0.5\parskip}{\small$\begin{array}[t]{@{}r@{\;}l@{}}&
 \forall [ x,x_{1}]\; ( \\ &
   proj\_xy(x)=proj\_xy(x_{1})
\\  \wedge  & proj\_z(x)=proj\_z(x_{1})
\\  \ra  & x=x_{1} )
		\end{array}$ } \vspace{3.5\parskip} \par  \eqd{u62}
		 Gleichheit
\par\vspace{-0.5\parskip}{\small$\begin{array}[t]{@{}r@{\;}l@{}}&
 \forall [ aa]\; (aa=aa)
		\end{array}$ } \vspace{3.5\parskip} \par  \eqd{u66}
		 Lineare Ordnungsrelation
\par\vspace{-0.5\parskip}{\small$\begin{array}[t]{@{}r@{\;}l@{}}&
 \forall [ aa,bb]\; (\neg  bb \.\leq aa \lra  aa<bb)
		\end{array}$ } \vspace{3.5\parskip} \par  \eqd{u66b}
\par\vspace{-0.5\parskip}{\small$\begin{array}[t]{@{}r@{\;}l@{}}&
 \forall [ aa,bb]\; (aa<bb \ra  aa \.\leq bb)
		\end{array}$ } \vspace{3.5\parskip} \par  \eqd{u66c}
\par\vspace{-0.5\parskip}{\small$\begin{array}[t]{@{}r@{\;}l@{}}&
 \forall [ aa,bb,cc]\; (aa< bb <cc \ra  aa<cc)
		\end{array}$ } \vspace{3.5\parskip} \par  \eqd{u66d}
\par\vspace{-0.5\parskip}{\small$\begin{array}[t]{@{}r@{\;}l@{}}&
 \forall [ aa,bb,cc]\; (aa \.\leq  bb <cc \ra  aa<cc)
		\end{array}$ } \vspace{3.5\parskip} \par  \eqd{u66f}
\par\vspace{-0.5\parskip}{\small$\begin{array}[t]{@{}r@{\;}l@{}}&
 \forall [ aa,bb,cc]\; (aa< bb  \.\leq cc \ra  aa<cc)
		\end{array}$ } \vspace{3.5\parskip} \par  \eqd{u66e}
\par\vspace{-0.5\parskip}{\small$\begin{array}[t]{@{}r@{\;}l@{}}&
 \forall [ aa,bb,cc]\; (aa \.\leq  bb  \.\leq cc \ra  aa \.\leq cc)
		\end{array}$ } \vspace{3.5\parskip} \par  \eqd{u61}
		 Archimedischer K\"orper
\par\vspace{-0.5\parskip}{\small$\begin{array}[t]{@{}r@{\;}l@{}}&
 \forall [ aa,bb]\; (aa*(-1) < bb*(-1) \lra  bb < aa)
		\end{array}$ } \vspace{3.5\parskip} \par  \eqd{u67}
		 Polarkoordinaten
\par\vspace{-0.5\parskip}{\small$\begin{array}[t]{@{}r@{\;}l@{}}&
 \forall [ x,x_{1},x_{2}]\; ( \\ &
   winkel\_xy(x_{2},x)=winkel\_xy(x_{2},x_{1})
\\  \wedge  & dist\_xy(x_{2},x)=dist\_xy(x_{2},x_{1})
\\  \ra  & proj\_xy(x)=proj\_xy(x_{1}) )
		\end{array}$ } \vspace{3.5\parskip} \par  \eqd{u69}
		 Arithmetik
\par\vspace{-0.5\parskip}{\small$\begin{array}[t]{@{}r@{\;}l@{}}&
270-90=180
		\end{array}$ } \vspace{3.5\parskip} \par  \eqd{u69a}
		 Aussagenlogik
\par\vspace{-0.5\parskip}{\small$\begin{array}[t]{@{}r@{\;}l@{}}&
aaa \wedge  bbb \ra  aaa
		\end{array}$ } \vspace{3.5\parskip} \par  \eqd{u69c}
\par\vspace{-0.5\parskip}{\small$\begin{array}[t]{@{}r@{\;}l@{}}&
aaa \wedge  bbb \ra  bbb
		\end{array}$ } \vspace{3.5\parskip} \par  \eqd{u69d}
\par\vspace{-0.5\parskip}{\small$\begin{array}[t]{@{}r@{\;}l@{}}&
aaa \ra  (\neg  aaa \ra  bbb)
\end{array}$ } \vspace{1.0cm} \par {\bf Steuerung
} \vspace{-0.3cm} \par {\small $\begin{array}[t]{@{}r@{\;}l@{}} &
		\end{array}$ } \vspace{3.5\parskip} \par  \eqd{u70}
		 Steigende Flanke
\par\vspace{-0.5\parskip}{\small$\begin{array}[t]{@{}r@{\;}l@{}}&
 \forall [ c,t]\; ( \\ &
    up(c,t)
\\  \lra  & val(c,t)=1
\\  \wedge  &   \exists [ t_{1}]\; (t_{1}<t \wedge   \forall [ t_{2}]\; (t_{1}< t_{2} <t \ra  \neg  val(c,t_{2})=1)) )
		\end{array}$ } \vspace{3.5\parskip} \par  \eqd{u71}
		  Flanken--getriggertes Flip--Flop	{\tt	\\
		  \verb" ca  ___------------___ "		\\
		  \verb" cb  ___________--_____ "		\\
		 \verb" out ___--------_______ "		}
\par\vspace{-0.5\parskip}{\small$\begin{array}[t]{@{}r@{\;}l@{}}&
 \forall [ ca,cb,t_{2}]\; ( \\ &
    val(dff(ca,cb),t_{2})=1
\\  \lra  &  \exists [ t]\; (   t \.\leq t_{2}
	     \wedge  up(ca,t)
	     \wedge   \forall [ t_{1}]\; (t \.\leq  t_{1} <t_{2} \ra  \neg up(cb,t_{1}))) )
		\end{array}$ } \vspace{3.5\parskip} \par  \eqd{u72}
		  Zeitverz\"ogerung			{\tt	\\
		  \verb" c   ___---____________ "		\\
		  \verb" out ____________---___ "		\\
		  \verb"        |        |      "		\\
		 \verb"        t        t+d    "		}
\par\vspace{-0.5\parskip}{\small$\begin{array}[t]{@{}r@{\;}l@{}}&
 \forall [ c,d,t]\; val(mff(c,d),t+d)=val(c,d)
		\end{array}$ } \vspace{3.5\parskip} \par  \eqd{u73}
		  Trigger				{\tt	\\
		  \verb" c   _-_   _-_   _-	  ___ v ___"	\\
		  \verb"        -_-   -_-    "			\\
								\\
		 \verb" out ___---___---___ "			}
\par\vspace{-0.5\parskip}{\small$\begin{array}[t]{@{}r@{\;}l@{}}&
 \forall [ c,v,t]\; ( \\ &
    val(trigger(c,v),t)=1
\\  \lra  & val(c,t)<v )
		\end{array}$ } \vspace{3.5\parskip} \par  \eqd{u74}
		  Inverter				{\tt	\\
		  \verb" c   ___---___--- "			\\
		 \verb" out ---___---___ "			}
\par\vspace{-0.5\parskip}{\small$\begin{array}[t]{@{}r@{\;}l@{}}&
 \forall [ c,t]\; ( \\ &
    val(neg(c),t)=1
\\  \lra  & \neg  val(c,t) = 1 )
		\end{array}$ } \vspace{3.5\parskip} \par  \eqd{u75}
		  Oder--Gatter				{\tt	\\
		  \verb" ca  ______------ "			\\
		  \verb" cb  ___---___--- "			\\
		 \verb" out ___--------- "			}
\par\vspace{-0.5\parskip}{\small$\begin{array}[t]{@{}r@{\;}l@{}}&
 \forall [ ca,cb,t]\; ( \\ &
    val(or(ca,cb),t)=1
\\  \lra  & val(ca,t) = 1
  \vee  val(cb,t) = 1)
		\end{array}$ } \vspace{3.5\parskip} \par  \eqd{u76}
		  Und--Gatter				{\tt	\\
		  \verb" ca  ______------ "			\\
		  \verb" cb  ___---___--- "			\\
		 \verb" out _________--- "			}
\par\vspace{-0.5\parskip}{\small$\begin{array}[t]{@{}r@{\;}l@{}}&
 \forall [ ca,cb,t]\; ( \\ &
    val(and(ca,cb),t)=1
\\  \lra  & val(ca,t) = 1
\\  \wedge  &  val(cb,t) = 1)
		\end{array}$ } \vspace{3.5\parskip} \par  \eqd{u77}
		 Verst\"arker
\par\vspace{-0.5\parskip}{\small$\begin{array}[t]{@{}r@{\;}l@{}}&
 \forall [ c,v,t]\; (val(ampl(c,v),t) = val(c,t)*v)
\end{array}$ }
\end{itemize}

\begin{itemize}
\item[]
{\small $\begin{array}[t]{@{}r@{\;}l@{}}&
\end{array}$ } \vspace{1.0cm} \par {\bf Modul 1 der Robotersteuerung
} \vspace{-0.3cm} \par {\small $\begin{array}[t]{@{}r@{\;}l@{}} &
		\end{array}$ } \vspace{3.5\parskip} \par  \eqd{r1}
\par\vspace{-0.5\parskip}{\small$\begin{array}[t]{@{}r@{\;}l@{}}&
 up(ci,t_0)	\\
		\end{array}$ } \vspace{3.5\parskip} \par  \eqd{r2}
\par\vspace{-0.5\parskip}{\small$\begin{array}[t]{@{}r@{\;}l@{}}&
 ort(s_0,t_0)=d_3	\\
		\end{array}$ } \vspace{3.5\parskip} \par  \eqd{r3}
\par\vspace{-0.5\parskip}{\small$\begin{array}[t]{@{}r@{\;}l@{}}&
pos_1(r,t_0)=d_3	\\
		\end{array}$ } \vspace{3.5\parskip} \par  \eqd{r4}
\par\vspace{-0.5\parskip}{\small$\begin{array}[t]{@{}r@{\;}l@{}}&
winkel(s_0,t_0)=winkel\_xy(d_4,d_3)-90	\\
		\end{array}$ } \vspace{3.5\parskip} \par  \eqd{r6}
\par\vspace{-0.5\parskip}{\small$\begin{array}[t]{@{}r@{\;}l@{}}&
dist\_xy(d_4,d_5) \leq dist\_xy(d_4,pos_1(r,t_0))	\\
		\end{array}$ } \vspace{3.5\parskip} \par  \eqd{r7}
\par\vspace{-0.5\parskip}{\small$\begin{array}[t]{@{}r@{\;}l@{}}&
minlg_1 \leq dist\_xy(d_4,d_5)	\\
		\end{array}$ } \vspace{3.5\parskip} \par  \eqd{r8}
\par\vspace{-0.5\parskip}{\small$\begin{array}[t]{@{}r@{\;}l@{}}&
winkel\_xy(d_4,d_5) \leq 270	\\
		\end{array}$ } \vspace{3.5\parskip} \par  \eqd{r9}
\par\vspace{-0.5\parskip}{\small$\begin{array}[t]{@{}r@{\;}l@{}}&
winkel\_xy(d_4,pos_1(r,t_0)) \leq winkel\_xy(d_4,d_5)	\\
		\end{array}$ } \vspace{3.5\parskip} \par  \eqd{r10}
\par\vspace{-0.5\parskip}{\small$\begin{array}[t]{@{}r@{\;}l@{}}&
\forall[t,r] \;	\\
& (   t_0 \leq t \leq trv_1(r,270,t_0)	\\
\ra & \neg val(cfa_1,t)=1 \wedge \neg val(cfe_1,t)=1	\\
\wedge & \neg val(cdrv,t)=1 \wedge \neg val(cgr_1,t)=1 )	\\
		\end{array}$ } \vspace{3.5\parskip} \par  \eqd{r11}
\par\vspace{-0.5\parskip}{\small$\begin{array}[t]{@{}r@{\;}l@{}}&
roboter(r(c_1,c_2,c_3,c_4,c_5,c_6,c_7,c_8,s_1,s_2,s_3),d_4)	\\
		\end{array}$ } \vspace{3.5\parskip} \par  \eqd{r20}
BEWEISZIEL:
\par\vspace{-0.5\parskip}{\small$\begin{array}[t]{@{}r@{\;}l@{}}&
proj\_xy(pos_1(r,t))=proj\_xy(d_5)	\\
\wedge & ort(s_0,t)=pos_1(r,t)	\\
\wedge & winkel(s_0,t)=180	\\
\wedge & up(co,t)	\\
\end{array}$ }
\end{itemize}

\clearpage

\addtocontents{toc}
{\protect\contentsline {section}{\protect\numberline{C}%
Deduktive Synthese eines einzelnen Steuerungsmoduls}{\thepage}}

\newpage

$\;$

\newpage

\scriptsize

\section{Deduktive Synthese eines einzelnen Steuerungsmoduls}

%

\end{document}